\documentclass{aa}  

\usepackage{graphicx}

\usepackage{txfonts}

\usepackage{hyperref} 
\hypersetup{colorlinks = true,  
                linkcolor = blue,
                filecolor = magenta,
                urlcolor = cyan,
                citecolor = blue
            }

\usepackage{amssymb}
\usepackage{natbib} 
\bibpunct{(}{)}{;}{a}{}{,} 
\usepackage{multirow, multicol}
\usepackage[table,xcdraw]{xcolor}
\usepackage{wrapfig}
\usepackage{subcaption}
\usepackage{hhline}
\usepackage[labelfont = bf]{caption}

\usepackage{lscape}

\begin{document}

   \title{Gas-phase Elemental abundances in Molecular cloudS (GEMS) }
   \subtitle{XI. The evolution of HCN, HNC, and N$_2$H$^+$ isotopic ratios in starless cores}

   \author{A. Tasa-Chaveli\inst{1}
          \and
          A. Fuente\inst{1}
          \and
          G. Esplugues\inst{2}
          \and
          D. Navarro-Almaida\inst{1}
          \and
          Liton Majumdar\inst{3,4}
          \and
          Prathap Rayalacheruvu\inst{3,4}
          \and
          P. Rivi{\`e}re-Marichalar\inst{2}
                    \and
          M. Rodr{\'i}guez-Baras\inst{2}
}

   \institute{Centro de Astrobiolog{\'i}a (CSIC-INTA), Ctra. de Ajalvir, km 4, Torrej{\'o}n de Ardoz, 28850 Madrid, Spain 
   \and
   Observatorio Astron{\'o}mico Nacional (OAN), Alfonso XII, 3, 28014 Madrid, Spain
   \and Exoplanets and Planetary Formation Group, School of Earth and Planetary Sciences, National Institute of Science Education and Research, Jatni 752050, Odisha, India
   \and
   Homi Bhabha National Institute, Training School Complex, Anushaktinagar, Mumbai 400094, India
}

   \date{Received }
 
  \abstract 
   {Isotopic ratios have been used as chemical diagnostics to investigate the origin of the material in the Solar System. These isotopic ratios depend on the physical conditions at formation but can be altered during the star formation process through different physical and chemical processes.}
   {Our aim is to determine the HCN, HNC, and N$_2$H$^+$ isotopic ratios and the chemical age in a large sample of starless cores located in different environments.
   }
   {This work uses IRAM 30m data to constrain
the D/H isotopic ratios of HCN, HNC, and N$_2$H$^+$ as well as the $^{14}$N/$^{15}$N ratio of HCN and HNC. We also modeled the deuterium fractions with the chemical code \texttt{DNAUTILUS 2.0}.}
   {Deuterated compounds are detected in all of our sample cores, with average DNC/HNC, DCN/HCN, and N$_2$D$^+$/N$_2$H$^+$ values of 0.054$\pm$0.019, 0.036$\pm$0.033, and 0.15$\pm$0.11, respectively. The deuterium fractions (D$_{frac}$) show a weak correlation with temperature and a large scatter that reflects that other factors such as core evolution could also play a significant role. Our chemical model is able to reproduce all the observed values with 0.2-0.3 Myr in Taurus and 0.3-0.5 Myr in Perseus and Orion. The $^{14}$N/$^{15}$N isotopic ratio is found to be different between HCN/HC$^{15}$N (430$\pm$120) and HNC/H$^{15}$NC (296$\pm$64). 
   We find no correlation between these ratios and the deuterium fractions, but we report a weak correlation with temperature was found.}
   {The deuterium fractions of  HCN, HNC, and N$_2$H$^+$ can be used as evolutionary tracers of starless cores as long as the physical parameters are well constrained. The HCN/HC$^{15}$N and HNC/H$^{15}$NC ratios are not correlated with D$_{frac}$, suggesting that the detected variations are not correlated with the core evolutionary stage. The average value of the  HCN/HC$^{15}$N ratio in our sample is significantly higher than the values measured in protostars and protoplanetary disks, 
   possibly indicating that nitrogen fractionation processes are taking place during the protostellar phase. }


   \maketitle

\section{Introduction} \label{sect:1}

Isotopic ratios provide invaluable insights into the connection between Solar System objects and the Galactic interstellar medium \citep{Aleon2010}. One example of the great astrobiological interest in isotopic ratios concerns the elevated HDO/H$_2$O ratio measured in the terrestrial oceans (Vienna Standard Mean Ocean Water  = 1.5 $\times$ 10$^{-4}$; \citealp{Lecuyer1998}), which is one order of magnitude larger than the cosmic value (D/H = 1.6 $\times$ 10$^{-5}$, \citealp{Linsky2007}). This discrepancy has been interpreted as being clear evidence that at least part of Earth's water was delivered by comets from the outer Solar System, bearing a chemical composition akin to interstellar grains in starless cores. The $^{14}$N/$^{15}$N ratio has also been proposed as a tracer of the evolution of material from molecular clouds to planetary systems. Different $^{14}$N/$^{15}$N values have been found in the Solar System: $\sim$440 \citep{Owen2001, Fouchet2004} in the solar wind and Jupiter (also considered as a representative ratio of the protosolar nebula), $\sim$270 on Earth \citep{Aleon2010}, and $\sim$136 in comets \citep{comets-Nratio}. From this, we may infer that there was an enrichment of $^{15}$N during the formation of the Solar System \citep{EvelynFury2015}. Despite the increasing number of observations, the origin of this enrichment is still unknown, having found a wide range of $^{14}$N/$^{15}$N ratios through different molecules: 190-783 for HNC, HCN, CN, or 180-1300 for  N$_2$H$^+$ in high-mass star-forming cores at different galactocentric distances \citep{Colzi2018, Fontani2015}. Understanding the chemical processes that alter these ratios is a requisite to disentangling the intricate physical and chemical processes occurring during the formation of a low-mass star and to eventually determining the origin of these molecules in the Solar System.

Deuterium fractionation is widely used as a chemical diagnostic to establish the evolutionary stage of starless cores, i.e., dense ($>$ 10$^4$ cm$^{-3}$) and cold ($\sim$ 10 K) regions of molecular gas without compact luminous sources, where stars may eventually be formed (\citealp{Crapsi&Caselli}, \citealp{Navarro_Linking}, \citealp{GEMS-VI-GiselaH2CS}, \citealp{GEMS-IX-MarinaH2S}). At these temperatures, deuterium fractionation is primarily driven by the reaction of H$_{3}^{+}$ with HD to form H$_2$D$^+$ and H$_2$ \citep{Millar1989}. The formation of H$_2$D$^+$ proceeds without a reaction barrier, whereas the backward reaction has an endothermicity of $\sim$232K when H$_2$ is in para J = 0 level, and therefore it does not proceed at these cold temperatures, as far as para-H$_2$ is concerned \citep{Watson1974, Caselli2012, Albertsson2013}. The backward reaction can occur, and deuterium fractionation does not proceed if the ortho-H$_2$ makes a significant contribution \citep{Pagani2009}, but the ortho-to-para ratio is less than 0.01 in cold molecular clouds \citep[][]{Dislaire2012}. Therefore, at these
low temperatures there is an enhancement of H$_2$D$^+$ abundance and the H$_2$D$^+$/H$_{3}^{+}$ ratio becomes larger than the D/H elemental abundance ratio \citep{Pagani1992}. Successive reactions of H$_2$D$^+$ with HD form D$_2$H$^{+}$ and D$_{3}^{+}$, and all of these ions transfer deuterium to other molecules in the gas phase. The enhanced atomic D/H ratio in the gas phase is also transferred to grains, promoting the deuteration of the molecules formed on their surfaces such as CH$_3$OH, H$_2$O, and H$_2$S \citep{Ambrose2021,GEMS-IX-MarinaH2S}. 
Although H$_3^+$ is the main ion driving deuteration at low temperatures, HD may also react with CH$^{+}_3$ and C$_2$H$^{+}_2$ since it is the main pathway for deuteration at high temperatures. The endothermicity of the reverse reactions is much larger \citep{Parise2009}, which leads to large enhancements of deuterated ions \citep[e.g., CH$_2$D$^+$, C$_2$HD$^+$,][]{Millar2005, Roueff2007, Roueff2015}, and this could also transfer deuterium to other compounds.  Another important factor in deuterium fractionation is given by CO depletion. CO plays a significant role in regulating the H$_3^+$ abundance through destruction reactions. In cold dense regions, CO freezes onto dust grains, leading to an increase in the H$_{3}^{+}$ abundance and enhancing deuterium fractionation \citep{Dalgarno1984, Caselli2002, Roberts2003}. 

The efficiency of the deuterium enrichment under these conditions is now well established with the detection of single and multiply deuterated molecules in starless cores and young protostars \citep{Roueff2000,  Parise2002, Bacmann2003, Melosso2020}. 
Although there seems to be a general trend with deuteration increasing at low temperatures for all species in cold regions, different species probe different regions along the line of sight
and probe different layers of the molecular cloud. In particular, N-bearing species such as NH$_3$, N$_2$H$^+$, and nitriles (HCN, HNC, CN) are still abundant in regions where most molecules are highly depleted \citep{Tafalla2006, Sipila2018, Kim2020}. The observation of N-bearing species and their deuterated compounds are therefore expected to provide essential information on the physical conditions and the evolutionary stage of starless cores. 

The dominant mechanism for the nitrogen fractionation in the interstellar medium is not clear. Similar to D-enrichment,
isotopic exchange reactions at low temperatures can produce $^{15}$N-enrichment \citep{Rodgers2008}. However, this mechanism has been challenged in the case of N$_2$H$^+$ because of the barriers between the interacting molecules found in theoretical works \citep{Roueff2015}. This is not consistent with observational results toward starless cores, and later studies have proposed that the isotope-selective photodissociation might be the key \citep{Spezzano}. In this case, $^{14}$N$_2$ self-shielding is more effective against photodissociation than that of $^{14}$N$^{15}$N, leading to a higher availability of free $^{15}$N to form $^{15}$N-bearing molecules.
Comparison between chemical models and observations of nitriles (CN, HCN, HNC) is difficult because carbon and nitrogen chemistry are interdependent. Observed $^{14}$N/$^{15}$N isotopic ratios are usually based on the observation of the $^{13}$C isotopologs to avoid opacity problems. This implies assuming a fixed $^{12}$C/$^{13}$C ratio, with  $^{12}$C/$^{13}$C  =  68 being the canonical value \citep{12C-13Cratio}. However, this value is not held for all molecular compounds because isotopic fractionation reactions change it \citep{Roueff2015}. For example, the ratio of HCCCN/H$^{13}$CCCN is 79 \citep{Takano1998}, CCS/$^{13}$CCS is 230 \citep{Sakai2007}, and CCH/$^{13}$CCH is greater than 250 \citep{Sakai2010}. The isotopic $^{12}$C/$^{13}$C and $^{14}$N/$^{15}$N ratios both need to be known to be able to interpret observations.

In this paper, we use data from the IRAM
30m large program Gas phase Elemental abundances in Molecular CloudS
\citep[GEMS; PI Asunción Fuente,][]{GEMS-I-Asun} observations to investigate the deuterium fraction of HCN, HNC, and N$_2$H$^+$ as well as the  $^{12}$C/$^{13}$C and $^{14}$N/$^{15}$N isotopic ratios in a wide sample of starless cores. In particular, we selected 23 low-mass cores out of the GEMS sample located in three different molecular clouds: Taurus, Perseus, and Orion. Additional observations were taken with the IRAM 30m telescope to complete the dataset in these sources.  We did not consider in our study TMC 1- CP, TMC1-C, and  NGC 1333-C7-1, which are also included in GEMS,  because these objects were previously studied by \citealp{Navarro2021} and  \citealp{Navarro_Linking}. The paper is organized as follows: The full observational dataset is described in Sect.~\ref{sect:2}. In Sect.~\ref{sect:3} the sample is briefly described, while in Sect.~\ref{sect:4} we describe the density of H$_2$ and molecular column densities derivation. Sect. \ref{sect:5} provides our isotopic ratio results. We discuss the relationship between the observed isotopic ratios and the evolutionary stage and the cores of our sample as well as their chemical and physical histories in Sect.~\ref{sect:6}. Finally, we summarize our conclusions in Sect.~\ref{sect:conclusions}.

\section{Observations} \label{sect:2}

This work is based on two IRAM 30m observational projects\footnote{The data underlying this article are available at \url{https://github.com/atasa-chaveli/aa54121-25/}}: GEMS\footnote{GEMS home page: \url{https://www.oan.es/gems/doku.php?id = start}} (No. 006-17) and the project  011-20. The first one is an IRAM 30m Large Program run between summer 2017 and 2019. A detailed description of these observations is given by \citealp{GEMS-I-Asun}. Intense lines of more than  12 species lied  in the GEMS setup. Among them,  the ground-state J  =  1$\rightarrow$0 lines of HCN, H$^{13}$CN, HC$^{15}$N, HNC, H$^{15}$NC and N$_2$H$^+$  that we are using. Observations of the project No. 011-20  were carried out in 2020 with the IRAM 30m telescope.These  observations targeted the  HN$^{13}$C J  =  1$\rightarrow$0, 
DN$^{13}$C J  =  1$\rightarrow$0, DNC  J  =  1$\rightarrow$0 and  J  =  2$\rightarrow$1, DCN  J  =  1$\rightarrow$0, N$_2$H$^+$  J  =  1$\rightarrow$0, and N$_2$D$^+$  J  =  2$\rightarrow$1
lines. Table \ref{tab:transitions} shows the observed transitions and their spectroscopic information that were taken from the CDMS catalog \citep{CDMS_Muller2001, CDMS_Moller2005, CDMS}) except for H$^{15}$NC and HN$^{13}$C 1-0 lines (obtained from the JPL catalog, \citealp{JPL}) and DN$^{13}$C 1-0 (from SLAIM, \citealp{SLAIM}). 

For both projects, the Eight MIxer Receivers (EMIR) and the Fourier Transform Spectrometer (49 kHz resolution) were deployed using frequency-switching observing mode for 2 and 3 mm observations. Main-beam temperature scale T$_{\rm MB}$ is the intensity scale, related with T$^*_{\rm A}$ by T$_{\rm MB}$  =  (F$_{\rm eff}$/B$_{\rm eff}$) T$^*_{\rm A}$. The telescope forward and main beam efficiencies F$_{\rm eff}$, B$_{\rm eff}$, as well as the beam width at the frequencies of interest are also included in Table \ref{tab:transitions}. They were obtained by interpolating 
the efficiencies and the beam widths measured by EMIR in 2009 \footnote{IRAM 30m efficiencies and beam widths: \url{https://publicwiki.iram.es/Iram30mEfficiencies}} since they vary as $\theta_{\rm MB}('') = (2460/\nu (GHz))$.

Data reduction was carried out using the CLASS package from IRAM GILDAS software.\footnote{GILDAS home page: \url{https://www.iram.fr/IRAMFR/GILDAS/}} We used the Ruze equation which is included in this package, so as to get the conversion between the before-mentioned temperatures. 

\begin{table*}[h!]  
\caption[l]{Source sample: Low-mass cores studied in GEMS and 011-20 projects.} \label{tab:source}
\centering
\begin{tabular}{lllccccc}
\hhline{ =  =  =  =  =  =  =  = }
MC Complex & Filament & Core & \multicolumn{2}{c}{Coordinates} &  T$_{\rm k}$\tablefootmark{a}  & A$_{\nu}$\tablefootmark{a} & n(H$_2$)\tablefootmark{b}
\\
 &  &  & RA (J2000) & Dec (J2000) & (K) & (mag) & ($\times 10^5$ cm$^{-3}$) \\ \hline
\multirow{8}{*}{Taurus} & \multicolumn{1}{l}{B213} & C1 & \multicolumn{1}{c}{04:17:41.8} & +28:08:47.0 &10.9     &26.9 & 0.6 $\pm$ 0.1\\  
 & \multicolumn{1}{l}{} & C2-1 & \multicolumn{1}{c}{04:17:50.6} & +27:56:01.0 &11.0  &        20.9 & 0.8 $\pm$ 0.2\\  
 & \multicolumn{1}{l}{} & C5-1 & \multicolumn{1}{c}{04:18:03.8} & +28:23:06.0 & 11.9 &        23.6 & 1.0 $\pm$ 0.2 \\  
 & \multicolumn{1}{l}{} & C6-1 & \multicolumn{1}{c}{04:18:08.4} & +28:05:12.0  &10.9 &        22.2 & 1.2 $\pm$ 0.3\\  
 & \multicolumn{1}{l}{} & C7-1 & \multicolumn{1}{c}{04:18:11.5} & +27:35:15.0 & 11.0 &        20.2 & 0.6 $\pm$ 0.4\\  
 & \multicolumn{1}{l}{} & C10-1 & \multicolumn{1}{c}{04:19:37.6} & +27:15:31.0 & 11.2 &        20.7 & 1.2 $\pm$ 0.4\\  
 & \multicolumn{1}{l}{} & C12-1 & \multicolumn{1}{c}{04:19:51.7} & +27:11:33.0 & 10.6 &        22.1 & 1.0 $\pm$ 0.3 \\  
 & \multicolumn{1}{l}{} & C16-1 & \multicolumn{1}{c}{04:21:21.0} & +27:00:09.0 & 10.3 &        24.8 & 1.0 $\pm$ 0.3\\ \hline
\multirow{12}{*}{Perseus} & L1448 & C1 & \multicolumn{1}{c}{03:25:49.0} & +30:42:24.6 & 15.1      & 28.4 & 0.9 $\pm$ 0.3\\  
 & NGC 1333 & C1-1 & \multicolumn{1}{c}{03:29:18.2} & +31:28:02.0  & 17.7 & 5.3 & 1.9 $\pm$ 0.6\\  
 &  & C2-1 & \multicolumn{1}{c}{03:29:08.6} & +31:06:02.0  & 14.8 &     3.8 & 2.1 $\pm$ 0.7\\  
 &  & C3-1 & \multicolumn{1}{c}{03:29:11.0} & +31:18:27.4  & 17.7 &     30.3 & 3 $\pm$ 1 \\  
 &  & C3-14 & \multicolumn{1}{c}{03:28:39.4} & +31:18:27.4 & 15.9 &     26.8 & 1.2 $\pm$ 0.4\\   
 &  & C4-1 & \multicolumn{1}{c}{03:29:08.8} & +31:15:18.1  & 16.8 &     39.9  & 2.1 $\pm$ 0.9\\  
 &  & C5-1 & \multicolumn{1}{c}{03:29:04.5} & +31:20:59.1  & 19.2 &  27.5  & 3 $\pm$ 1 \\  
 &  & C6-1 & \multicolumn{1}{c}{03:29:18.2} & +31:25:10.8  & 19.9 &     16.1 & 5 $\pm$ 3 \\  
 &  & C7-1 & \multicolumn{1}{c}{03:29:25.5} & +31:28:18.1  & 17.3 &     17.6 & 2.6 $\pm$ 0.8\\  
 & Barnard 5 & 79-C1-1 & \multicolumn{1}{c}{03:47:38.9} & +32:52:15.0 & 15.4    & 19.9 & 2.2 $\pm$ 0.9\\  
 & IC 348 & C1-1 & \multicolumn{1}{c}{03:44:01.0} & +32:01:54.8  &21.7 &        21.8 & 3 $\pm$ 1 \\  
 &  & C1-10 & \multicolumn{1}{c}{03:44:05.7} & +32:01:53.5  &17.5 & 20.1 & 0.9 $\pm$ 0.3 \\ \hline
\multirow{3}{*}{Orion} & Orion A & C1-2 & \multicolumn{1}{c}{05:35:19.5} & $-$05:00:41.5 & 18.1& 173.0 & 4 $\pm$ 2\\  
 &  & C2-3 & \multicolumn{1}{c}{05:35:08.1} & $-$05:35:41.5 & 22.8 & 56.5 & 1.3 $\pm$ 0.3\\
 &  & C3-1 & \multicolumn{1}{c}{05:35:23.6} & $-$05:12:31.8  &21.4 &    129.7 & 4 $\pm$ 3 \\ \hline
\end{tabular}
\tablefoot{
\tablefoottext{a}{Dust temperature and visual extinction were obtained from Herschel and \textit{Planck} data \citep{Hatchell2005, Malinen2012, Lombardi2014, Zari2016}, as explained in \citealp{GEMS-VI-GiselaH2CS}.}
\tablefoottext{b}{The density n(H$_2$) was obtained from DNC (1-0) and DNC (2-1), see Sect. \ref{sect:4}.}}
\end{table*}

\section{Source sample} \label{sect:3}
We studied 23 low-mass starless cores. These cores are located in three archetypal molecular clouds with different star formation activity: Taurus, Perseus, and Orion. In the following, we briefly discuss our sample (listed in Table \ref{tab:source}).

The Taurus molecular cloud (TMC) is one of the closest molecular clouds ($\sim$ 141 pc, \citealp{Pers&Ori-d}) and is a well-thought-out archetypal low-mass star-forming region ($\sim10^4 M_\odot$, \citealp{Goldsmith2008}), whose structure and evolution have been analyzed in several studies (\citealp{B213}, \citealp{Hacar2013}). In this work, we have studied B 213, a dense prominent filament that could be accreting material along striations from its surroundings \citep{B213}. B 213 ends up in the L1495 cloud, which contains several Barnard dark nebulae. In particular, we considered eight starless cores (see Table \ref{tab:source}; numeration by \citealp{Hacar2013}). 

Perseus is a low- and intermediate-mass star-forming molecular cloud ($\sim$400 young stellar objects), at a distance of $\sim$294 pc \citep{Pers&Ori-d}. Unlike Taurus, many of its protostars are associated in protoclusters. Cores located in NGC 1333 and IC 348 regions are immersed in a more active environment due to their proximity to these clusters, while the studied core in Barnard 5 may be affected by an energetic outflow \citep{Barnard5}. On the contrary, L1448 (in the western Perseus) is a more quiescent region.

Orion A is the most active nearby complex, located at $\sim$ 432 pc \citep{Pers&Ori-d}. We studied three low-mass starless cores in three different clouds: ORI-C3 (OMC-2), ORI-C1 (OMC-3), and ORI-C2 (OMC-4), the first one being the closest to the HII region and the last one the least luminous and turbulent of the region.

\section{Density and molecular column density estimates} \label{sect:4}

In this section, we carry out a multi-transitional study of the lines we observed (Table \ref{tab:transitions}) to obtain the H$_2$ gas density and the column density of the molecular species listed in Table \ref{tab:transitions}. 
We first fitted the observed lines using the \textit{minimize} routine of the CLASS-GILDAS software. The line profiles of transitions with unresolved hyperfine structures are shown in Fig. \ref{fig:radex-fits}, and their properties (radial velocity, linewidth, intensity and flux) are listed in Table \ref{tab:gaussian_fits}. In the case of  molecules with resolved hyperfine structure, we used the \textit{HFS} method of GILDAS-CLASS in order to obtain line opacities (see Table \ref{tab:hfs_fits}).
\subsection{DNC} 
We observed two rotational lines of DNC (J  =  1$\rightarrow$0 and 2$\rightarrow$1), which allowed us to run RADEX for different pairs of values of n(H$_2$) and N(DNC) for each starless core.  
In our calculations, we assumed that T$_{\rm k}$  =  T$_{\rm dust}$ and that the emission fills the beam (beam filling factor  =  1): typical angular sizes of pre-stellar cores are $\sim$10000 AU \citep{Ceccarelli_2014Adrians}, corresponding to 73$''$ for cores in Taurus (d = 141 pc, \citealp{Pers&Ori-d}), 35$''$ in the case of Perseus (d = 294 pc) and 24$''$ in Orion (d = 432 pc), whereas the beam widths for our transitions are 15$''$-32$''$ (see Table \ref{tab:transitions}). Dust temperatures were taken from the calculations by  \citealp{Palmeirim2013}, \citealp{Lombardi2014}, and \citealp{Zari2016}  on the basis of the \textit{Herschel} Gould Belt Survey \citep{Andre2010} and \textit{Planck} data  \citep{Bernard2010}.
Following this method, the program was run with column density values ranging from $5 \times 10^{11}$ to $10^{15}$ cm$^{-2}$ and H$_2$ densities from $5 \times 10^{3}$ to $10^{6}$ cm$^{-3}$ (typical values for starless cores). The results are shown in Table \ref{tab:source}. 
 
\begin{figure}[h!]
    \centering
    \includegraphics[width = \hsize]{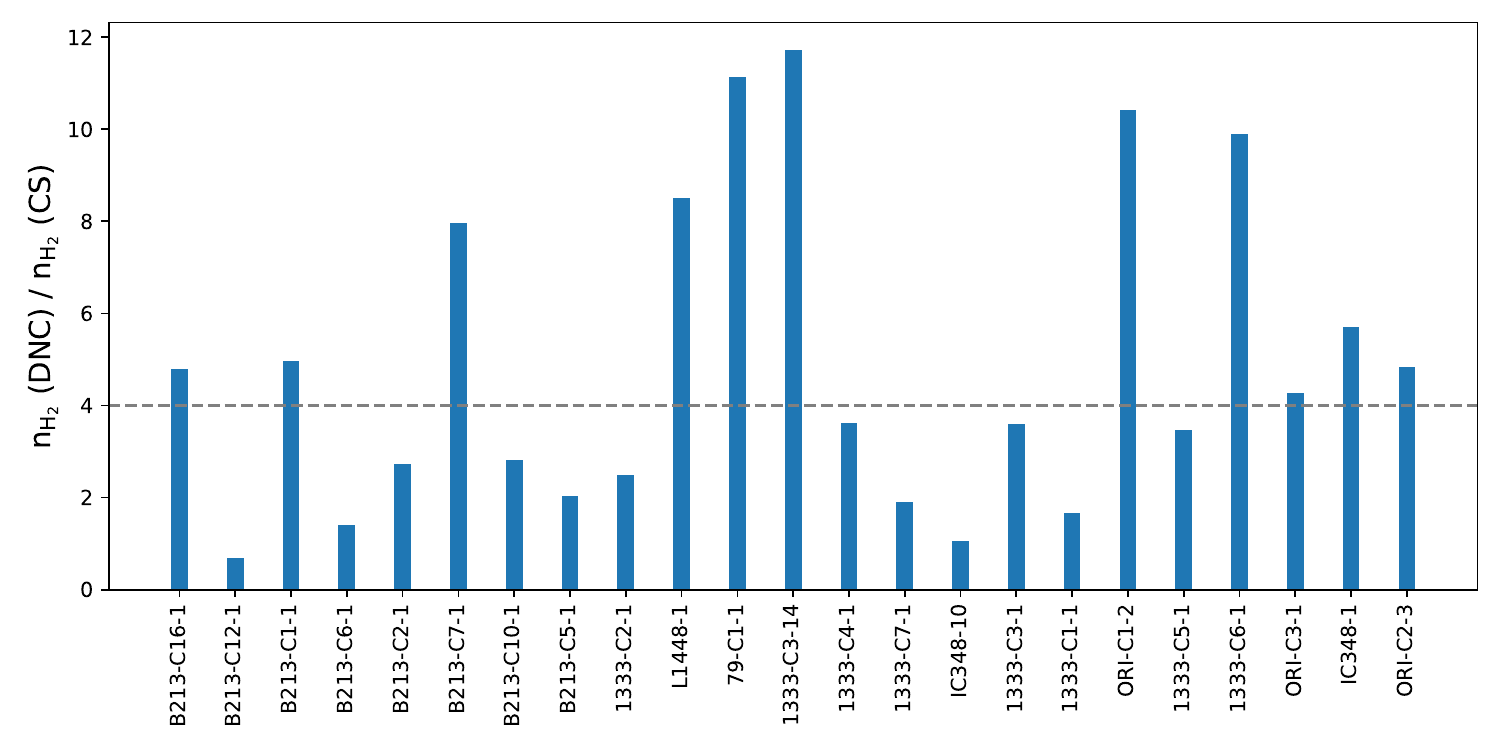}
    \caption{Comparison between molecular hydrogen densities calculated from DNC in this work and from CS in \citealp{GEMS-IV}. The dashed horizontal line indicates the separation between the more and less evolved cores (see Sect. \ref{sect:6}).}
    \label{fig:comp_nH2}
\end{figure}

In Fig.~\ref{fig:comp_nH2} we compare the obtained densities with those derived with CS in \citealp{GEMS-IV}. In that case, densities were obtained from two transitions (J  =  2$\rightarrow$1 and 3$\rightarrow$2) of CS, C$^{34}$S, and $^{13}$CS using a Monte Carlo Markov chain methodology with a Bayesian inference approach.
Considering the uncertainties, 8 out of 23 starless cores present compatible results. For the rest of sources, the densities obtained with DNC are greater than those with CS, up to a factor of 11 higher in 79-C1-1 and 1333-C3-14. This is consistent with the interpretation of the DNC emission coming from the densest and coldest part of the starless cores in which CS is severely depleted. 
In this scenario, one would expect that the greatest difference between the densities calculated from DNC and CS data would be found in the more evolved starless cores. 
However, we do not find any correlation between the difference between the densities calculated from DNC and CS, and the absolute value of the density estimated by using DNC. This lack of correlation could be due to the still significant uncertainties in our density estimates (up to 30\%-50\%, see Table \ref{tab:source}). Moreover, the angular resolution is limited and the beam is larger than the densest region of the core; therefore, the derived densities might not represent the peak densities. Although a detailed discussion is presented later, we note that some of the cores with the highest n$_{\rm H_2}$(DNC)/n$_{\rm H_2}$(CS) appear to be among the most evolved ones based on trends that are explored with chemical models and deuterium fraction (Sect. \ref{sect:6})

Finally, we would like to remind that we assumed that the gas temperature is fixed and equal to the dust temperature. The densities derived using the DNC lines are sensitive to temperature. With the aim to estimate the uncertainty introduced by the fixed temperature values, we calculated the molecular hydrogen densities toward all starless cores again, introducing a kinetic temperature of 7 K. The cores located in Taurus (the coldest of those we have considered) would increase their density by a factor of 2. This factor increases for cores with higher temperatures, such as those in Orion (see in Table \ref{tab:source}), for which the density would be about one order of magnitude higher.
More accurate density calculations, and  based on higher spatial resolution observations are required to have a more accurate view of the starless core structure.

\subsection{HNC, HN$^{13}$C, DN$^{13}$C, H$^{15}$NC and HC$^{15}$N} 
In order to calculate the intended column densities of the other molecules with unresolved hyperfine structure components (HNC, HN$^{13}$C, DN$^{13}$C, H$^{15}$NC, and HC$^{15}$N), we used RADEX with the densities obtained with DNC and the collisional coefficients without hyperfine structure listed in Table \ref{tab:refs_col}. 

The uncertainties in the column density estimates shown in Table \ref{tab:N-nohfs} are defined as the maximum between the error in the integrated intensity and the calibration error, which is assumed to be as 10$\%$. 
Apart from calibration, one of the main sources of uncertainty to consider is the low angular resolution of the telescope (between $\sim$14$''$ and $\sim$29$''$ depending on frequency), which implies that the emission from the inner regions is blended with the outer cold areas. This mainly affects Orion and Perseus cores at low frequencies. Also, the possible volume density gradients along the line of sight may influence the results as well. 

The obtained column densities are included in Table \ref{tab:N-nohfs} and the observed line profiles are shown in Fig. \ref{fig:radex-fits}. The HN$^{13}$C column density N(HN$^{13}$C) ranges from 3.59 $ \times$ 10$^{11}$ (in B213-C5) to 2.14 $ \times$ 10$^{12}$ cm$^{-2}$ (in 1333-C4). In these starless cores, we also detected the lowest and highest H$^{15}$NC column densities, 5.75 $ \times$ 10$^{10}$ and 1.53 $ \times$ 10$^{12}$, respectively. This pattern is also obtained for DN$^{13}$C and HC$^{15}$N, with non-detection in the first core and the highest molecular column densities in the second one, suggesting that these two cores are in an early and late phase, respectively. 

\begin{table}[h!]
\caption[l]{References for the collisional rate coefficients.} \label{tab:refs_col}
\centering
\begin{tabular}{ll} 
\hhline{ =  = }
Molecule & Reference \\ \hline
HCN & \citealp{col_HCN}\tablefootmark{a} \\
H$^{13}$CN & \citealp{Magalh_es_2018-H13CNcolcoef}\tablefootmark{a} \\
HC$^{15}$N &  \citealp{Navarro_Linking}\tablefootmark{a} \\
DCN & \citealp{Giers_2023-DCNcolcoef}\tablefootmark{a}\\
HNC & \citealp{col_HNC}\tablefootmark{b} \\
HN$^{13}$C &  \citealp{Navarro_Linking}\tablefootmark{a} \\
H$^{15}$NC &  \citealp{Navarro_Linking} \\
DNC &  \citealp{Navarro_Linking}\tablefootmark{a} \\
DN$^{13}$C & \citealp{Navarro_Linking} \\
N$_2$H$^+$ & \citealp{col_N2H+_hfs}\tablefootmark{b} \\
N$_2$D$^+$ & \citealp{col_N2H+}\tablefootmark{b} \\ \hline 
\end{tabular}
\tablefoot{Collisional data were taken from the \tablefoottext{a}{EMAA database (\url{https://emaa.osug.fr})} and \tablefoottext{b}{LAMDA database (\url{https://home.strw.leidenuniv.nl/~moldata/})}
}
\end{table}

\subsection{HCN, H$^{13}$CN, DCN, N$_2$H$^+$, and N$_2$D$^+$} 

We used the collisional coefficients listed in Table \ref{tab:refs_col} and the densities obtained with DNC (Table \ref{tab:source}) to derive the column densities of the molecules with resolved hyperfine structure. In the case of H$^{13}$CN, DCN and N$_2$H$^+$, the collisional coefficient files take into account the hyperfine structure. For these species, we ran RADEX over a range of column density values from 10$^{11}$ to 10$^{14}$ cm$^{-2}$. For each value, we compared the observed fluxes of each hyperfine component with the corresponding synthetic ones by evaluating $\epsilon^2$:
\begin{equation}
\epsilon^2=\sum_{i}\left( \frac{flux_{model,i}-flux_{obs,i}}{flux_{obs,i}}\right)^2 
.\end{equation}The optimal column density was determined as the value that minimized that $\epsilon^2$. 

In the case of HCN, this method failed to reproduce observations accurately, due to its lines being optically thick: trying to fit the main component led to a significant underestimation of the column densities. Instead, we follow a different method: we first estimated a partial column density by fitting the integrated intensity of the weakest hyperfine component with RADEX. The total column density of the species was then obtained scaling the partial column density by the spectroscopic integrated intensity ratio between the total splitting and the weakest hyperfine component. However, the weakest hyperfine line of HCN 1-0 is optically thick in almost all the targets (see opacities in Table \ref{tab:hfs_fits}), which still leads to an underestimation of HCN column densities; thus we consider its $^{13}$C isotopolog from now on. 

Regarding N$_2$D$^+$, given that its transitions were optically thin in all targets (see Table \ref{tab:hfs_fits}) and that we did not spectrally resolve the hyperfine components, we did not consider hyperfine splitting in the RADEX calculations. We ran RADEX until the resulting integrated intensity was compatible with the sum of fluxes of all components obtained in the Gaussian fits. 

The results of the column densities that were eventually used for our subsequent discussion are shown in Table \ref{tab:N-hfs}. Once again, B213-C5 presents one of the lowest column densities: N(H$^{13}$CN) = (3.16 $\pm$ 0.44) $ \times 10^{11}$, N(DCN) = (6.87 $\pm$ 0.69) $ \times 10^{11}$, and N(N$_2$H$^+$) = (3.80 $\pm$ 0.38) $ \times 10^{12}$. The core 1333-C4 exhibits the highest ones: N(H$^{13}$CN) = (4.94 $\pm$ 0.49) $ \times 10^{12}$ and N(N$_2$D$^+$) =  (4.16 $\pm$ 0.42) $ \times 10^{12}$.   

\section{Isotopic and isomeric ratios} \label{sect:5}
\subsection{$^{12}$C/$^{13}$C}

Before going deeper into our discussion of the deuterium fractionation and the $^{14}$N/$^{15}$N isotopic ratio, we relay the information regarding our investigation of the $^{12}$C/$^{13}$C and isomeric ratios. Rare isotopologs are often used as representative of the most abundant ones because of their generally optically thin transitions. 
Nevertheless, isotopic ratios are usually uncertain since they depend on a great variety of factors. For instance, the results of \citealp{Roueff2015} suggest that the H$^{12}$CN/H$^{13}$CN and  HN$^{12}$C/HN$^{13}$C ratios can be up to twice higher than the solar isotopic ratio depending on the chemical age. 

\begin{figure}[h!]
    \centering
    \includegraphics[width = \hsize]{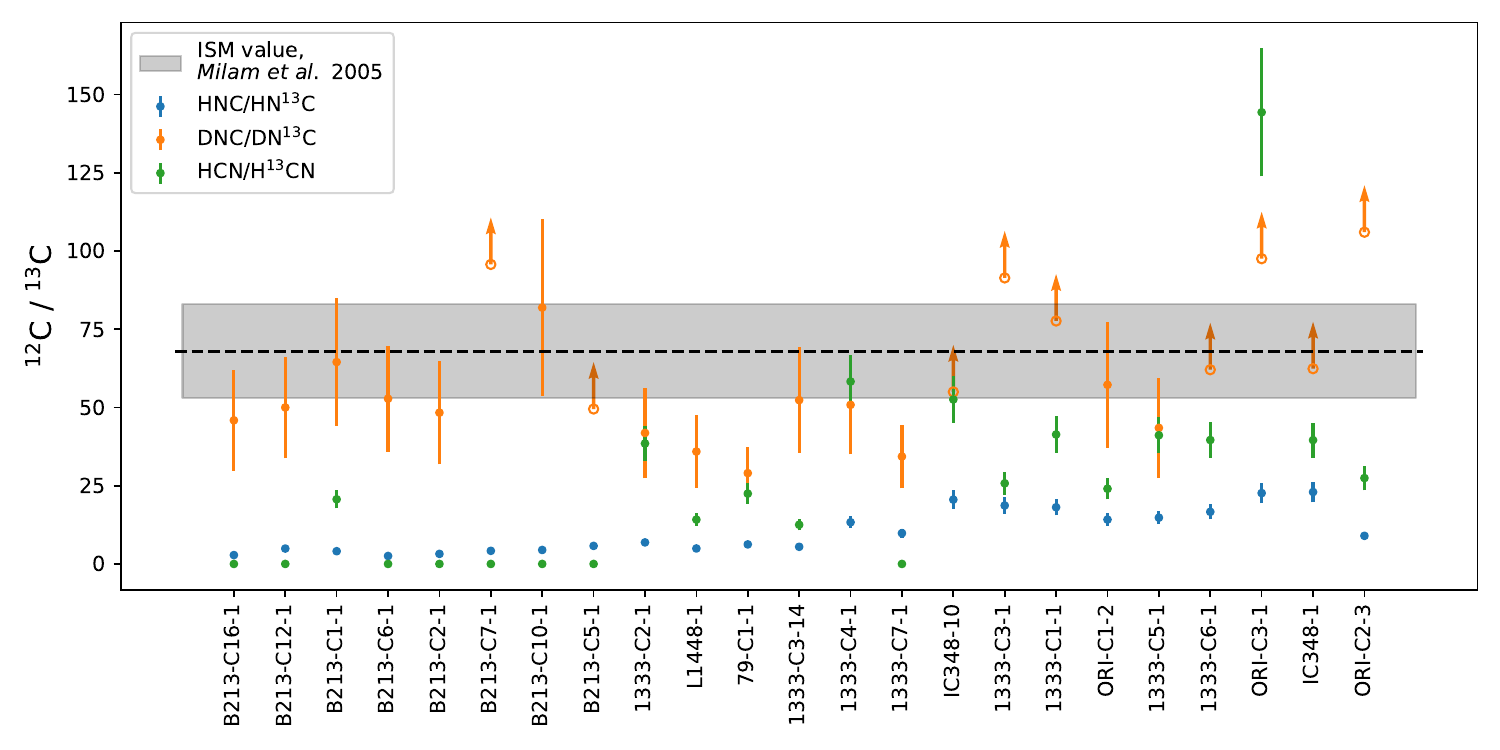}
    \caption{Comparison between our $^{12}$C/$^{13}$C ratio results (represented with dots; open dots are for the lower limits) and the ISM canonical value of 68 $\pm$ 15 \citep{12C-13Cratio} (shown with a horizontal band).}
    \label{fig:12C_13C_ISM}
\end{figure}

We obtained $^{12}$C/$^{13}$C for HCN, HNC, and DNC (see Table \ref{tab:12-13Cratio+isomers}). The resulting average of the HCN/H$^{13}$CN ratio, $\sim$ 33, is lower than the ISM value $^{12}$C/$^{13}$C =  68 $\pm$ 15 \citep{12C-13Cratio}. However, our results are not reliable because of the high opacities (listed in Table \ref{tab:hfs_fits}). 
This low average value could be explained by the underestimation of N(HCN) due to the high optical thickness of the HCN 1-0 line. ORI-C3-1, a target where HCN 1-0 was found to be optically thin, is the exception, as HCN/H$^{13}$CN = 144 $\pm$ 20.
The opacity problem is even more remarkable for the HNC 1-0 line, and thus the low observed value of the HNC/HN$^{13}$C ratio, with an average of $\sim$ 10.
DN$^{12}$C/DN$^{13}$C is not expected to be as affected by opacity, and therefore it is more adequate for a comparison with the standard value of 68 $\pm$ 15 in Fig.~\ref{fig:12C_13C_ISM}. Our results seem to be consistent with that value for almost all starless cores (11 out of 14 where DN$^{13}$C was detected). Additionally, the obtained $^{12}$C/$^{13}$C ratios in ORI-C1-2 and ORI-C2-3 are compatible with the results found in Orion (in the coldest regions of Orion KL) by \citealp{Esplugues_2013}. Some cores lie out of the 68 $\pm$ 15 shadowed area in  Fig.~\ref{fig:12C_13C_ISM}, but never more than a factor of 2, as predicted by \citealp{Roueff2015}.

In the following study, we adopted the canonical value of 68 \citep{12C-13Cratio} as the $^{12}$C/$^{13}$C ratio in our sample. 

\subsection{Isomeric ratios: HCN/HNC, H$^{13}$CN/HN$^{13}$C, HC$^{15}$N/H$^{15}$NC, and DCN/DNC}
The HCN/HNC ratio has been interpreted as a gas kinetic temperature tracer \citep{Hacar2020_isomers}. In starless cores, under cold and dense conditions, HCN/HNC is expected to be $\sim$1 \citep{F.Daniel2013}, increasing at higher temperatures. As a result of underestimations again (even greater for N(HNC)), HCN/HNC was found to be greater than 1.
We consider that the isomeric ratios of the isotopologs are more reliable. In general, our results (listed in Table \ref{tab:12-13Cratio+isomers}) are close to 1, being H$^{13}$CN/HN$^{13}$C  =  1.24 $\pm$  0.44 and  HC$^{15}$N/H$^{15}$NC  =  0.89 $\pm$ 0.30.  
We checked any possible correlation with the gas temperature by using a linear regression fit. The slope and correlation coefficient obtained for the H$^{13}$CN/HN$^{13}$C ratio demonstrate the lack of correlation between these isomeric ratios and gas temperatures in starless cores (shown in Fig.~\ref{fig:isom-T}). A weak correlation is observed for HC$^{15}$N/H$^{15}$NC, with strong statistical significance (p-value=0.017).
We might think that these results are due to analyzing different star-forming regions. However, after doing linear fits to individual regions, the resulting R$^2$ remained low, with high p-values, suggesting a random distribution. Even so, 
this is still  consistent with results reported by \citep{Hacar2020_isomers} since the temperature range in our sample (10 K - 22 K) is too small to confirm or reject this trend. 

\subsection{Deuterium fractionation} \label{sect:Dfrac}
\begin{figure}[h!]
    \centering
    \includegraphics[width = 0.49\textwidth]{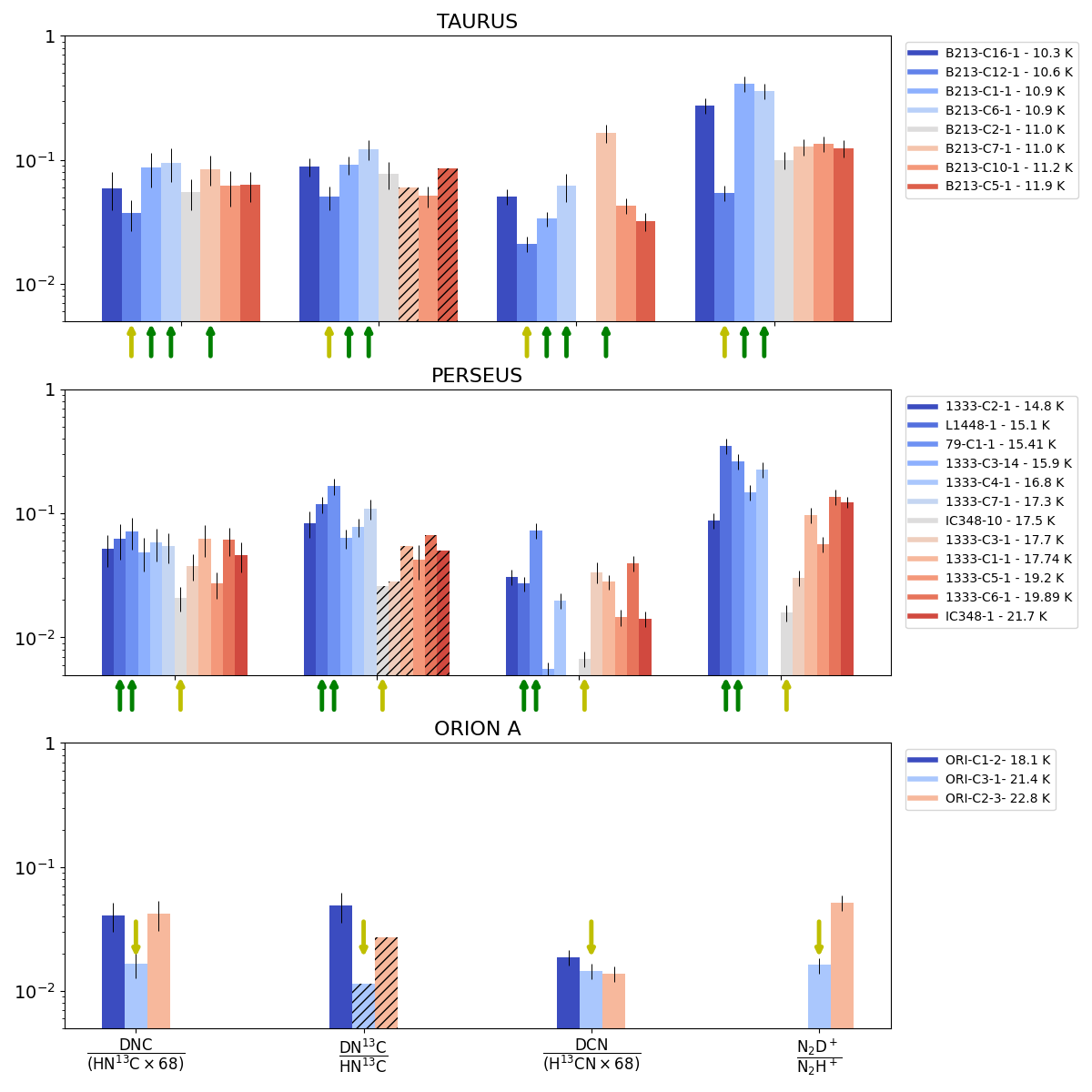}
    \caption{Deuterium fraction ratios in the starless core-sample in the three molecular clouds considered: Taurus, Perseus, and Orion. The sources are ordered by increasing kinetic temperature. Dashed bars indicate upper limits. Green and yellow arrows point out the highest and lowest values of D$_{frac}$, respectively.}
    \label{fig:Dfrac_comparation}
\end{figure}

The deuterium fractionation is defined as the abundance ratio between a deuterated molecule and its hydrogen-bearing counterpart. Significant deuterium fraction enhancements can be detected in some sources. In particular, for starless cores (under cold and dense conditions), the more evolved they are, the higher D$_{frac}$ is expected until  the protostar forms. After this collapse, the young stellar object starts to heat the gas and dust in the surroundings, leading to a D$_{frac}$ drop. Thus, deuterium fraction can be used to trace the evolutionary stage.
\\\\ 
The value of D$_{frac}$ for HNC, HN$^{13}$C, HCN, and N$_2$H$^+$ was obtained from the column densities shown in Table \ref{tab:N-nohfs} and Table \ref{tab:N-hfs}. We calculated DNC/HNC and DCN/HCN in two different ways: a direct method through the use of their column densities, and an indirect one, using the H$^{13}$CN and HN$^{13}$C isotopologs and assuming the canonical $^{12}$C/$^{13}$C isotopic ratio \citep{12C-13Cratio}.
The first method can lead to an overestimation of D$_{frac}$ due to the higher opacity of the main isotopolog line. The second method solves this problem since we expect the emission of H$^{13}$CN and HN$^{13}$C to be optically thin, although we should also consider its downsides: the HCN/H$^{13}$CN and HNC/HN$^{13}$C could differ from the ISM value as much as a factor of 2 due to isotopic fractionation reactions (see \citealp{Roueff2015}). 
When DN$^{13}$C is detected, the DN$^{13}$C/HN$^{13}$C value presents higher accuracy. Since we have obtained DNC/DN$^{13}$C ratio consistent with the canonical value, we considered that the indirect method can provide better estimates of D$_{frac}$ than the direct one. The values of D$_{frac}$ obtained for the studied molecules are shown in Table \ref{tab:deut} and Fig.~\ref{fig:Dfrac_comparation}.

The compound DN$^{13}$C is the only deuterated molecule that was not detected in all starless cores, with a rate of successful detections of 60.9 $\%$. Most of the non-detections occur in starless cores that are immersed in the most active environments: B213-C5, which is in a northern region containing the largest number of evolved objects in B213; the two cores that were considered in IC 348 (C1-1, C1-10) as well as three cores located in NGC 1333 (C1-1, C3-1, C6-1), which are regions associated with protostar clusters in Perseus; and ORI-C3-1, nearby an HII region. On the opposite side, the  DN$^{13}$C line was detected in almost all the cores of Taurus. In Perseus, the highest column densities of D-bearing molecules are always found in 1333-C4, usually followed by 1333-C3-14. In fact, 1333-C3-14 has been proposed as a Class 0 object, and our 1333-C4 results might be contaminated by a protostar candidate \citep{YSOcandidate} located at a distance of 12.57$''$, which is lower than the beam size. Regarding Orion A, ORI-C1-2 is the only starless core where DN$^{13}$C was detected, and it has the highest D$_{frac}$ ratios, due it being the core with the highest extinction (A$_{\nu}$ = 173.0 mag) and the lowest temperature (T$_{k}$ = 18.1 K), so that deuteration can be enhanced because of CO depletion. The trend observed for D$_{frac}$ (HNC) is similar to those found for the rest of species we considered. In general terms, the starless cores that present the highest (indicated by green arrows in Fig.~\ref{fig:Dfrac_comparation}) and lowest values of D$_{frac}$  (indicated by dark yellow arrows in Fig.~\ref{fig:Dfrac_comparation}) are the same for HCN, HNC, and N$_2$H$^+$, with some variations that are discussed in details in the following. 

The deuterium fraction of HNC turned out to be slightly higher than that of HCN: average values of 0.085 $\pm$ 0.034 and 0.054 $\pm$ 0.019 for DN$^{13}$C/HN$^{13}$C and DNC/(HN$^{13}$C $\times$ 68), respectively, compared to 0.036 $\pm$ 0.033 for DCN/(H$^{13}$CN $\times$ 68). Comparing our results with those obtained by other authors, \citealp{Turner_2001} reported slightly lower vales of DNC/HNC and DCN/HCN  (0.018 and 0.008, respectively) in the prototypical starless core TMC 1, again with D$_{frac}$ (HNC) $>$  D$_{frac}$ (HCN).
\citealp{Navarro_Linking} obtained D$_{frac}$(HNC) = 0.094 $\pm$ 0.033 and D$_{frac}$(HCN) = 0.075 $\pm$ 0.027 toward TMC 1-C and  D$_{frac}$ (HNC) = 0.095 $\pm$ 0.034 and D$_{frac}$(HCN) = 0.032 $\pm$0.011 toward 1333-C7-1. Similar results are obtained in protostars: DNC/HNC $\sim$ 0.08 and DCN/HCN $\sim$ 0.07 in HH211 \citep{K.Giers2023} or DNC/HNC $\sim$ 0.11 and DCN/HCN $\sim$ 0.03 in B1-b  \citep{F.Daniel2013}. Although the difference between the deuterium fraction of the two isomers is of about a factor of $\sim$2, roughly within the uncertainties, this result seems to be systematic and can reflect a change in the HCN/HNC ratio along the line of sight, or in the efficiency of the deuteration mechanisms.

The highest values of D$_{frac}$ were measured for N$_2$H$^+$, which presents the highest mean value  (0.15), but also the highest dispersion (0.11).
Moreover, this scattering of D$_{frac}$ (N$_2$H$^+$) values is not related with the different molecular cloud complexes, but stands within each considered region.  In particular, the average N$_2$D$^+$/N$_2$H$^+$ is 0.20, 0.14 and 0.03 with a standard deviation of 0.12, 0.10, and 0.018 in Taurus, Perseus, and Orion, respectively. Similar values of  D$_{frac}$  (N$_2$H$^+$) were obtained by \citealp{Crapsi_2005}: average value of 0.11 $\pm$ 0.10 from a different sample of 29 starless cores. Although this ratio is expected to decrease after the collapse, comparable values were observed for 20 protostars by \citealp{M.Emprechtinger2009}, with an average ratio of 0.10 $\pm$ 0.07. Nevertheless, the molecular lines considered mainly trace the cold envelope, whose physical and chemical composition remains similar to the pre-stellar phase. 

In Fig.~\ref{fig:Dfrac-T}, we explore the correlation of the observed values of deuterium fractionation with the average temperature along the line of sight. Comparing the values of D$_{frac}$ (HNC),  D$_{frac}$ (HCN), and D$_{frac}$ (N$_2$H$^+$),  we concluded that D$_{frac}$ (HNC) and  D$_{frac}$ (HCN) exhibit a higher correlation with temperature than D$_{frac}$ (N$_2$H$^+$), with a strong statistical significance (p-value $\leq$ 0.03). The adopted temperature was derived from \textit{Herschel} observations and represent an averaged value along the line of sight, and it is expected to be higher than the value in the densest part of the core. We can reasonably assume that DNC and DCN are tracing a more external part of the core than N$_2$D$^+$ with an averaged temperature closer to that obtained from \textit{Herschel} observations.
Previous observational works suggest that the abundance of the deuterated compound N$_2$D$^+$ is more affected by the density and the chemical time, both related with the evolutionary stage of the starless core.
Using a chemical model, \citealp{Caselli2002a} concluded that D$_{frac}$ can be used as a chemical clock in star forming regions. \citealp{Navarro2021} also used the ratio to determine the evolutionary stage and the collapse timescale for TMC1-CP and TMC1-C using simple models of collapse \citep{Priestley2018} and the chemical model DNautilus \citep{Majumdar2017}, which includes nuclear spin states chemistry.  Accordingly, we propose that the large scattering in the values of  D$_{frac}$ (N$_2$H$^+$) is more likely related with a higher dependence on the evolutionary stages of the starless cores.

Comparing cores within the same region is more useful for detecting the effect of the starless core evolution on the chemistry, thus avoiding the impact of the environmental conditions. Focusing on each molecular complex, some trends are observed in Fig.~\ref{fig:Dfrac_comparation}. In Taurus, the lowest D$_{frac}$ is always found in B213-C12-1. On the contrary, B213-C6-1 (followed by C1 and C7 in most cases) has the highest values.  These starless cores are located in the northern end of the filament, while B213-C12 is in the southern part, indicating that cores located in the north of B 213 are more chemically evolved than the cores in the south. This trend was also found by \citealp{GEMS-VI-GiselaH2CS} and \citealp{GEMS-IX-MarinaH2S} for  D$_{frac}$ (H$_2$CS) and D$_{frac}$ (H$_2$S), respectively. A similar conclusion was also drawn in \citealp{GEMS-V}, based on methanol emission studies.

In Perseus molecular complex, IC 348-10 is the starless core with the lowest D$_{frac}$, very likely because its proximity to a cluster containing a pre-main sequence star. The cores L1448-1 and 79-C1-1 have the highest deuteration, reaching values of N$_2$D$^+$/N$_2$H$^+$  =  0.35 $\pm$ 0.05 and 0.26 $\pm$ 0.04, respectively. Both regions present one of the lowest temperatures in Perseus. In NGC 1333, the core 1333-C5-1 presents the lowest deuterium fraction, suggesting that this is a less evolved object. Low values are also obtained in 
1333-C3-1, which could be affected by nearby star formation activity. 

Regarding Orion, ORI-C3-1 has lower ratios than the other cores as it is the closest to the HII region, which is heating its surroundings. Our limited core sample in this region prevents us from extracting firm conclusions.

\begin{figure}[h!]
    \centering
    \includegraphics[width = \hsize]{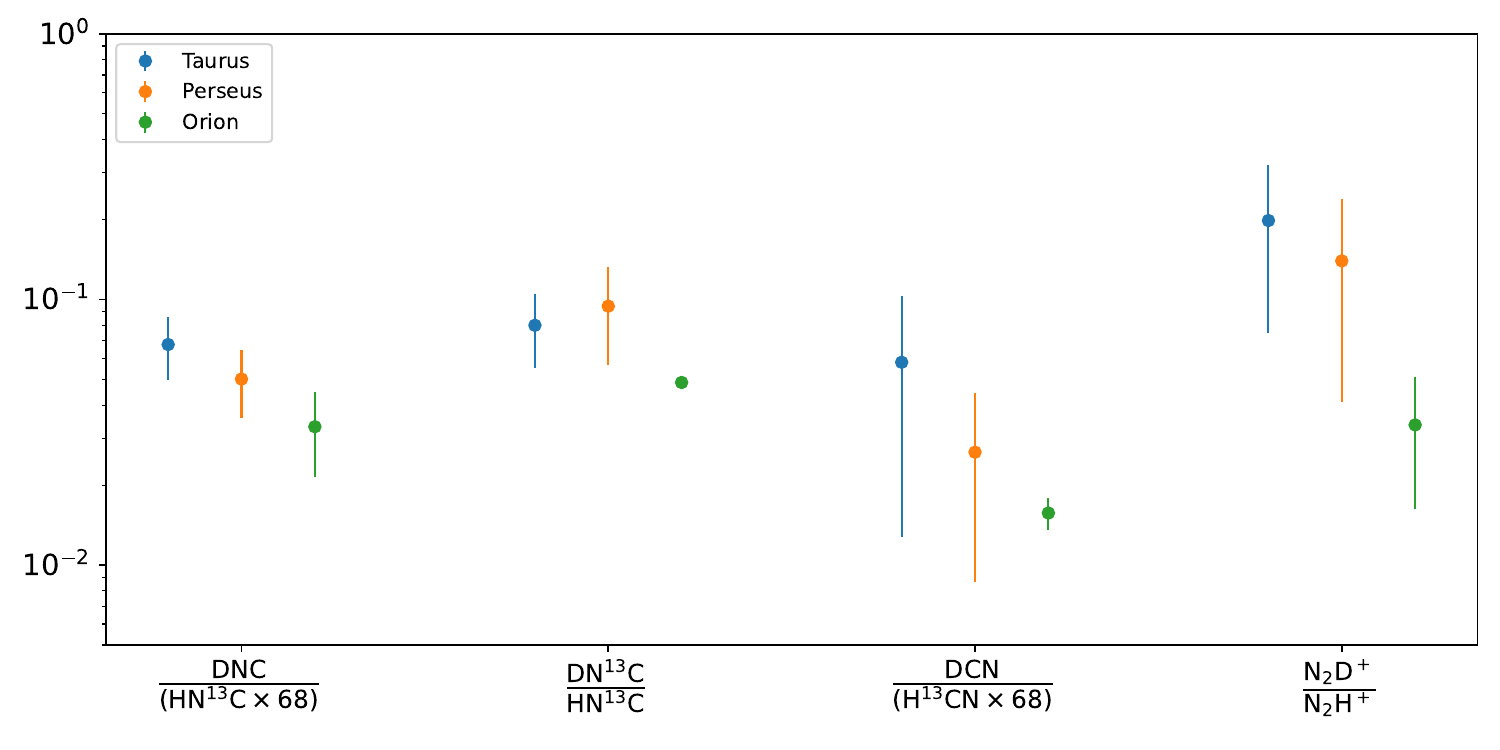}
    \caption{Mean deuterium fraction ratios for each molecular cloud. Error bars represent the standard deviations.}
    \label{fig:Dfrac_comp_MC}
\end{figure}

Finally, we compare the average values of D$_{frac}$ toward the three different molecular clouds, which allows us to focus on differences driven by environmental conditions rather than interpreting them as evolutionary trends. Fig.~\ref{fig:Dfrac_comp_MC} shows that Taurus has the highest D$_{frac}$ for almost all ratios, being the region with the coldest cores, followed by Perseus and finally Orion A.  However, the mean values in Taurus and Perseus remain compatible. One could think that this is the effect of the increasing star formation activity, ambient UV field, and dust temperature from Taurus to Perseus and Orion, as testified by the slight correlation with temperature shown in 
Fig.~\ref{fig:Dfrac-T}, but we need also to keep in mind the different distances of the molecular cloud complexes. Orion is located at a greater distance: $\sim$ 432 pc. Assuming a typical starless core size of 2 $ \times$ 10$^4$ au ($\sim$ 0.01 pc) it is equivalent to an angular size of $\sim$ 5$''$ at the distance of Orion, which is a factor of $\sim$ 5 smaller than the beam of our observations. Thus, we are not able to distinguish if the line emission comes from the external or the densest zone, and the emission of deuterated compounds could be more diluted than those of the hydrogenated compounds. In addition, the smaller sample size in this region may be introducing a bias in the mean value of D$_{frac}$ as we could be missing the most evolved objects. Higher spatial resolution observations and a larger sample in this molecular cloud are required for a more reliable comparison.

\subsection{$^{14}$N/$^{15}$N }

Nitrogen isotopic ratios are key tools to trace the chemical history from molecular clouds to planetary systems \citep{CaselliCeccarelli2012}. They allow us to study and link properties in the interstellar medium and Solar System objects. Different $^{14}$N/$^{15}$N values have been found along the different bodies in the Solar System. The ratio measured for the protosolar nebula from the solar wind is $\sim$ 440 \citep{Marty2011}, similar to the one measured from NH$_3$ in the atmosphere of Jupiter ($\sim$ 450 \citealp{Fouchet2004}), while an enrichment in $^{15}$N is found toward the atmosphere of Earth ($\sim$ 270 for N$_2$, \citealp{Marty2011}) and comets ($\sim$ 172 for HCN, \citealp{comets-Nratio}).
One could think that the low values of the $^{14}$N/$^{15}$N ratio observed in Earth and comets are due to isotopic fractionation processes occurred during the star and planet formation process. Understanding the isotopic fractionation of $^{14}$N/$^{15}$N in star forming regions can therefore shed some light on the origin of N-bearing compounds of our Solar System. In particular, studying sources similar to the environment where the Sun was born is of paramount importance to disentangle the detailed processes that can lead to the  $^{14}$N/$^{15}$N values observed in the solar neighborhood.

Our data allowed us to calculate the  $^{14}$N/$^{15}$N  in a wide sample of starless cores that can be considered as representative of the initial conditions of the formation of a solar mass star. The values of HCN/HC$^{15}$N and HNC/H$^{15}$NC were obtained by both direct and indirect methods: the direct one makes use of estimated HCN, HNC and their $^{15}$N-isotoplog column densities, while the indirect method uses $^{13}$C-isotopolog emission lines are that supposed to be optically thin. For this second method, N(H$^{13}$CN)/N(HC$^{15}$N)  =  $^{14}$N/$^{15}$N $\times$ $^{12}$C/$^{13}$C where we assume $^{12}$C/$^{13}$C  =  68. This second method avoids the uncertainties due to opacity effects but it is affected by possible variations of the  $^{12}$C/$^{13}$C ratio.

The derived nitrogen isotopic ratio in the sampled starless cores are shown in Table \ref{tab:14-15N}. The isotopolog  HC$^{15}$N was detected in all cores except B213-C5-1, for which a lower limit of H$^{13}$CN/HC$^{15}$N is provided. The HNC/H$^{15}$NC ratios that were obtained using the direct method are not reliable because of the high opacities of the HNC 1$\rightarrow$0 line. The fitting of the hyperfine lines of HCN allowed us to estimate the opacity and thus test the reliability of our calculations.
Toward ORI-C3-1, 1333-C4-1, and 1333-C5-1, the HCN 1$\rightarrow$0 line has low or moderate opacities (see Table \ref{tab:hfs_fits}). 
For 1333-C4-1 and 1333-C5-1, the values obtained using the direct and indirect methods are compatible. However, the value derived in ORI-C3-1, using the direct method (850 $\pm$ 130) is a factor of $\sim$2 larger than the value obtained using the indirect method \citep{14-15N_highRatio}, suggesting that the  $^{12}$C/$^{13}$C ratio can differ from the canonical value in this source (as obtained with HCN/H$^{13}$CN and DNC/DN$^{13}$C, see Table \ref{tab:12-13Cratio+isomers}).
Using the indirect method, we obtained average ratios of (HN$^{13}$C$\times$ 68)/H$^{15}$NC = 296 and (H$^{13}$CN$\times$ 68)/HC$^{15}$N = 430. The first one present uniform values across our sources ($\sigma$ = 64) while variations are detected for the second one ($\sigma$ = 120). In the following, we comment the variations of the (H$^{13}$CN$\times$ 68)/HC$^{15}$N ratio in our sample.

In B 213, (H$^{13}$CN$\times$ 68)/HC$^{15}$N is, on average, higher than in any other molecular complex we observed, being the average ratio 490 and the standard deviation 130.
In B213-C12-1 and B213-C10-1 we obtained high values (700 $\pm$ 170 and 490 $\pm$ 90, respectively), while the cores B213-C1-1 and B213-C2-1 present lower ratios (420 $\pm$ 90 and 260 $\pm$ 40, respectively). Therefore, we might find a slight increase from north to south of $^{14}$N/$^{15}$N (and then $^{15}$N raising toward the north), but uncertainties are too large to draw firm conclusions. Nevertheless, this would be in agreement with the interpretation that there is a $^{15}$N enrichment in the more evolved cores. The cores located in the north are those toward which we obtained higher values of  D$_{frac}$ and also the ones where \citealp{GEMS-VI-GiselaH2CS} and \citealp{GEMS-IX-MarinaH2S} also found higher values for the deuterium fractions  of H$_2$CS and H$_2$S, respectively, suggesting that these are in a more evolved evolutionary stage.

Perseus presents high variations in the values of HC$^{14}$N/HC$^{15}$N, being the standard deviation 120 and the average value 410. The maximum values of (H$^{13}$CN$\times$ 68)/HC$^{15}$N are found in 1333-C4-1 (660 $\pm$ 100), which could be contaminated by a nearby protostar,  and 1333-C3-14 (540 $\pm$ 90), the most evolved core in our sample, which is not consistent with the scenario of low values of  $^{14}$N/$^{15}$N in the more evolved cores. However, we do find a low value (310 $\pm$ 50) toward 79-C1-1, which presents also high values of D$_{frac}$ in all the studied molecules (this work, \citealp{GEMS-VI-GiselaH2CS}, \citealp{GEMS-IX-MarinaH2S}). In this case,
isotope-selective photodissociation might also be at work, since this core is located in Barnard 5, an active star formation associated with energetic outflows \citep{Barnard5}.

Regarding Orion, a uniform $^{14}$N/$^{15}$N was found, whose average is 407 and its standard deviation is 19. The maximum ratio (400 $\pm$ 60) was reached in ORI-C3-1,  while the other two cores present similar values ($\sim$ 350).

Two main mechanisms have been proposed to explain low values of  $^{14}$N/$^{15}$N: (i) isotopic fractionation reactions that would change the $^{14}$N/$^{15}$N ratios with time  \citep{Roueff2015}. Since we are using the indirect method for our calculations, our results would follow the H$^{13}$CN/HC$^{15}$N ratio that is predicted to decrease for longer times. And (ii) selective photodissociation of N$_2$ that would lead to an increase in $^{15}$N atoms, and therefore a $^{15}$N enrichment of nitriles in regions bathed in enhanced UV fields \citep{Spezzano, Hily-Blant2019}. We explored which is the dominant mechanism for the variations we found in the  H$^{13}$CN/HC$^{15}$N ratio by exploring the correlation of this quantity with dust temperature and D$_{frac}$ (N$_2$H$^+$) (see Figs. \ref{fig:14N/15N-T}, \ref{fig:Dfrac-14N/15N}).  If  selective photodissociation were driving the $^{14}$N/$^{15}$N ratio, one would expect some correlation between this isotopic ratio and the dust temperature. Indeed, we find weak correlation of  H$^{13}$CN/HC$^{15}$N with temperature (p-value=0.09), although the correlation is mainly due to the higher values of  H$^{13}$CN/HC$^{15}$N in B 213. On the contrary, if the $^{14}$N/$^{15}$N ratio were driven by the core evolution, one would expect a tighter correlation with D$_{frac}$ (N$_2$H$^+$), which is not found (p-value=0.2). Our results suggest that although local differences of starless cores in the same region can be explained in terms of core evolution and/or selective photodissociation, none of these mechanisms can explain the results in all the studied regions. The large scattering of the values obtained and the uncertainties in the $^{12}$C/$^{13}$C ratio make it difficult to extract firm conclusions.

\section{Chemical modeling: Deuterium fractionation}
\label{sect:6}
In order to explore the ability of deuterium fractionation to probe the chemical and dynamical evolution of starless cores, we modeled the chemistry of our targets using the latest version of the three-phase chemical code, DNAUTILUS 2.0 \citep{Kashyap2024}. DNAUTILUS 2.0 has the capability to investigate deuterium fractionation in both two-phase (only gas and grain surface) and three-phase (gas, grain surface, and grain mantle) modes, with the option to include or exclude the ortho- and para-spin states of key hydrogenated species such as H$_2$, H$_2^+$, H$_3^+$, and their isotopologs. In this study, we utilized the three-phase configuration of DNAUTILUS 2.0, incorporating spin-state chemistry for light hydrogen-bearing species (H$_2$, H$_2^+$, H$_3^+$) and their isotopologs, as described in \citealp{Majumdar2017}. DNAUTILUS 2.0 computes the time evolution of chemical abundances while considering the gas, grain surface, and grain mantle phases, as well as their interactions. In our calculations, we assume an initial deuterium abundance of $1.6 \times 10^{-5}$ \citep{Linsky2007}, and the initial ortho-to-para ratio of H$_2$ is set to its statistical value of 3 by following \citealp{Majumdar2017}. 

\subsection{Exploring the parameter space}
\label{explor}

To test the sensitivity of deuterium fraction ratios to different physical parameters (density, gas and dust temperature and cosmic ray ionization), we considered a grid of zero-dimensional (0D) DNAUTILUS models adopting different initial conditions, as summarized in Table \ref{tab:params_models}. In all models we adopted the elemental initial abundances as in \citealp{Majumdar2017}  except for sulfur. \citealp{GEMS-VII-Asun2023} estimated a sulfur depletion of a factor of $\sim$10 in Taurus and Perseus,  and of $\sim$1 in Orion, based on the fitting of the molecular abundances derived from GEMS data. We  checked the impact of the assumed initial elemental abundances in our calculations and the changes in D$_{frac}$ estimates. Varying the initial S$^+$, the differences observed were negligible; therefore we adopted S/H  =  1.5 $\times$ 10$^{-6}$ in all our runs (depletion of a factor of 10). We also tested the sensitivity to the initial form of carbon (C$^+$ or CO) and oxygen and found no substantial variation in D$_{frac}$ for t$>$0.1 Myr (typical ages of starless cores). In the case of ortho-to-para ratio of H$_2$, its value always thermalizes to 3 $\times$ 10$^{-5}$. A model with a low initial ratio (e.g., 10$^{-4}$) reaches steady state more quickly than one with a value of 3 (see Fig. \ref{fig:o-pH2}). However, both models converge to the same D$_{frac}$ predictions for t $>$ 0.3 Myr.

Regarding gas density and temperature, we considered representative physical conditions for starless cores: n$_{\rm H_2}$ = 10$^5$ cm$^{-3}$ and 10$^6$ cm$^{-3}$,  and T  =  7 K, 10 K,  and 15 K \citep{GEMS-I-Asun, Navarro2021, GEMS-IX-MarinaH2S}.  Cosmic rays play a leading role in the chemistry of the interstellar medium. In absence of other ionization agents (X-rays, UV photons, and J-type shocks), the ionization fraction is proportional to the square root of the cosmic ray ionization rate per molecular hydrogen, $\zeta_{H_2}$, which becomes the key parameter in the molecular cloud evolution \citep{McKee1989,Caselli2002,Wakelam2010}. Values ranging from a few 10$^{-18}$ s$^{-1}$ to a few  10$^{-17}$ s$^{-1}$  have been measured in diffuse and dense interstellar clouds (see, e.g., \citealp{Padovani2009}), the lower values being measured toward regions with high molecular hydrogen column densities.
 \citealp{GEMS-VII-Asun2023} estimated  $\zeta_{H_2}$ $\sim$ 1$-$ 50 $\times$ 10$^{-17}$ s$^{-1}$ in the dense cores within the GEMS sample located in Taurus, Perseus, and Orion. These estimates were in agreement with previous estimates in Barnard 1b by \citealp{Asun2016} and in TMC 1 by \citealp{Navarro2021}.  According to these results, we decided to consider two values: $\zeta_{H_2}$ = 1.3$\times$ 10$^{-17}$ and 1.3$\times$ 10$^{-16}$ s$^{-1}$, which are representative of the physical conditions in our targets.  In total, we ran the 5 models shown in Table~\ref{tab:params_models}.

\begin{table}[h!]
\caption{Initial relative abundances with respect to hydrogen and the model input parameters.} \label{tab:params_models}
\centering
\begin{tabular}{cccccc}
\hhline{ =  =  =  =  =  = }
Parameter      & A        & B        & C        & D  & E            \\ \hline
He  & \multicolumn{5}{c}{9.0$\times$10$^{-2}$}  \\
N   & \multicolumn{5}{c}{6.2$\times$10$^{-5}$}  \\
O   & \multicolumn{5}{c}{2.4$\times$10$^{-4}$}  \\
C$^+$ & \multicolumn{5}{c}{1.7$\times$10$^{-4}$}  \\
S$^+$ & \multicolumn{5}{c}{1.5$\times$10$^{-6}$}  \\
Si$^+$& \multicolumn{5}{c}{8.0$\times$10$^{-9}$}  \\
Fe$^+$& \multicolumn{5}{c}{3.0$\times$10$^{-9}$}  \\
Na$^+$& \multicolumn{5}{c}{2.0$\times$10$^{-9}$}  \\
Mg$^+$& \multicolumn{5}{c}{7.0$\times$10$^{-9}$}  \\
P$^+$ & \multicolumn{5}{c}{2.0$\times$10$^{-10}$}  \\
Cl$^+$& \multicolumn{5}{c}{1.0$\times$10$^{-9}$}  \\
F   & \multicolumn{5}{c}{6.68$\times$10$^{-9}$} \\
HD  & \multicolumn{5}{c}{1.6$\times$10$^{-5}$}  \\
o-H$_2$/p-H$_2$      & \multicolumn{5}{c}{3}           \\
A$_V$ (mag)    &  \multicolumn{5}{c}{15}  \\ \hline
$\zeta$ ($\times$10$^{-17}$  s$^{-1}$) & 1.3 & 13 & 1.3 &  1.3 & 1.3   \\
T (K)      & 7              & 7      & 10        & 15             & 7                   \\ 
n$_{\rm H_2}$ (cm$^{-3}$)  & 10$^5$         & 10$^5$         & 10$^5$   & 10$^5$  & 10$^6$              \\ 
 \hline
\end{tabular}
\end{table}

\begin{figure}[h!]
\centering
    \begin{subfigure}{0.48\textwidth}
        \centering
        \includegraphics[width = \hsize]{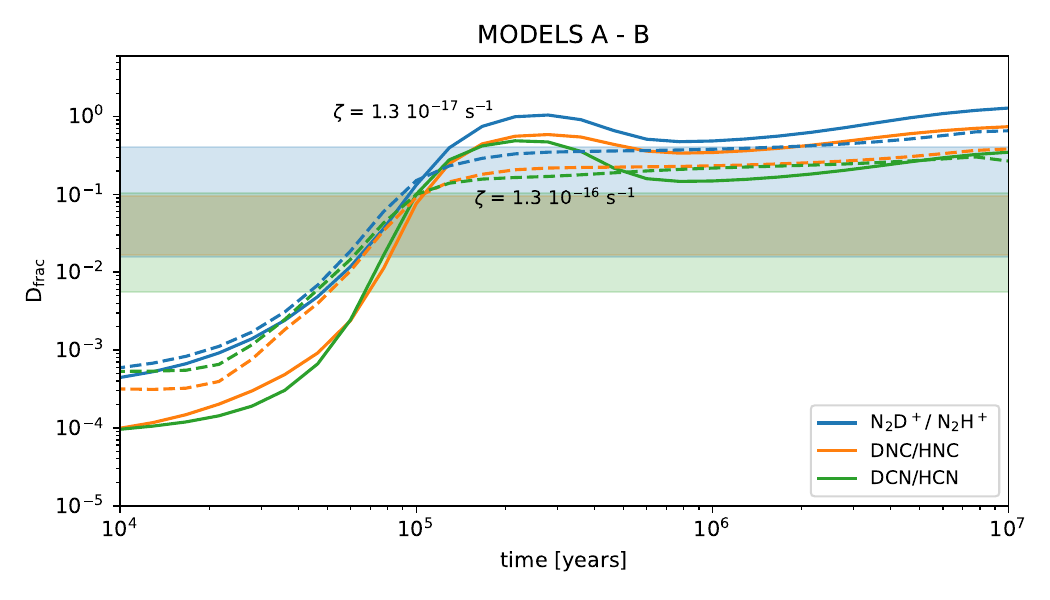}
    \end{subfigure}
    \hfill
     \begin{subfigure}{0.48\textwidth}
        \centering
        \includegraphics[width = \hsize]{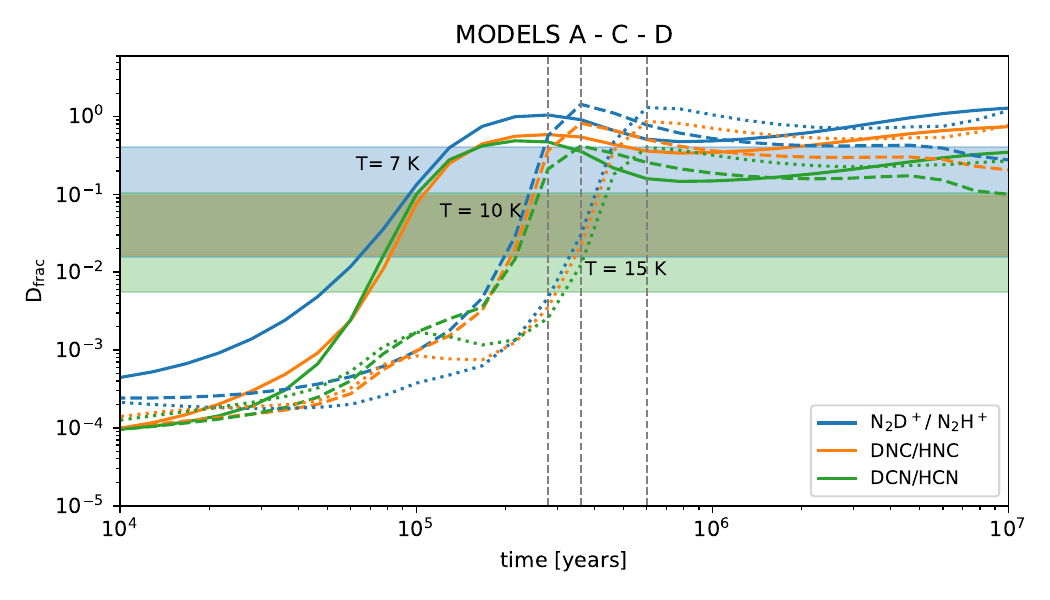}
    \end{subfigure}
    \hfill
     \begin{subfigure}{0.48\textwidth}
        \centering
        \includegraphics[width = \hsize]{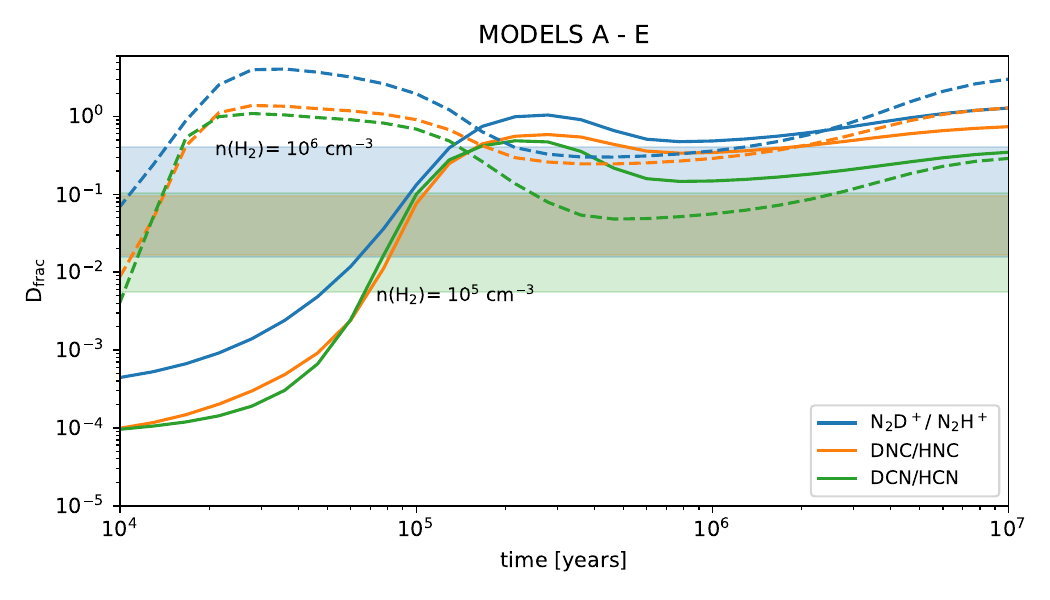}
    \end{subfigure}
    \caption{Comparison of the evolution of D$_{frac}$ up to 10 Myr according to the different parameters of the models described in Table \ref{tab:params_models}. Shaded horizontal bands indicate the observed range of the deuteration values for N$_2$H$^+$ (blue), HNC (orange), and HCN (green). Vertical dashed lines indicate the time when steady state is reached.} \label{fig:models_comp}
\end{figure}

In Fig.~\ref{fig:models_comp} we show the predicted D$_{frac}$ for N$_2$H$^+$, HNC and HCN over the course of 10 Myr for our set of models. The comparison between models A and B allowed us to study the influence of cosmic ray ionization ($\zeta_{H_2}$) on D$_{frac}$. Large values of the $\zeta_{H_2}$ produce slightly lower values of D$_{frac}$ in steady state, which is reached at t $\geq$ 0.1 Myr with these physical conditions.
This is consistent with N$_2$D$^+$ being mainly destroyed by dissociative recombination (N$_2$D$^+$ + e$^-$ $\rightarrow$ D + N$_2$) for t $\geq$ 0.2 Myr (see Table~\ref{tab:model-reactions}).
 The value of $\zeta_{H_2}$ also have an impact on the DNC/DCN isomeric ratio as suggested by \citealp{Navarro2021}. The isomeric ratio, DNC/DCN is predicted to be closer to 1 with model B than with model A.

Gas and grain temperatures (assumed equal in our modeling) have a strong influence on deuteration, determining the time at which steady-state values are achieved, i.e,  the deuteration timescale  (see Fig.~\ref{fig:models_comp}). Assuming T  =  7 K (model A), the values of D$_{frac}$ increase until t $\sim$ 0.3 Myr, when the steady-state is achieved. For model D, with T = 15 K, steady-state values are achieved at t $\sim$ 0.6 Myr, which means a timescale 2 times longer.  With  T  =  10 K, we find an intermediate situation. 
For temperatures $\leq$ 10 K, deuteration mainly proceeds through ion-neutral reactions of H$_2$D$^+$, D$_2$H$^+$, and D$_3^+$ with different compounds (see Table~\ref{tab:model-reactions}). All these deuterated ions are formed by H-D substitution reactions of H$_3^+$, and then H$_2$D$^+$, with HD, producing a progressive deuteration. These reactions are favored at low temperatures. Moreover, the parent ion, H$_3^+$, is rapidly destroyed through reactions with CO. The CO depletion in these cold regions increases the H$_3^+$ abundance and consequently, the formation of its deuterated forms. 
At moderate temperatures ($\sim$ 15 K), there are other isotopic fractionation reactions that can contribute to increase the value of D$_{frac}$. This is the case of the formation of N$_2$D$^+$ in model C. At T = 15 K,  the reaction N$_2$H$^+$ + D $ \rightarrow$ N$_2$D$^+$ + H is an important N$_2$D$^+$ formation route (see Table~\ref{tab:model-reactions}).
The high sensitivity of the  D$_{frac}$  to temperature conditions is used as a tracer of the evolution of dense gas, since an accurate chemical age estimation requires precise knowledge of the physical conditions.

Finally, we explore the influence of density on the deuteration of HCN, HNC, and N$_2$H$^+$ by comparing models A and E. Higher values of density have a high impact on the highest values of D$_{frac}$ obtained and  also, on the time at which these values are achieved. Changing the density from 10$^5$ cm$^{-3}$ (model A) to 10$^6$ cm$^{-3}$ (model E) leads to an increase in the values of D$_{frac}$ by a factor of a few. Moreover, these high values of D$_{frac}$ are achieved in a short timescale,  t $\sim$10$^4$ yr.  Again this shorter timescale is related to the CO depletion time, which enhances the abundances of H$_3^+$ and their deuterated compounds. According to our results in Sect.~\ref{sect:4}, 
the densities in our sample have an n$_{\rm H}$ of a few 10$^5$ cm$^{-3}$. Thus, our physical conditions would be closer to model A, B, C, and D. 

\begin{figure*}[h!]
    \centering
\includegraphics[width = \textwidth]{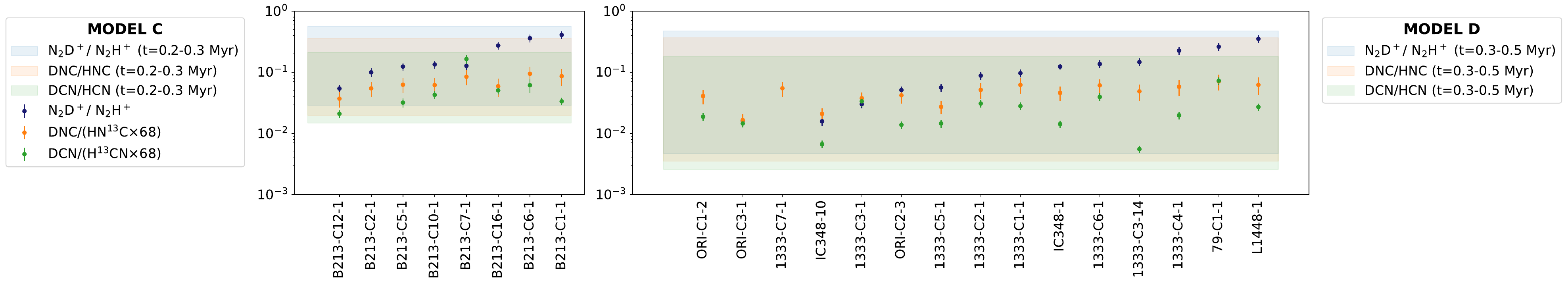}
    \caption{
    Comparison between observations and models C and D to constrain the chemical age of the starless cores. Blue, orange, and green dots represent the different observed deuterium fraction ratios across the core sample. The shadowed areas show the predicted values of the ratios for specific time ranges. The starless cores are ordered within each molecular cloud according to their evolutionary stage.}
    \label{fig:modelsAC-obs}
\end{figure*}

\subsection{Comparison with observations}
In this section, we check the ability of D$_{frac}$ to trace the evolutionary stage of our starless cores by comparing our observational results with the different models.  As a first step, we show in Fig.~\ref{fig:models_obs}  the comparison between the values of  D$_{frac}$ for all our models in reasonable evolutionary times in between 0.1 and 10 Myr. The predictions of our models in the time range are shown as a color band in each panel, while the  values of D$_{frac}$ estimated from our observations of HCN, HNC, and N$_2$H$^+$ are indicated by dots. There are two models, model E and model B, that perform poorly when reproducing our observations. This is not unexpected in the case of model E since the densities derived for our DNC observations are significantly lower than the density considered in this run (n$_H$  =  10$^6$ cm$^{-3}$). Moreover, model B assumes a relatively high value of $\zeta_{H_2}$ of 1.3$\times$ 10$^{-16}$ s$^{-1}$.  

Models A, C, and D are able to reproduce all our observations. However, different evolutionary times are derived depending on which model is used. This is the consequence of the dependence of the deuteration timescale with temperature. As commented in Sect.~\ref{explor} and shown in Fig.~\ref{fig:models_comp}, the main effect of increasing temperature is to delay the deuteration of DCN, DNC, and N$_2$D$^+$ to longer times. We need to fix the temperature to have reliable evolutionary time estimates. Temperatures $\sim$ 10 K are measured in B 213 using dust continuum and spectroscopic observations \citep{Hacar2013,Palmeirim2013, Marsh2014}.

Assuming T = 10 K, all our values of D$_{frac}$ in cores located in Taurus can be reproduced with t $\sim$ 0.2 $-$ 0.3 Myr (see Fig. \ref{fig:modelsAC-obs}). Model D, with T = 15 K, is more adequate for Perseus and Orion \citep{Lombardi2014, Zari2016}, which are more active star formation regions. Using this value, we can reproduce the values measured in all cores of our sample with t $\sim$ 0.3 $-$ 0.5 Myr. 

Within these ranges, deuteration increases very fast, and these are the only temporal windows that reproduce the observed values in each region. With this simple approximation (0D, assumed temperature and density), since deuteration rises with time, it can be used as an evolutionary tracer. In Fig. \ref{fig:modelsAC-obs}, the cores are ordered according to their evolutionary stage based on D$_{frac}$ (N$_2$H$^+$), which is the deuteration with the largest variations and therefore is more suitable for tracing evolution. The most evolved cores in Taurus appear to be B213-C1-1, B213-C6-1, and B213-C16-1, while in Perseus, they seem to be L1448-1, 79-C1-1, 1333-C4-1, and 1333-C3-14. We note that once the steady state is reached, core evolution could no longer be assessed with these deuterated species, and it is only possible to give lower time limits, but the cores in our sample have not yet reached this state. Given that the derived ages are temperature-dependent, more complex models would be required to accurately constrain the age. Nevertheless, the cores classified as more evolved based on deuteration are those with higher n(DNC)/n(CS) (see cores with n(DNC)/n(CS) $>$ 4 in Fig. \ref{fig:comp_nH2}). Therefore, although our model is simple, the derived evolutionary scale is supported by additional indicators, suggesting that it is not far from reality.

To summarize, our modeling is able to successfully reproduce the values of  D$_{frac}$ of HCN, HNC, and N$_2$H$^+$ for the 23 cores of our sample located in Taurus, Perseus, and Orion. Not unexpected, our results confirm the goodness of D$_{frac}$ to estimate the evolutionary time of starless cores as long as temperature and density are accurately known. Using D$_{frac}$ (HCN),  D$_{frac}$ (HNC), and  D$_{frac}$ (N$_2$H$^+$),  we estimate ages of t$\sim$  0.2 $-$ 0.3 Myr  in the cores located in the filament B 213. In the case of Perseus and Orion, where higher gas and dust temperatures are expected, we need  t $\sim$  0.3 $-$  0.5 Myr to account for observations. These ages are consistent with the ones derived for Taurus by other authors using chemical diagnostics \citep{Loison2020, GEMS-V} and the results obtained in other low-mass star forming regions (\citealp{Konyves2015} for densities $\sim$ 10$^5$ cm$^{-3}$, \citealp{Wakelam2021}).
The differences in the starless core ages found in the different regions are affected by the adopted temperatures and densities. This introduces uncertainties in the derived ages.
Despite this, the good agreement between model predictions and observations, in such a large sample of starless cores, located in regions with different star formation activity, suggests that our model is robust and successfully accounts for the processes involved in molecular deuteration.

\section{Changes in the $^{14}$N/$^{15}$N ratio during star formation}

\begin{figure*}[h!]
    \centering
    \includegraphics[width = \textwidth]{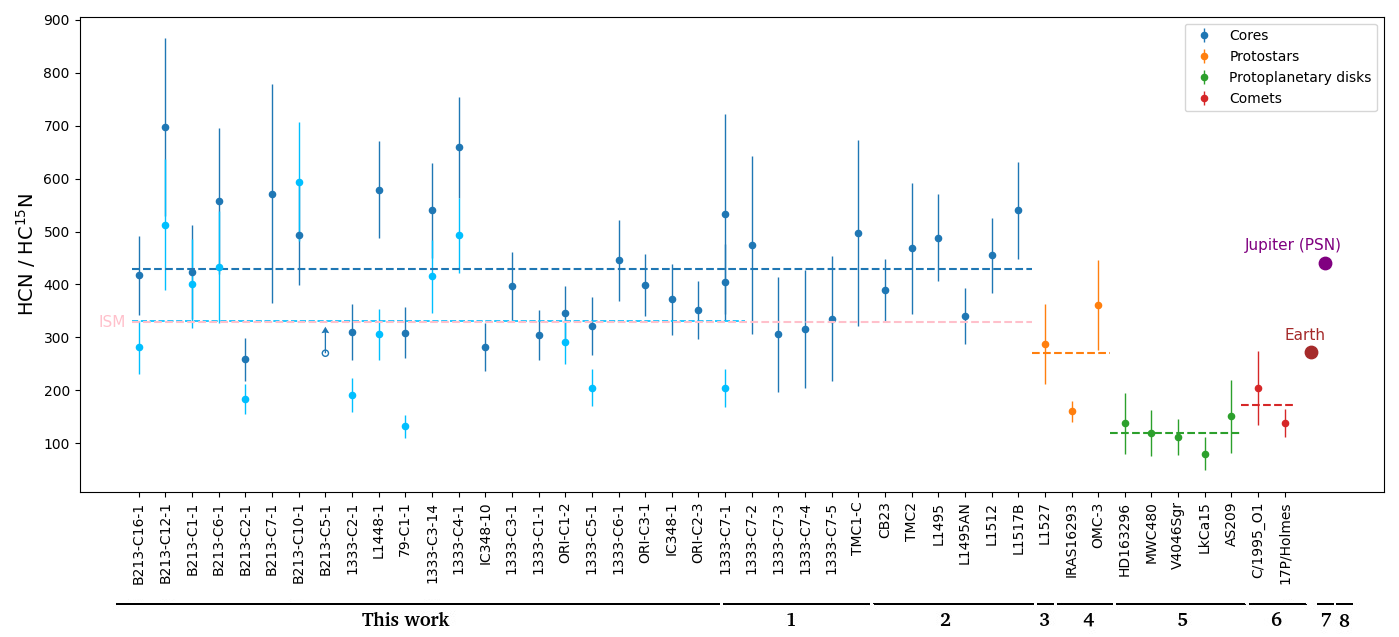}
    \caption{$^{14}$N/$^{15}$N from (68 $\times$ H$^{13}$CN)/HC$^{15}$N at different evolutionary stages, from cores to protoplanetary disks. Measures from comets, Earth, and Jupiter are also considered. Light blue dots represent our results using the $^{12}$C/$^{13}$C derived from DNC/DN$^{13}$C instead of the fixed ratio of 68. Dashed lines show the average values of each state. The local ISM value of $\sim$ 330 \citep{Hily-Blant2019} is represented with pink dashed lines. \textbf{References:} (1) \citealp{Navarro_Linking}; (2) \citealp{Jensen}; (3) \citealp{14N15N_3yoshida}; (4) \citealp{14N15N_4W}; (5) \citealp{14N15N_5G}; (6) \citealp{14N15N_6B}; (7) \citealp{Fouchet2004}; (8) \citealp{Owen2001}.}
    \label{fig:nitro_1415N}
\end{figure*}

The interest in nitrogen fractionation processes has increased in recent years, as a tool to establishing the degree to which planetary systems inherit their chemical
composition from their parent interstellar clouds. The $^{14}$N/$^{15}$N isotopic ratios measured in the Solar System present a complex picture,
with values ranging from 441 in Jupiter and the proto-Sun, down to approximately 50 in micron-sized inclusions in chondrites
\citep{Bonal2010}. Despite the observational and theoretical efforts aimed to answer this question, the origin of nitrogen isotopic ratios  measured in our 
planetary system remains elusive \citep{Hily-Blant2013, Furi2015}. Moreover, care must be taken when comparing the 4.6 Gyr-old
protosolar nebula to present-day star- and planet-forming regions since stellar nucleosynthesis, or the migration of the Solar System
from a different Galactocentric radius \citep{Minchev2013, Barbosa2015}, could change the isotopic ratio.
Isotopic ratios well below from $\sim$ 441, to as low as $\sim$ 270, have been measured in the diffuse \citep{Lucas1998} and dense
\citep{Adande2012, Hily-Blant2013, Daniel2013} molecular gas in the  local interstellar medium.  
\citealp{Hily-Blant2017} and \citealp{Hily-Blant2019} proposed that the isotopic ratio in the local interstellar medium is $\sim$ 330, i.e., lower than the protosolar value of 441. 
A low isotopic ratio in the solar neighborhood has received further observational support from observations in massive star forming regions \citep{Kahane2018, Colzi2018a, Colzi2018b}. 
Most of these observations, however, come from active star forming regions where the gas has been heavily processed.
Yet, measurements of the $^{14}$N/$^{15}$N ratio in starless and pre-stellar cores remain scarce, clearly insufficient to establish firm conclusions on the initial
value of nitrogen fractionation in the star formation process. 

In this paper we present observations of the HCN/HC$^{15}$N and  HNC/H$^{15}$NC ratios toward 23 starless cores. On average, we obtained HCN/HC$^{15}$N  =  430 $\pm$ 120 and HNC/H$^{15}$NC  =  296 $\pm$ 64. The value of HNC/H$^{15}$NC is quite uniform along our sample within the dispersion. However, the value of HCN/HC$^{15}$N is weakly correlated with the temperature  (see Fig.~\ref{fig:14N/15N-T}). Thus, our data suggest that isotopic fractionation could depend on the environmental conditions. 
 In Fig.~\ref{fig:nitro_1415N}, we compared the R$_{^{14}N/^{15}N}$ defined as (68$\times$H$^{13}$CN)/HC$^{15}$N in our sample with the values obtained in previous works, as well as those found toward low-mass protostars, protoplanetary disks and Solar System bodies. In order to make this comparison as reliable as possible, in this figure we used the published column densities of  H$^{13}$CN and HC$^{15}$N to estimate R$_{^{14}N/^{15}N}$ instead of using the value of the nitrogen fractions given by the authors. In this way, we are using the same approximation in all the objects.  We find a large dispersion in the values of  R$_{^{14}N/^{15}N}$ thus estimated, ranging from $\sim$700 in B213-C12-1 to
$\sim$260 in B213-C2-1, the later being more consistent with the values measured in protostars. Interestingly, eight cores present  R$_{^{14}N/^{15}N}$ $>$ 441, and half of them are located in B 213. One could think that these surprisingly high values of  R$_{^{14}N/^{15}N}$ are the consequence of assumed value of the $^{12}$C/$^{13}$C ratio.  Indeed, some starless core in B 213 present DNC/DN$^{13}$C $<$ 68 $\pm$15 (see Fig.~\ref{fig:12C_13C_ISM}). In order to test this possibility, we re-calculated R$_{^{14}N/^{15}N}$ assuming that HCN/H$^{13}$CN ratio is the same as the estimated DNC/DN$^{13}$C in the cores of our sample. The average value of this R$_{^{14}N/^{15}N}$ for the cores where DN$^{13}$C is detected is 332, similar to the ISM value of 330 \citep{Hily-Blant2019}.

Gas fractionation reactions, selective photodissociation, and selective adsorption and diffusion on the grain surfaces can alter the isotopic fractionation.   
  \citealp{Hily-Blant2020} presented time dependent calculations of the evolution of the $^{14}$N/$^{15}$N isotopic ratio of several  species during the early phases of the gravitational collapse. They concluded that enrichment of $^{15}$N-containing species can occur during the collapse due to the higher mass of $^{15}$N species and hence a lower rate of thermal collisions with grains and therefore adsorption to grains. However, the impact of gas-phase fractionation reactions in the $^{14}$N/$^{15}$N isotopic ratio should be low during gravitational collapse because the timescales for isotopic fractionation seem to be longer than those for core collapse \citep{Hily-Blant2020}. This is consistent with the lack of correlation between  R$_{^{14}N/^{15}N}$ and D$_{frac}$ (N$_2$H$^+$) in our sample as shown in Fig.~\ref{fig:Dfrac-14N/15N}. 
  Selective photodissociation has also been proposed to explain low values of the  $^{14}$N/$^{15}$N ratio in starless cores. Indeed, selective photodissociation might explain the variation of the  HC$^{14}$N/HC$^{15}$N in B 213. The HC$^{14}$N/HC$^{15}$N ratio is lower in the northern part of the filament where the protostars density is higher. However, when we considered the whole of our sample, we find a weak correlation of the HC$^{14}$N/HC$^{15}$N ratio  with temperature, which is contrary to that expected if selective photodissociation were driving R$_{^{14}N/^{15}N}$. This suggests that selective photodissociation is not the main agent driving  R$_{^{14}N/^{15}N}$ in our sample of starless cores.

As shown in Fig.~\ref{fig:nitro_1415N}, R$_{^{14}N/^{15}N}$ decreases not only during the collapse but also during the protostellar phase: starless cores present the highest values of R$_{^{14}N/^{15}N}$, with a mean value of  429; in the next phase (protostars) R$_{^{14}N/^{15}N}$ was reduced by 37$\%$ to an average value of 270, and even more so in protoplanetary disks (by 72$\%$), where the mean value is 120.  This trend is different from that observed in the case of deuteration, for which the D/H values observed in protostars are similar to those in starless cores, suggesting that it is settled during the gravitational collapse
(see Fig.~\ref{fig:deut_comp_sour}). We should bear in mind that the protostars considered in Fig.~\ref{fig:deut_comp_sour} are mostly Class 0, i. e., objects embedded in a cold and dense envelope, still unaffected by the recently born star; a greater decrease in deuterium fraction would be expected in more evolved protostars.  On the contrary, R$_{^{14}N/^{15}N}$  evolves during the whole process of star and planet formation with a variation of a factor of $\sim$ 4 between starless cores and protoplanetary disks. However, it should be noted that the observations of  R$_{^{14}N/^{15}N}$ in Class 0 and I protostars and protoplanetary disks are still scarce and could be biased by regions, insufficient for a reliable statistics.  Also important, the $^{14}$N/$^{15}$N  ratio is expected to vary for different species depending on their formation routes and the chemical history of the cloud \citep{Spilla2023}.
High angular observations of the  $^{14}$N/$^{15}$N ratio in different species  toward young protostars and protoplanetary disks are needed to disentangle the underlying chemistry. In addition, 
detailed chemo-dynamical models fully accounting for the surface chemistry are required to further understand the evolution of nitrogen fractionation and the possible link between the values measured in molecular clouds and those in the Solar System bodies.

\section{Summary and conclusions} \label{sect:conclusions}
In this work, we have presented a chemical study of 23 cores in the star-forming regions of Taurus, Perseus, and Orion. Using millimeter IRAM 30m observations and the radiative transfer code RADEX, we obtained the column densities of HCN, HNC, N$_2$H$^+$, and their isotopologs to study their evolution and chemistry through molecular isotopic and isomeric ratios. Our conclusions are outlined as follows:  
\begin{itemize}
    \item[-] We have obtained an average $^{12}$C/$^{13}$C ratio from DNC/DN$^{13}$C of 49 $\pm$ 13, which is consistent with the ISM value of 68 $\pm$ 15 as measured from CN.   

    \item[-] Our results for the isomeric ratios H$^{13}$CN/HN$^{13}$C,  HC$^{15}$N/H$^{15}$NC, and DCN/DNC are in good agreement with the expected value of $\sim 1$ for cold cores.     

    \item[-] 
    We studied deuterium fractionation through DNC/HNC, DCN/HCN, and N$_2$D$^+$/N$_2$H$^+$, and their average values are 0.054 $\pm$ 0.019, 0.036 $\pm$ 0.033, and 0.15 $\pm$ 0.11, respectively. The deuterium fractions of the three species show a weak correlation with temperature, 
    with D$_{frac}$ (N$_2$H$^+$) presenting the greatest amount of scattering. The starless cores that present the highest (lowest) deuteration are B213-C1-1 and B213-C6-1 (B213-C12-1) in Taurus and L1448-1 and 79-C1-1 (IC348-10) in Perseus. When comparing the three different molecular clouds instead, we shifted our attention to the environmental effects. We found the highest deuterations in Taurus, with more quiescent properties than Perseus and Orion.   
    \item[-] We found that the $^{14}$N/$^{15}$N is different between HCN/HC$^{15}$N ($\sim$ 430) and HNC/H$^{15}$NC ($\sim$ 300). No correlation was found between these ratios and deuterium fractions. While the local variations among starless cores within the same region can be attributed to core evolution and/or selective photodissociation, these mechanisms alone cannot explain the results in all the studied regions.
   \item[-] We used DNAUTILUS 2.0 to model the deuteration of HCN, HNC, and N$_2$H$^+$ in our sample.  
   In Taurus, a model with n$_{\rm H_2}$=10$^5$ cm$^{-3}$, T = 10 K, and $\zeta_{H_2}$ =  1.3 10$^{-17}$ s$^{-1}$ reproduces the observed D$_{frac}$ values in the cores located in Taurus for t $\sim$0.2-0.3 Myr. In Perseus and Orion, where temperatures are slightly higher (T = 15 K), we obtained a chemical age of 0.3-0.5 Myr. As the derived ages are temperature-dependent with this 0D model, more complex models would be required to accurately constrain the age.   Nevertheless, since deuteration increases with time, we used D$_{frac}$ (N$_2$H$^+$) as an evolutionary tracer, and the cores classified as more evolved (B213-C1-1, B213-C6-1, and B213-C16-1 in Taurus and L1448-1, 79-C1-1, 1333-C4-1, and 1333-C3-14 in Perseus) also present higher n(DNC)/n(CS). 
    \item[-] When comparing our HCN/HC$^{15}$N results (using the indirect method, $\sim$ 430) with those obtained in the literature in sources at different evolutionary stages, we observed an apparent decreasing trend toward protostars ($\sim$ 270) and protoplanetary disks ($\sim$ 120). In contrast, deuterium fractions in protostars remain similar to those in starless cores. 
    This suggests that the $^{15}$N enrichment of HCN proceeds during the protostellar phase until the formation of the protoplanetary disk, whereas D$_{frac}$ is established during gravitational collapse. Selective adsorption on grain surfaces during the Class 0 phase and selective photodissociation in the more evolved Class I protostars could contribute to this HC$^{15}$N enhancement.
\end{itemize}

Isotopic ratios have been used as chemical diagnostics to investigate the origin of material in the Solar System. In this work, we used data from the large IRAM 30m program GEMS and an additional IRAM 30m project to constrain the D/H isotopic ratios of HCN, HNC, and N$_2$H$^+$ as well as the $^{14}$N/$^{15}$N ratio of HCN and HNC in a sample of 23 cores located in the star forming regions of Taurus, Perseus, and Orion. We used the chemical code DNautilus to determine the evolutionary stage of the cores from the observed values. Our large sample size allowed us to investigate the possible influence of the environment on the isotopic ratios. 
Since the deuterium fractionation of HCN, HNC, and N$_2$H$^+$ depends on the evolutionary stage of the core and the environment, 
we can use the deuterium fractions as chemical clocks in starless cores with similar initial conditions. The HNC/H$^{15}$NC and HCN/HC$^{15}$N are not correlated with D$_{frac}$, suggesting that the detected variations are not correlated with the evolutionary stage of the core. Although, with a large dispersion, the average value of the HCN/HC$^{15}$N ratio in our sample is significantly higher than the values measured in protostars and protoplanetary disks, suggesting that nitrogen fractionation processes take place during the protostellar phase. 

\begin{acknowledgements}
This project has received funding from the European Research Council (ERC) under the European Union$’$s Horizon Europe research and innovation programme ERC-AdG-2022 (GA No. 101096293). Funded by the European Union. Views and opinions expressed are however those of the author(s) only and do not necessarily reflect those of the European Union or the European Research Council Executive Agency. Neither the European Union nor the granting authority can be held responsible for them. AF, GE, and MRB also thanks project PID2022-137980NB-I00 funded by the Spanish Ministry of Science and Innovation/State Agency of Research MCIN/AEI/10.13039/501100011033 and by $"$ERDF A way of making Europe$"$. L.M. acknowledges the financial support provided by DAE and the DST-SERB research grant (MTR/2021/000864) from the Government of India for this work. This research has made use of spectroscopic and collisional data from the EMAA database (\url{https://emaa.osug.fr} and \url{https://dx.doi.org/10.17178/EMAA}). EMAA is supported by the Observatoire des Sciences de l’Univers de Grenoble (OSUG).
\end{acknowledgements}

\bibliographystyle{aa} 
\bibliography{refs} 

\begin{thebibliography}{114}
\expandafter\ifx\csname natexlab\endcsname\relax\def\natexlab#1{#1}\fi

\bibitem[{{Adande} \& {Ziurys}(2012)}]{Adande2012}
{Adande}, G.~R. \& {Ziurys}, L.~M. 2012, \apj, 744, 194

\bibitem[{{Albertsson} {et~al.}(2013){Albertsson}, {Semenov}, {Vasyunin},
  {Henning}, \& {Herbst}}]{Albertsson2013}
{Albertsson}, T., {Semenov}, D.~A., {Vasyunin}, A.~I., {Henning}, T., \&
  {Herbst}, E. 2013, \apjs, 207, 27

\bibitem[{{Al{\'e}on}(2010)}]{Aleon2010}
{Al{\'e}on}, J. 2010, \apj, 722, 1342

\bibitem[{{Ambrose} {et~al.}(2021){Ambrose}, {Shirley}, \&
  {Scibelli}}]{Ambrose2021}
{Ambrose}, H.~E., {Shirley}, Y.~L., \& {Scibelli}, S. 2021, \mnras, 501, 347

\bibitem[{{Andr{\'e}} {et~al.}(2010){Andr{\'e}}, {Men'shchikov}, {Bontemps},
  {et~al.}}]{Andre2010}
{Andr{\'e}}, P., {Men'shchikov}, A., {Bontemps}, S., {et~al.} 2010, \aap, 518,
  L102

\bibitem[{{Bacmann} {et~al.}(2003){Bacmann}, {Lefloch}, {Ceccarelli},
  {Steinacker}, {Castets}, \& {Loinard}}]{Bacmann2003}
{Bacmann}, A., {Lefloch}, B., {Ceccarelli}, C., {et~al.} 2003, \apjl, 585, L55

\bibitem[{{Bernard} {et~al.}(2010){Bernard}, {Paradis}, {Marshall}, {Montier},
  {Lagache}, {Paladini}, {Veneziani}, {Brunt}, {Mottram}, {Martin},
  {Ristorcelli}, {Noriega-Crespo}, {Compi{\`e}gne}, {Flagey}, {Anderson},
  {Popescu}, {Tuffs}, {Reach}, {White}, {Benedettini}, {Calzoletti},
  {Digiorgio}, {Faustini}, {Juvela}, {Joblin}, {Joncas}, {Mivilles-Deschenes},
  {Olmi}, {Traficante}, {Piacentini}, {Zavagno}, \& {Molinari}}]{Bernard2010}
{Bernard}, J.~P., {Paradis}, D., {Marshall}, D.~J., {et~al.} 2010, \aap, 518,
  L88

\bibitem[{{Bizzocchi} {et~al.}(2013){Bizzocchi}, {Caselli}, {Leonardo}, \&
  {Dore}}]{14-15N_highRatio}
{Bizzocchi}, L., {Caselli}, P., {Leonardo}, E., \& {Dore}, L. 2013, \aap, 555,
  A109

\bibitem[{{Bockel{\'e}e-Morvan} {et~al.}(2008){Bockel{\'e}e-Morvan}, {Biver},
  {Jehin}, {Cochran}, {Wiesemeyer}, {Manfroid}, {Hutsem{\'e}kers}, {Arpigny},
  {Boissier}, {Cochran}, {Colom}, {Crovisier}, {Milutinovic}, {Moreno},
  {Prochaska}, {Ramirez}, {Schulz}, \& {Zucconi}}]{14N15N_6B}
{Bockel{\'e}e-Morvan}, D., {Biver}, N., {Jehin}, E., {et~al.} 2008, \apjl, 679,
  L49

\bibitem[{{Bonal} {et~al.}(2010){Bonal}, {Huss}, {Krot}, {Nagashima}, {Ishii},
  \& {Bradley}}]{Bonal2010}
{Bonal}, L., {Huss}, G.~R., {Krot}, A.~N., {et~al.} 2010, \gca, 74, 6590

\bibitem[{{Caselli}(2002)}]{Caselli2002a}
{Caselli}, P. 2002, \planss, 50, 1133

\bibitem[{{Caselli} \& {Ceccarelli}(2012{\natexlab{a}})}]{Caselli2012}
{Caselli}, P. \& {Ceccarelli}, C. 2012{\natexlab{a}}, \aapr, 20, 56

\bibitem[{{Caselli} \&
  {Ceccarelli}(2012{\natexlab{b}})}]{CaselliCeccarelli2012}
{Caselli}, P. \& {Ceccarelli}, C. 2012{\natexlab{b}}, \aapr, 20, 56

\bibitem[{{Caselli} {et~al.}(2002){Caselli}, {Walmsley}, {Zucconi}, {Tafalla},
  {Dore}, \& {Myers}}]{Caselli2002}
{Caselli}, P., {Walmsley}, C.~M., {Zucconi}, A., {et~al.} 2002, \apj, 565, 344

\bibitem[{Ceccarelli {et~al.}(2014)Ceccarelli, Caselli, Bockelée-Morvan,
  {et~al.}}]{Ceccarelli_2014Adrians}
Ceccarelli, C., Caselli, P., Bockelée-Morvan, D., {et~al.} 2014, Deuterium
  Fractionation: The Ariadne’s Thread from the Precollapse Phase to
  Meteorites and Comets Today (University of Arizona Press)

\bibitem[{{Colzi} {et~al.}(2018{\natexlab{a}}){Colzi}, {Fontani}, {Caselli},
  {Ceccarelli}, {Hily-Blant}, \& {Bizzocchi}}]{Colzi2018a}
{Colzi}, L., {Fontani}, F., {Caselli}, P., {et~al.} 2018{\natexlab{a}}, \aap,
  609, A129

\bibitem[{{Colzi} {et~al.}(2018{\natexlab{b}}){Colzi}, {Fontani}, {Rivilla},
  {S{\'a}nchez-Monge}, {Testi}, {Beltr{\'a}n}, \& {Caselli}}]{Colzi2018b}
{Colzi}, L., {Fontani}, F., {Rivilla}, V.~M., {et~al.} 2018{\natexlab{b}},
  \mnras

\bibitem[{{Colzi} {et~al.}(2018{\natexlab{c}}){Colzi}, {Fontani}, {Rivilla},
  {et~al.}}]{Colzi2018}
{Colzi}, L., {Fontani}, F., {Rivilla}, V.~M., {et~al.} 2018{\natexlab{c}},
  \mnras, 478, 3693

\bibitem[{Crapsi {et~al.}(2005)Crapsi, Caselli, Walmsley, Myers, Tafalla, Lee,
  \& Bourke}]{Crapsi_2005}
Crapsi, A., Caselli, P., Walmsley, C.~M., {et~al.} 2005, The Astrophysical
  Journal, 619, 379

\bibitem[{{Crapsi} {et~al.}(2007){Crapsi}, {Caselli}, {Walmsley}, \&
  {Tafalla}}]{Crapsi&Caselli}
{Crapsi}, A., {Caselli}, P., {Walmsley}, M.~C., \& {Tafalla}, M. 2007, \aap,
  470, 221

\bibitem[{{Dalgarno} \& {Lepp}(1984)}]{Dalgarno1984}
{Dalgarno}, A. \& {Lepp}, S. 1984, \apjl, 287, L47

\bibitem[{{Daniel} {et~al.}(2005){Daniel}, {Dubernet}, {Meuwly}, {Cernicharo},
  \& {Pagani}}]{col_N2H+_hfs}
{Daniel}, F., {Dubernet}, M.~L., {Meuwly}, M., {Cernicharo}, J., \& {Pagani},
  L. 2005, \mnras, 363, 1083

\bibitem[{{Daniel} {et~al.}(2013{\natexlab{a}}){Daniel}, {G{\'e}rin}, {Roueff},
  {Cernicharo}, {Marcelino}, {Lique}, {Lis}, {Teyssier}, {Biver}, \&
  {Bockel{\'e}e-Morvan}}]{F.Daniel2013}
{Daniel}, F., {G{\'e}rin}, M., {Roueff}, E., {et~al.} 2013{\natexlab{a}}, \aap,
  560, A3

\bibitem[{{Daniel} {et~al.}(2013{\natexlab{b}}){Daniel}, {G{\'e}rin}, {Roueff},
  {Cernicharo}, {Marcelino}, {Lique}, {Lis}, {Teyssier}, {Biver}, \&
  {Bockel{\'e}e-Morvan}}]{Daniel2013}
{Daniel}, F., {G{\'e}rin}, M., {Roueff}, E., {et~al.} 2013{\natexlab{b}}, \aap,
  560, A3

\bibitem[{{Dislaire} {et~al.}(2012){Dislaire}, {Hily-Blant}, {Faure}, {Maret},
  {Bacmann}, \& {Pineau Des For{\^e}ts}}]{Dislaire2012}
{Dislaire}, V., {Hily-Blant}, P., {Faure}, A., {et~al.} 2012, \aap, 537, A20

\bibitem[{{Dumouchel} {et~al.}(2011){Dumouchel}, {K{\l}os}, \&
  {Lique}}]{col_HNC}
{Dumouchel}, F., {K{\l}os}, J., \& {Lique}, F. 2011, Physical Chemistry
  Chemical Physics (Incorporating Faraday Transactions), 13, 8204

\bibitem[{{Emprechtinger} {et~al.}(2009){Emprechtinger}, {Caselli}, {Volgenau},
  {Stutzki}, \& {Wiedner}}]{M.Emprechtinger2009}
{Emprechtinger}, M., {Caselli}, P., {Volgenau}, N.~H., {Stutzki}, J., \&
  {Wiedner}, M.~C. 2009, \aap, 493, 89

\bibitem[{{Endres} {et~al.}(2016){Endres}, {Schlemmer}, {Schilke}, {Stutzki},
  \& {M{\"u}ller}}]{CDMS}
{Endres}, C.~P., {Schlemmer}, S., {Schilke}, P., {Stutzki}, J., \&
  {M{\"u}ller}, H. S.~P. 2016, Journal of Molecular Spectroscopy, 327, 95

\bibitem[{{Esplugues} {et~al.}(2022){Esplugues}, {Fuente}, {Navarro-Almaida},
  {et~al.}}]{GEMS-VI-GiselaH2CS}
{Esplugues}, G., {Fuente}, A., {Navarro-Almaida}, D., {et~al.} 2022, \aap, 662,
  A52

\bibitem[{Esplugues {et~al.}(2013)Esplugues, Cernicharo, Viti, Goicoechea,
  Tercero, Marcelino, Palau, Bell, Bergin, Crockett, \& Wang}]{Esplugues_2013}
Esplugues, G.~B., Cernicharo, J., Viti, S., {et~al.} 2013, \aap, 559, A51

\bibitem[{{Fontani} {et~al.}(2015){Fontani}, {Caselli}, {Palau}, {Bizzocchi},
  \& {Ceccarelli}}]{Fontani2015}
{Fontani}, F., {Caselli}, P., {Palau}, A., {Bizzocchi}, L., \& {Ceccarelli}, C.
  2015, \apjl, 808, L46

\bibitem[{{Foster} {et~al.}(2015){Foster}, {Cottaar}, {Covey}, {Arce}, {Meyer},
  {Nidever}, {Stassun}, {Tan}, {Chojnowski}, {da Rio}, {Flaherty}, {Rebull},
  {Frinchaboy}, {Majewski}, {Skrutskie}, {Wilson}, \&
  {Zasowski}}]{YSOcandidate}
{Foster}, J.~B., {Cottaar}, M., {Covey}, K.~R., {et~al.} 2015, \apj, 799, 136

\bibitem[{{Fouchet} {et~al.}(2004){Fouchet}, {Irwin}, {Parrish}, {Calcutt},
  {Taylor}, {Nixon}, \& {Owen}}]{Fouchet2004}
{Fouchet}, T., {Irwin}, P. G.~J., {Parrish}, P., {et~al.} 2004, \icarus, 172,
  50

\bibitem[{{Fuente} {et~al.}(2016){Fuente}, {Cernicharo}, {Roueff}, {Gerin},
  {Pety}, {Marcelino}, {Bachiller}, {Lefloch}, {Roncero}, \&
  {Aguado}}]{Asun2016}
{Fuente}, A., {Cernicharo}, J., {Roueff}, E., {et~al.} 2016, \aap, 593, A94

\bibitem[{{Fuente} {et~al.}(2019){Fuente}, {Navarro}, {Caselli},
  {et~al.}}]{GEMS-I-Asun}
{Fuente}, A., {Navarro}, D.~G., {Caselli}, P., {et~al.} 2019, \aap, 624, A105

\bibitem[{{Fuente} {et~al.}(2023){Fuente}, {Rivi{\`e}re-Marichalar},
  {Beitia-Antero}, {Caselli}, {Wakelam}, {Esplugues}, {Rodr{\'\i}guez-Baras},
  {Navarro-Almaida}, {Gerin}, {Kramer}, {Bachiller}, {Goicoechea},
  {Jim{\'e}nez-Serra}, {Loison}, {Ivlev}, {Mart{\'\i}n-Dom{\'e}nech},
  {Spezzano}, {Roncero}, {Mu{\~n}oz-Caro}, {Cazaux}, \&
  {Marcelino}}]{GEMS-VII-Asun2023}
{Fuente}, A., {Rivi{\`e}re-Marichalar}, P., {Beitia-Antero}, L., {et~al.} 2023,
  \aap, 670, A114

\bibitem[{{F{\"u}ri} \& {Marty}(2015{\natexlab{a}})}]{EvelynFury2015}
{F{\"u}ri}, E. \& {Marty}, B. 2015{\natexlab{a}}, Nature Geoscience, 8, 515

\bibitem[{{F{\"u}ri} \& {Marty}(2015{\natexlab{b}})}]{Furi2015}
{F{\"u}ri}, E. \& {Marty}, B. 2015{\natexlab{b}}, Nature Geoscience, 8, 515

\bibitem[{{Giers} {et~al.}(2023){Giers}, {Spezzano}, {Caselli}, {Wirstr{\"o}m},
  {Sipil{\"a}}, {Pineda}, {Redaelli}, {Bop}, \& {Lique}}]{K.Giers2023}
{Giers}, K., {Spezzano}, S., {Caselli}, P., {et~al.} 2023, \aap, 676, A78

\bibitem[{Giers {et~al.}(2023)Giers, Spezzano, Caselli, WirstrÃ¶m, SipilÃ¤,
  Pineda, Redaelli, Bop, \& Lique}]{Giers_2023-DCNcolcoef}
Giers, K., Spezzano, S., Caselli, P., {et~al.} 2023, \aap, 676, A78

\bibitem[{{Goldsmith} {et~al.}(2008){Goldsmith}, {Heyer}, {Narayanan}, {Snell},
  {Li}, \& {Brunt}}]{Goldsmith2008}
{Goldsmith}, P.~F., {Heyer}, M., {Narayanan}, G., {et~al.} 2008, \apj, 680, 428

\bibitem[{{Guzm{\'a}n} {et~al.}(2017){Guzm{\'a}n}, {{\"O}berg}, {Huang},
  {Loomis}, \& {Qi}}]{14N15N_5G}
{Guzm{\'a}n}, V.~V., {{\"O}berg}, K.~I., {Huang}, J., {Loomis}, R., \& {Qi}, C.
  2017, \apj, 836, 30

\bibitem[{{Hacar} {et~al.}(2020){Hacar}, {Bosman}, \& {van
  Dishoeck}}]{Hacar2020_isomers}
{Hacar}, A., {Bosman}, A.~D., \& {van Dishoeck}, E.~F. 2020, \aap, 635, A4

\bibitem[{{Hacar} {et~al.}(2013){Hacar}, {Tafalla}, {Kauffmann},
  {et~al.}}]{Hacar2013}
{Hacar}, A., {Tafalla}, M., {Kauffmann}, J., {et~al.} 2013, \aap, 554, A55

\bibitem[{{Hatchell} {et~al.}(2005){Hatchell}, {Richer}, {Fuller},
  {Qualtrough}, {Ladd}, \& {Chandler}}]{Hatchell2005}
{Hatchell}, J., {Richer}, J.~S., {Fuller}, G.~A., {et~al.} 2005, \aap, 440, 151

\bibitem[{{Hily-Blant} {et~al.}(2013){Hily-Blant}, {Bonal}, {Faure}, \&
  {Quirico}}]{Hily-Blant2013}
{Hily-Blant}, P., {Bonal}, L., {Faure}, A., \& {Quirico}, E. 2013, \icarus,
  223, 582

\bibitem[{{Hily-Blant} {et~al.}(2017){Hily-Blant}, {Magalhaes}, {Kastner},
  {Faure}, {Forveille}, \& {Qi}}]{Hily-Blant2017}
{Hily-Blant}, P., {Magalhaes}, V., {Kastner}, J., {et~al.} 2017, \aap, 603, L6

\bibitem[{{Hily-Blant} {et~al.}(2019){Hily-Blant}, {Magalhaes de Souza},
  {Kastner}, \& {Forveille}}]{Hily-Blant2019}
{Hily-Blant}, P., {Magalhaes de Souza}, V., {Kastner}, J., \& {Forveille}, T.
  2019, \aap, 632, L12

\bibitem[{{Hily-Blant} {et~al.}(2020){Hily-Blant}, {Pineau des For{\^e}ts},
  {Faure}, \& {Flower}}]{Hily-Blant2020}
{Hily-Blant}, P., {Pineau des For{\^e}ts}, G., {Faure}, A., \& {Flower}, D.~R.
  2020, \aap, 643, A76

\bibitem[{{Jensen} {et~al.}(2024){Jensen}, {Spezzano}, {Caselli}, {Sipil{\"a}},
  {Redaelli}, {Giers}, \& {Ferrer Asensio}}]{Jensen}
{Jensen}, S.~S., {Spezzano}, S., {Caselli}, P., {et~al.} 2024, \aap, 685, A149

\bibitem[{{Kahane} {et~al.}(2018){Kahane}, {Jaber Al-Edhari}, {Ceccarelli},
  {L{\'o}pez-Sepulcre}, {Fontani}, \& {Kama}}]{Kahane2018}
{Kahane}, C., {Jaber Al-Edhari}, A., {Ceccarelli}, C., {et~al.} 2018, \apj,
  852, 130

\bibitem[{{Kashyap} {et~al.}(2024){Kashyap}, {Majumdar}, {Dutrey},
  {Guilloteau}, {Willacy}, {Chapillon}, {Teague}, {Semenov}, {Henning},
  {Turner}, {Sahai}, {K{\'o}sp{\'a}l}, {Coutens}, {Pi{\'e}tu}, {Gratier},
  {Ruaud}, {Phuong}, {Di Folco}, {Lee}, \& {Tang}}]{Kashyap2024}
{Kashyap}, P., {Majumdar}, L., {Dutrey}, A., {et~al.} 2024, \apj, 976, 258

\bibitem[{{Kim} {et~al.}(2020){Kim}, {Lee}, {Gopinathan}, {Tafalla}, {Sohn},
  {Kim}, {Kim}, {Soam}, \& {Myers}}]{Kim2020}
{Kim}, S., {Lee}, C.~W., {Gopinathan}, M., {et~al.} 2020, \apj, 891, 169

\bibitem[{{K{\"o}nyves} {et~al.}(2015){K{\"o}nyves}, {Andr{\'e}},
  {Men'shchikov}, {Palmeirim}, {Arzoumanian}, {Schneider}, {Roy}, {Didelon},
  {Maury}, {Shimajiri}, {Di Francesco}, {Bontemps}, {Peretto}, {Benedettini},
  {Bernard}, {Elia}, {Griffin}, {Hill}, {Kirk}, {Ladjelate}, {Marsh}, {Martin},
  {Motte}, {Nguy{\^e}n Luong}, {Pezzuto}, {Roussel}, {Rygl}, {Sadavoy},
  {Schisano}, {Spinoglio}, {Ward-Thompson}, \& {White}}]{Konyves2015}
{K{\"o}nyves}, V., {Andr{\'e}}, P., {Men'shchikov}, A., {et~al.} 2015, \aap,
  584, A91

\bibitem[{Langer {et~al.}(1996)Langer, Velusamy, \& Xie}]{Barnard5}
Langer, W.~D., Velusamy, T., \& Xie, T. 1996, The Astrophysical Journal, 468,
  L41

\bibitem[{{L{\'e}cuyer} {et~al.}(1998){L{\'e}cuyer}, {Gillet}, \&
  {Robert}}]{Lecuyer1998}
{L{\'e}cuyer}, C., {Gillet}, P., \& {Robert}, F. 1998, Chemical Geology, 145,
  249

\bibitem[{{Linsky}(2007)}]{Linsky2007}
{Linsky}, J.~L. 2007, \ssr, 130, 367

\bibitem[{{Loison} {et~al.}(2020){Loison}, {Wakelam}, {Gratier}, \&
  {Hickson}}]{Loison2020}
{Loison}, J.-C., {Wakelam}, V., {Gratier}, P., \& {Hickson}, K.~M. 2020,
  \mnras, 498, 4663

\bibitem[{{Lombardi} {et~al.}(2014){Lombardi}, {Bouy}, {Alves}, \&
  {Lada}}]{Lombardi2014}
{Lombardi}, M., {Bouy}, H., {Alves}, J., \& {Lada}, C.~J. 2014, \aap, 566, A45

\bibitem[{{Lucas} \& {Liszt}(1998)}]{Lucas1998}
{Lucas}, R. \& {Liszt}, H. 1998, \aap, 337, 246

\bibitem[{{Magalh{\~a}es} {et~al.}(2018){Magalh{\~a}es}, {Hily-Blant}, {Faure},
  {Hernandez-Vera}, \& {Lique}}]{Magalh_es_2018-H13CNcolcoef}
{Magalh{\~a}es}, V.~S., {Hily-Blant}, P., {Faure}, A., {Hernandez-Vera}, M., \&
  {Lique}, F. 2018, \aap, 615, A52

\bibitem[{{Majumdar} {et~al.}(2017){Majumdar}, {Gratier}, {Ruaud}, {Wakelam},
  {Vastel}, {Sipil{\"a}}, {Hersant}, {Dutrey}, \& {Guilloteau}}]{Majumdar2017}
{Majumdar}, L., {Gratier}, P., {Ruaud}, M., {et~al.} 2017, \mnras, 466, 4470

\bibitem[{{Malinen} {et~al.}(2012){Malinen}, {Juvela}, {Rawlings},
  {Ward-Thompson}, {Palmeirim}, \& {Andr{\'e}}}]{Malinen2012}
{Malinen}, J., {Juvela}, M., {Rawlings}, M.~G., {et~al.} 2012, \aap, 544, A50

\bibitem[{{Marsh} {et~al.}(2014){Marsh}, {Griffin}, {Palmeirim}, {Andr{\'e}},
  {Kirk}, {Stamatellos}, {Ward-Thompson}, {Roy}, {Bontemps}, {di Francesco},
  {Elia}, {Hill}, {K{\"o}nyves}, {Motte}, {Nguyen-Luong}, {Peretto}, {Pezzuto},
  {Rivera-Ingraham}, {Schneider}, {Spinoglio}, \& {White}}]{Marsh2014}
{Marsh}, K.~A., {Griffin}, M.~J., {Palmeirim}, P., {et~al.} 2014, \mnras, 439,
  3683

\bibitem[{{Mart{\'\i}nez-Barbosa} {et~al.}(2015){Mart{\'\i}nez-Barbosa},
  {Brown}, \& {Portegies Zwart}}]{Barbosa2015}
{Mart{\'\i}nez-Barbosa}, C.~A., {Brown}, A.~G.~A., \& {Portegies Zwart}, S.
  2015, \mnras, 446, 823

\bibitem[{{Marty} {et~al.}(2011){Marty}, {Chaussidon}, {Wiens}, {Jurewicz}, \&
  {Burnett}}]{Marty2011}
{Marty}, B., {Chaussidon}, M., {Wiens}, R.~C., {Jurewicz}, A.~J.~G., \&
  {Burnett}, D.~S. 2011, Science, 332, 1533

\bibitem[{{McKee}(1989)}]{McKee1989}
{McKee}, C.~F. 1989, \apj, 345, 782

\bibitem[{{Melosso} {et~al.}(2020){Melosso}, {Bizzocchi}, {Sipil{\"a}},
  {Giuliano}, {Dore}, {Tamassia}, {Martin-Drumel}, {Pirali}, {Redaelli}, \&
  {Caselli}}]{Melosso2020}
{Melosso}, M., {Bizzocchi}, L., {Sipil{\"a}}, O., {et~al.} 2020, \aap, 641,
  A153

\bibitem[{{Milam} {et~al.}(2005){Milam}, {Savage}, {Brewster}, {Ziurys}, \&
  {Wyckoff}}]{12C-13Cratio}
{Milam}, S.~N., {Savage}, C., {Brewster}, M.~A., {Ziurys}, L.~M., \& {Wyckoff},
  S. 2005, \apj, 634, 1126

\bibitem[{{Millar}(2005)}]{Millar2005}
{Millar}, T.~J. 2005, Astronomy and Geophysics, 46, 2.29

\bibitem[{{Millar} {et~al.}(1989){Millar}, {Bennett}, \& {Herbst}}]{Millar1989}
{Millar}, T.~J., {Bennett}, A., \& {Herbst}, E. 1989, \apj, 340, 906

\bibitem[{{Minchev} {et~al.}(2013){Minchev}, {Chiappini}, \&
  {Martig}}]{Minchev2013}
{Minchev}, I., {Chiappini}, C., \& {Martig}, M. 2013, \aap, 558, A9

\bibitem[{{M{\"u}ller} {et~al.}(2005){M{\"u}ller}, {Schl{\"o}der}, {Stutzki},
  \& {Winnewisser}}]{CDMS_Moller2005}
{M{\"u}ller}, H. S.~P., {Schl{\"o}der}, F., {Stutzki}, J., \& {Winnewisser}, G.
  2005, Journal of Molecular Structure, 742, 215

\bibitem[{{M{\"u}ller} {et~al.}(2001){M{\"u}ller}, {Thorwirth}, {Roth}, \&
  {Winnewisser}}]{CDMS_Muller2001}
{M{\"u}ller}, H.~S.~P., {Thorwirth}, S., {Roth}, D.~A., \& {Winnewisser}, G.
  2001, \aap, 370, L49

\bibitem[{{Navarro-Almaida} {et~al.}(2023){Navarro-Almaida}, {Bop}, {Lique},
  {Esplugues}, {Rodr{\'\i}guez-Baras}, {Kramer}, {Romero}, {Fuente}, {Caselli},
  {Rivi{\`e}re-Marichalar}, {Kirk}, {Chac{\'o}n-Tanarro}, {Roueff},
  {Mroczkowski}, {Bhandarkar}, {Devlin}, {Dicker}, {Lowe}, {Mason}, {Sarazin},
  \& {Sievers}}]{Navarro_Linking}
{Navarro-Almaida}, D., {Bop}, C.~T., {Lique}, F., {et~al.} 2023, \aap, 670,
  A110

\bibitem[{{Navarro-Almaida} {et~al.}(2021){Navarro-Almaida}, {Fuente},
  {Majumdar}, {Wakelam}, {Caselli}, {Rivi{\`e}re-Marichalar},
  {Trevi{\~n}o-Morales}, {Cazaux}, {Jim{\'e}nez-Serra}, {Kramer},
  {Chac{\'o}n-Tanarro}, {Kirk}, {Ward-Thompson}, \& {Tafalla}}]{Navarro2021}
{Navarro-Almaida}, D., {Fuente}, A., {Majumdar}, L., {et~al.} 2021, \aap, 653,
  A15

\bibitem[{{Owen} {et~al.}(2001){Owen}, {Mahaffy}, {Niemann}, {Atreya}, \&
  {Wong}}]{Owen2001}
{Owen}, T., {Mahaffy}, P.~R., {Niemann}, H.~B., {Atreya}, S., \& {Wong}, M.
  2001, \apjl, 553, L77

\bibitem[{{Padovani} {et~al.}(2009){Padovani}, {Galli}, \&
  {Glassgold}}]{Padovani2009}
{Padovani}, M., {Galli}, D., \& {Glassgold}, A.~E. 2009, \aap, 501, 619

\bibitem[{Pagani {et~al.}(1992)Pagani, Salez, \& Wannier}]{Pagani1992}
Pagani, L., Salez, M., \& Wannier, P. 1992, \aap, 258,
  479

\bibitem[{{Pagani} {et~al.}(2009){Pagani}, {Vastel}, {Hugo}, {Kokoouline},
  {Greene}, {Bacmann}, {Bayet}, {Ceccarelli}, {Peng}, \&
  {Schlemmer}}]{Pagani2009}
{Pagani}, L., {Vastel}, C., {Hugo}, E., {et~al.} 2009, \aap, 494, 623

\bibitem[{{Palmeirim} {et~al.}(2013{\natexlab{a}}){Palmeirim}, {Andr{\'e}},
  {Kirk}, {Ward-Thompson}, {Arzoumanian}, {K{\"o}nyves}, {Didelon},
  {Schneider}, {Benedettini}, {Bontemps}, {Di Francesco}, {Elia}, {Griffin},
  {Hennemann}, {Hill}, {Martin}, {Men'shchikov}, {Molinari}, {Motte}, {Nguyen
  Luong}, {Nutter}, {Peretto}, {Pezzuto}, {Roy}, {Rygl}, {Spinoglio}, \&
  {White}}]{B213}
{Palmeirim}, P., {Andr{\'e}}, P., {Kirk}, J., {et~al.} 2013{\natexlab{a}},
  \aap, 550, A38

\bibitem[{{Palmeirim} {et~al.}(2013{\natexlab{b}}){Palmeirim}, {Andr{\'e}},
  {Kirk}, {Ward-Thompson}, {Arzoumanian}, {K{\"o}nyves}, {Didelon},
  {Schneider}, {Benedettini}, {Bontemps}, {Di Francesco}, {Elia}, {Griffin},
  {Hennemann}, {Hill}, {Martin}, {Men'shchikov}, {Molinari}, {Motte}, {Nguyen
  Luong}, {Nutter}, {Peretto}, {Pezzuto}, {Roy}, {Rygl}, {Spinoglio}, \&
  {White}}]{Palmeirim2013}
{Palmeirim}, P., {Andr{\'e}}, P., {Kirk}, J., {et~al.} 2013{\natexlab{b}},
  \aap, 550, A38

\bibitem[{{Parise} {et~al.}(2002){Parise}, {Ceccarelli}, {Tielens}, {Herbst},
  {Lefloch}, {Caux}, {Castets}, {Mukhopadhyay}, {Pagani}, \&
  {Loinard}}]{Parise2002}
{Parise}, B., {Ceccarelli}, C., {Tielens}, A.~G.~G.~M., {et~al.} 2002, \aap,
  393, L49

\bibitem[{{Parise} {et~al.}(2009){Parise}, {Leurini}, {Schilke}, {Roueff},
  {Thorwirth}, \& {Lis}}]{Parise2009}
{Parise}, B., {Leurini}, S., {Schilke}, P., {et~al.} 2009, \aap, 508, 737

\bibitem[{{Pickett} {et~al.}(1998){Pickett}, {Poynter}, {Cohen}, {Delitsky},
  {Pearson}, \& {M{\"u}ller}}]{JPL}
{Pickett}, H.~M., {Poynter}, R.~L., {Cohen}, E.~A., {et~al.} 1998, \jqsrt, 60,
  883

\bibitem[{{Priestley} {et~al.}(2018){Priestley}, {Viti}, \&
  {Williams}}]{Priestley2018}
{Priestley}, F.~D., {Viti}, S., \& {Williams}, D.~A. 2018, \aj, 156, 51

\bibitem[{{Remijan} {et~al.}(2007){Remijan}, {Markwick-Kemper}, \& {ALMA
  Working Group on Spectral Line Frequencies}}]{SLAIM}
{Remijan}, A.~J., {Markwick-Kemper}, A., \& {ALMA Working Group on Spectral
  Line Frequencies}. 2007, in American Astronomical Society Meeting Abstracts,
  Vol. 211, American Astronomical Society Meeting Abstracts, 132.11

\bibitem[{{Roberts} {et~al.}(2003){Roberts}, {Herbst}, \&
  {Millar}}]{Roberts2003}
{Roberts}, H., {Herbst}, E., \& {Millar}, T.~J. 2003, \apjl, 591, L41

\bibitem[{{Rodgers} \& {Charnley}(2008)}]{Rodgers2008}
{Rodgers}, S.~D. \& {Charnley}, S.~B. 2008, \mnras, 385, L48

\bibitem[{{Rodr{\'\i}guez-Baras} {et~al.}(2023){Rodr{\'\i}guez-Baras},
  {Esplugues}, {Fuente}, {et~al.}}]{GEMS-IX-MarinaH2S}
{Rodr{\'\i}guez-Baras}, M., {Esplugues}, G., {Fuente}, A., {et~al.} 2023, \aap,
  679, A120

\bibitem[{{Rodr{\'\i}guez-Baras} {et~al.}(2021){Rodr{\'\i}guez-Baras},
  {Fuente}, {Rivi{\'e}re-Marichalar}, {Navarro-Almaida}, {Caselli}, {Gerin},
  {Kramer}, {Roueff}, {Wakelam}, {Esplugues}, {Garc{\'\i}a-Burillo}, {Le Gal},
  {Spezzano}, {Alonso-Albi}, {Bachiller}, {Cazaux}, {Commercon}, {Goicoechea},
  {Loison}, {Trevi{\~n}o-Morales}, {Roncero}, {Jim{\'e}nez-Serra}, {Laas},
  {Hacar}, {Kirk}, {Lattanzi}, {Mart{\'\i}n-Dom{\'e}nech}, {Mu{\~n}oz-Caro},
  {Pineda}, {Tercero}, {Ward-Thompson}, {Tafalla}, {Marcelino}, {Malinen},
  {Friesen}, \& {Giuliano}}]{GEMS-IV}
{Rodr{\'\i}guez-Baras}, M., {Fuente}, A., {Rivi{\'e}re-Marichalar}, P.,
  {et~al.} 2021, \aap, 648, A120

\bibitem[{{Roueff} {et~al.}(2015){Roueff}, {Loison}, \& {Hickson}}]{Roueff2015}
{Roueff}, E., {Loison}, J.~C., \& {Hickson}, K.~M. 2015, \aap, 576, A99

\bibitem[{{Roueff} {et~al.}(2007){Roueff}, {Parise}, \& {Herbst}}]{Roueff2007}
{Roueff}, E., {Parise}, B., \& {Herbst}, E. 2007, \aap, 464, 245

\bibitem[{{Roueff} {et~al.}(2000){Roueff}, {Tin{\'e}}, {Coudert}, {Pineau des
  For{\^e}ts}, {Falgarone}, \& {Gerin}}]{Roueff2000}
{Roueff}, E., {Tin{\'e}}, S., {Coudert}, L.~H., {et~al.} 2000, \aap, 354, L63

\bibitem[{{Sakai} {et~al.}(2007){Sakai}, {Ikeda}, {Morita}, {Sakai}, {Takano},
  {Osamura}, \& {Yamamoto}}]{Sakai2007}
{Sakai}, N., {Ikeda}, M., {Morita}, M., {et~al.} 2007, \apj, 663, 1174

\bibitem[{{Sakai} {et~al.}(2010){Sakai}, {Saruwatari}, {Sakai}, {Takano}, \&
  {Yamamoto}}]{Sakai2010}
{Sakai}, N., {Saruwatari}, O., {Sakai}, T., {Takano}, S., \& {Yamamoto}, S.
  2010, \aap, 512, A31

\bibitem[{{Salinas} {et~al.}(2017){Salinas}, {Hogerheijde}, {Mathews},
  {{\"O}berg}, {Qi}, {Williams}, \& {Wilner}}]{Salinas2017}
{Salinas}, V.~N., {Hogerheijde}, M.~R., {Mathews}, G.~S., {et~al.} 2017, \aap,
  606, A125

\bibitem[{{Sch{\"o}ier} {et~al.}(2005){Sch{\"o}ier}, {van der Tak}, {van
  Dishoeck}, \& {Black}}]{col_N2H+}
{Sch{\"o}ier}, F.~L., {van der Tak}, F.~F.~S., {van Dishoeck}, E.~F., \&
  {Black}, J.~H. 2005, \aap, 432, 369

\bibitem[{{Shinnaka} {et~al.}(2016){Shinnaka}, {Kawakita}, {Jehin},
  {et~al.}}]{comets-Nratio}
{Shinnaka}, Y., {Kawakita}, H., {Jehin}, E., {et~al.} 2016, \mnras, 462, S195

\bibitem[{{Sipil{\"a}} \& {Caselli}(2018)}]{Sipila2018}
{Sipil{\"a}}, O. \& {Caselli}, P. 2018, \aap, 615, A15

\bibitem[{{Sipil{\"a}} {et~al.}(2023){Sipil{\"a}}, {Colzi}, {Roueff},
  {Caselli}, {Fontani}, \& {Wirstr{\"o}m}}]{Spilla2023}
{Sipil{\"a}}, O., {Colzi}, L., {Roueff}, E., {et~al.} 2023, \aap, 678, A120

\bibitem[{{Spezzano} {et~al.}(2022{\natexlab{a}}){Spezzano}, {Caselli},
  {Sipil{\"a}}, {et~al.}}]{Spezzano}
{Spezzano}, S., {Caselli}, P., {Sipil{\"a}}, O., {et~al.} 2022{\natexlab{a}},
  \aap, 664, L2

\bibitem[{{Spezzano} {et~al.}(2022{\natexlab{b}}){Spezzano}, {Fuente},
  {Caselli}, {Vasyunin}, {Navarro-Almaida}, {Rodr{\'\i}guez-Baras}, {Punanova},
  {Vastel}, \& {Wakelam}}]{GEMS-V}
{Spezzano}, S., {Fuente}, A., {Caselli}, P., {et~al.} 2022{\natexlab{b}}, \aap,
  657, A10

\bibitem[{{Tafalla} {et~al.}(2006){Tafalla}, {Santiago-Garc{\'\i}a}, {Myers},
  {Caselli}, {Walmsley}, \& {Crapsi}}]{Tafalla2006}
{Tafalla}, M., {Santiago-Garc{\'\i}a}, J., {Myers}, P.~C., {et~al.} 2006, \aap,
  455, 577

\bibitem[{{Takano} {et~al.}(1998){Takano}, {Masuda}, {Hirahara}, {Suzuki},
  {Ohishi}, {Ishikawa}, {Kaifu}, {Kasai}, {Kawaguchi}, \&
  {Wilson}}]{Takano1998}
{Takano}, S., {Masuda}, A., {Hirahara}, Y., {et~al.} 1998, \aap, 329, 1156

\bibitem[{Turner(2001)}]{Turner_2001}
Turner, B.~E. 2001, The Astrophysical Journal Supplement Series, 136, 579

\bibitem[{{Vera} {et~al.}(2014){Vera}, {Kalugina}, {Denis-Alpizar},
  {Stoecklin}, \& {Lique}}]{col_HCN}
{Vera}, M.~H., {Kalugina}, Y., {Denis-Alpizar}, O., {Stoecklin}, T., \&
  {Lique}, F. 2014, \jcp, 140, 224302

\bibitem[{{Wakelam} {et~al.}(2021){Wakelam}, {Gratier}, {Ruaud}, {Le Gal},
  {Majumdar}, {Loison}, \& {Hickson}}]{Wakelam2021}
{Wakelam}, V., {Gratier}, P., {Ruaud}, M., {et~al.} 2021, \aap, 647, A172

\bibitem[{{Wakelam} {et~al.}(2010){Wakelam}, {Herbst}, {Le Bourlot}, {Hersant},
  {Selsis}, \& {Guilloteau}}]{Wakelam2010}
{Wakelam}, V., {Herbst}, E., {Le Bourlot}, J., {et~al.} 2010, \aap, 517, A21

\bibitem[{{Wampfler} {et~al.}(2014){Wampfler}, {J{\o}rgensen}, {Bizzarro}, \&
  {Bisschop}}]{14N15N_4W}
{Wampfler}, S.~F., {J{\o}rgensen}, J.~K., {Bizzarro}, M., \& {Bisschop}, S.~E.
  2014, \aap, 572, A24

\bibitem[{{Watson}(1974)}]{Watson1974}
{Watson}, W.~D. 1974, \apj, 188, 35

\bibitem[{{Yoshida} {et~al.}(2019){Yoshida}, {Sakai}, {Nishimura}, {Tokudome},
  {Watanabe}, {Sakai}, {Takano}, \& {Yamamoto}}]{14N15N_3yoshida}
{Yoshida}, K., {Sakai}, N., {Nishimura}, Y., {et~al.} 2019, \pasj, 71, S18

\bibitem[{{Zari} {et~al.}(2016){Zari}, {Lombardi}, {Alves}, {Lada}, \&
  {Bouy}}]{Zari2016}
{Zari}, E., {Lombardi}, M., {Alves}, J., {Lada}, C.~J., \& {Bouy}, H. 2016,
  \aap, 587, A106

\bibitem[{{Zucker} {et~al.}(2019){Zucker}, {Speagle}, {Schlafly}, {Green},
  {Finkbeiner}, {Goodman}, \& {Alves}}]{Pers&Ori-d}
{Zucker}, C., {Speagle}, J.~S., {Schlafly}, E.~F., {et~al.} 2019, \apj, 879,
  125

\end{thebibliography}

\onecolumn
\begin{appendix}

\onecolumn
\section{Additional tables} \label{apend:Tab}
\counterwithin{table}{section}
\setcounter{table}{0}

\begin{table}[h!]
\caption{Spectroscopic information of the observed transitions in each project.} \label{tab:transitions}
\centering
\begin{tabular}{cccccccccc} \hhline{ =  =  =  =  =  =  =  =  =  = }
 Species     & $\nu$     & Resolved QNs: J & Log$_{10}$ A$_{\rm ij}$ & g$_{\rm u}$ & E$_{\rm u}$  & $\theta_{\rm MB}$ & F$_{\rm eff}$ & B$_{\rm eff}$ & Project  \\ 
     & (GHz)  &   &   &  & (K)  & ($''$)  & ($\%$) & ($\%$) & \\  \hline
\multirow{1}{*}{HC$^{15}$N} 
& 86.054966 & 1-0  & -4.6569 & 3 &  4.1300 & 29 & 95 & 81 & GEMS \\ \hline
\multirow{4}{*}{H$^{13}$CN} 
& 86.338735  &   1-0, F = 1-1  &  -4.6526 & 3 & 4.1436 &  \multirow{4}{*}{29} & \multirow{4}{*}{95} & \multirow{4}{*}{81} & \multirow{4}{*}{GEMS}  \\
& 86.339921\tablefootmark{a}
&   1-0  & -4.6526 & 9 & 4.1437 &  & & & \\
& 86.340167  &   1-0, F = 2-1   & -4.6526       &  5 & 4.1437  & & & & \\
& 86.342254  &   1-0, F = 0-1   &  -4.6525 &  1 &  4.1438 & & & & \\ \hline
\multirow{4}{*}{HCN} 
& 88.630416 &   1-0, F = 1-1  &-4.6184 &        3 & 4.2536 & \multirow{4}{*}{28} & \multirow{4}{*}{95} & \multirow{4}{*}{81} & \multirow{4}{*}{GEMS}  \\
\hline
\end{tabular}
\tablefoot{We provide rest frequencies (in GHz), resolved quantum numbers, Einstein A coefficients and degeneracies g$_{\rm u}$ and energies E$_{\rm u}$ (in Kelvin) of the upper levels. IRAM 30m efficiencies and beam widths are also included.\tablefoottext{a}{Weighted average if hyperfine structure were not taken into account.}}
\end{table}

\newpage 
\begin{table}[]
\caption{Parameters obtained from Gaussian fits.} \label{tab:gaussian_fits}
\resizebox{\textwidth}{!}{
\begin{tabular}{lcccclcccclcccclcccc}
\hhline{ =  =  =  =  =  =  =  =  =  =  =  =  =  =  =  =  =  =  =  = }
 & \multicolumn{4}{c}{DNC 1 $\rightarrow$ 0} &  & \multicolumn{4}{c}{DNC 2 $\rightarrow$ 1} &  &  \multicolumn{4}{c}{DN$^{13}$C 1 $\rightarrow$ 0} &  & \multicolumn{4}{c}{HN$^{13}$C 1   $\rightarrow$ 0}\\ \cline{2-5} \cline{7-10} \cline{12-15} \cline{17-20}
Starless   core & V$_{LSR}$ & $\Delta$V & $T_{MB}$ &  $\int T_{MB} \ d\nu$ &  & V$_{LSR}$ & $\Delta$V &  $T_{MB}$  & $\int T_{MB} \ d\nu$ &  & V$_{LSR}$ & $\Delta$V &$T_{MB}$ & $\int T_{MB} \ d\nu$ &  & V$_{LSR}$ & $\Delta$V &  $T_{MB}$   & $\int T_{MB} \ d\nu$ \\
 & (km s$^{-1}$) & (km s$^{-1}$) & (K) &  (K km s$^{-1}$) &  & (km s$^{-1}$) & (km s$^{-1}$) & (K)  & (K km s$^{-1}$) &  & (km s$^{-1}$) & (km s$^{-1}$) & (K) & (K km s$^{-1}$) &  & (km s$^{-1}$) & (km s$^{-1}$) &  (K) & (K km s$^{-1}$) \\ \hline 
 B213-C1-1 & 5.98 $\pm$ 0.01 & 1.03 $\pm$ 0.01 & 1.65  $\pm$  0.01 & 1.80 $\pm$ 0.01 &  & 5.99 $\pm$ 0.01 & 0.74 $\pm$ 0.01 & 0.98  $\pm$  0.01 & 0.77 $\pm$ 0.01 &  & 5.69 $\pm$ 0.03 & 0.75 $\pm$ 0.07 & 0.06  $\pm$  0.01 & 0.05 $\pm$ 0.01 &  & 6.01 $\pm$ 0.01 & 0.65 $\pm$ 0.02 & 0.68  $\pm$  0.02 & 0.47 $\pm$ 0.01 \\
B213-C2-1 & 7.07 $\pm$ 0.01 & 0.94 $\pm$ 0.01 & 1.46  $\pm$  0.01 & 1.46 $\pm$ 0.01 &  & 6.73 $\pm$ 0.01 & 0.74 $\pm$ 0.02 & 0.85  $\pm$  0.03 & 0.67 $\pm$ 0.02 &  & 6.88 $\pm$ 0.07 & 0.69 $\pm$ 0.16 & 0.06  $\pm$  0.02 & 0.04 $\pm$ 0.01 &  & 7.12 $\pm$ 0.01 & 0.61 $\pm$ 0.01 & 0.80  $\pm$  0.02 & 0.52 $\pm$ 0.01 \\
B213-C5-1 & 6.46 $\pm$ 0.01 & 0.98 $\pm$ 0.01 & 0.60  $\pm$  0.01 & 0.63 $\pm$ 0.01 &  & 6.44 $\pm$ 0.01 & 0.63 $\pm$ 0.02 & 0.47  $\pm$  0.02 & 0.31 $\pm$ 0.01 &  &  &  &  & $<$ 0.02 &  & 6.52 $\pm$ 0.02 & 0.71 $\pm$ 0.04 & 0.24  $\pm$  0.02 & 0.19 $\pm$ 0.01 \\
\hline 
\end{tabular}}
\end{table}

\begin{table}[]
\resizebox{0.76\textwidth}{!}{
\begin{tabular}{lcccclcccclcccc}
\hhline{ =  =  =  =  =  =  =  =  =  =  =  =  =  =  = }
 & \multicolumn{4}{c}{HNC 1 $\rightarrow$ 0} &  & \multicolumn{4}{c}{H$^{15}$NC 1 $\rightarrow$ 0} &  &  \multicolumn{4}{c}{HC$^{15}$N 1 $\rightarrow$ 0} \\ \cline{2-5} \cline{7-10} \cline{12-15}
Starless   core & V$_{LSR}$ & $\Delta$V & $T_{MB}$ &  $\int T_{MB} \ d\nu$ &  & V$_{LSR}$ & $\Delta$V &  $T_{MB}$  & $\int T_{MB} \ d\nu$ &  & V$_{LSR}$ & $\Delta$V &$T_{MB}$ & $\int T_{MB} \ d\nu$ \\
 & (km s$^{-1}$) & (km s$^{-1}$) & (K) &  (K km s$^{-1}$) &  & (km s$^{-1}$) & (km s$^{-1}$) & (K)  & (K km s$^{-1}$) &  & (km s$^{-1}$) & (km s$^{-1}$) & (K) & (K km s$^{-1}$) \\ \hline 
 B213-C1-1 & 5.67 $\pm$ 0.01& 0.86 $\pm$ 0.01 & 1.52 $\pm$ 0.02 & 1.39 $\pm$ 0.01 &  & 5.98 $\pm$ 0.01 & 0.47 $\pm$ 0.02 & 0.18 $\pm$ 0.01 & 0.09 $\pm$ 0.01&  & 5.98 $\pm$ 0.03 & 0.35 $\pm$ 0.06 & 0.09 $\pm$ 0.02 & 0.03 $\pm$ 0.01 \\
B213-C2-1 & 7.24 $\pm$ 0.01 & 0.83 $\pm$ 0.01 & 1.49 $\pm$ 0.03 & 1.32 $\pm$ 0.01 &  & 7.11 $\pm$ 0.01& 0.41 $\pm$ 0.01 & 0.28 $\pm$ 0.01 & 0.12 $\pm$ 0.01&  & 7.04 $\pm$ 0.01 & 0.35 $\pm$ 0.03 & 0.13 $\pm$ 0.01 & 0.05 $\pm$ 0.01\\
B213-C5-1 & 6.28 $\pm$ 0.01 & 0.98 $\pm$ 0.01 & 0.88 $\pm$ 0.01 & 0.92 $\pm$ 0.01 &  & 6.42 $\pm$ 0.03 & 0.40 $\pm$ 0.06 & 0.06 $\pm$ 0.01 & 0.03 $\pm$ 0.01&  &  &  &  &  $<$ 0.02  \\
B213-C6-1 & 7.01 $\pm$ 0.01 & 1.33 $\pm$ 0.02 & 0.84 $\pm$ 0.01 & 1.19 $\pm$ 0.01 &  & 7.00 $\pm$ 0.01 & 0.40 $\pm$ 0.02 & 0.23 $\pm$ 0.01 & 0.10 $\pm$ 0.01&  & 7.08 $\pm$ 0.03 & 0.28 $\pm$ 0.05 & 0.07 $\pm$ 0.02 & 0.02 $\pm$ 0.01\\
B213-C7-1 & 7.03 $\pm$ 0.01 & 0.94 $\pm$ 0.01 & 0.89 $\pm$ 0.02 & 0.89 $\pm$ 0.01 &  & 6.97 $\pm$ 0.01 & 0.37 $\pm$ 0.03 & 0.14 $\pm$ 0.01 & 0.06 $\pm$ 0.01&  & 6.84 $\pm$ 0.06 & 0.36 $\pm$ 0.12 & 0.04 $\pm$ 0.02 & 0.02 $\pm$ 0.01 \\
\hline 
\end{tabular}}
\end{table}
\begin{table}[h!]
\caption{Parameters obtained from hyperfine structure fits.} \label{tab:hfs_fits}
\resizebox{0.76\textwidth}{!}{
\begin{tabular}{lcccclcccclcccc}
\hhline{ =  =  =  =  =  =  =  =  =  =  =  =  =  =  = }
 & \multicolumn{4}{c}{H$^{13}$CN 1 $\rightarrow$ 0} &  & \multicolumn{4}{c}{DCN 2 $\rightarrow$ 1} &  &  \multicolumn{4}{c}{HCN 1 $\rightarrow$ 0} \\ \cline{2-5} \cline{7-10} \cline{12-15} 
Starless core & (T$_{ex}$ - T$_{bg}$) $ \cdot \ \tau $ & V$_{LSR}$ & $\Delta$V & $\tau_{main}$ &  & (T$_{ex}$ - T$_{bg}$) $ \cdot \ \tau $ & V$_{LSR}$ & $\Delta$V & $\tau_{main}$ &  & (T$_{ex}$ - T$_{bg}$) $ \cdot \ \tau $ & V$_{LSR}$ & $\Delta$V & $\tau_{main}$ \\
 & (K) & (km s$^{-1}$) & (km s$^{-1}$) &  &  & (K) & (km s$^{-1}$) & (km s$^{-1}$) &  &  & (K) & (km s$^{-1}$) & (km s$^{-1}$) &  \\ \hline 
B213-C1-1 & 0.28 $\pm$ 0.04 & 5.99 $\pm$ 0.01 & 0.39 $\pm$ 0.02 & 0.46 $\pm$ 0.41 &  & 0.53 $\pm$ 0.06 & 5.82 $\pm$ 0.01 & 0.25 $\pm$ 0.02 & 1.54 $\pm$ 0.33 &  & 5.14 $\pm$ 0.24 & 4.99 $\pm$ 0.01 & 0.37 $\pm$ 0.01 & 10.30 $\pm$ 0.53 \\
B213-C2-1 & 0.50 $\pm$ 0.10 & 7.08 $\pm$ 0.01 & 0.30 $\pm$ 0.02 & 1.40 $\pm$ 0.61 &  &  &  &  &  &  &  &  &  & \tablefootmark{a}*** \\
B213-C5-1 & 0.07 $\pm$ 0.01 & 6.35 $\pm$ 0.03 & 0.53 $\pm$ 0.07 & 0.10 $\pm$ 0.80 &  &   0.07        $\pm$   0.03 &  5.92    $\pm$   0.04 &  0.26    $\pm$         0.14 &  0.10    $\pm$   0.32 &  &  &  &  & ***  \\
B213-C6-1 & 0.22 $\pm$ 0.04 & 6.97 $\pm$ 0.01 & 0.42 $\pm$ 0.05 & 0.11 $\pm$ 3.42 &  & 0.42 $\pm$ 0.07 & 6.42 $\pm$ 0.01 & 0.29 $\pm$ 0.03 & 0.57 $\pm$ 0.35 &  &  &  &  & *** \\
B213-C7-1 & 0.23 $\pm$ 0.07 & 6.95 $\pm$ 0.01 & 0.35 $\pm$ 0.04 & 1.17 $\pm$ 0.90 &  & 0.20 $\pm$ 0.01 & 8.60 $\pm$ 0.02 & 0.86 $\pm$ 0.05 & 0.10 $\pm$ 0.03 &  &  &  &  & *** \\
\hline
\end{tabular}}
\end{table}
\begin{table*}[h!]
\resizebox{0.52\textwidth}{!}{
\begin{tabular}{lccccccccc}
\hhline{ =  =  =  =  =  =  =  =  =  = }
 & \multicolumn{4}{c}{N$_2$H$^+$ 1 $\rightarrow$ 0} &  & \multicolumn{4}{c}{N$_2$D$^+$ 2 $\rightarrow$ 1}  \\ \cline{2-5} \cline{7-10} 
Starless core & (T$_{ex}$ - T$_{bg}$) $ \cdot \ \tau $ & V$_{LSR}$ & $\Delta$V & $\tau_{main}$ &  & (T$_{ex}$ - T$_{bg}$) $ \cdot \ \tau $ & V$_{LSR}$ & $\Delta$V & $\tau_{main}$ \\
 & (K) & (km s$^{-1}$) & (km s$^{-1}$) &  &  & (K) & (km s$^{-1}$) & (km s$^{-1}$) &  \\ \hline
B213-C1-1 & 3.35 $\pm$ 0.05 & 5.92 $\pm$ 0.01 & 0.42 $\pm$ 0.01 & 2.08 $\pm$ 0.06 &  & 0.81 $\pm$ 0.02 & 5.70 $\pm$ 0.01 & 0.54 $\pm$ 0.02 & 0.10 $\pm$ 0.02 \\
B213-C2-1 & 4.27 $\pm$ 0.01 & 7.04 $\pm$ 0.01 & 0.32 $\pm$ 0.01 & 2.73 $\pm$ 0.01 &  & 0.55 $\pm$ 0.03 & 6.51 $\pm$ 0.01 & 0.39 $\pm$ 0.03 & 0.10 $\pm$ 0.08 \\
B213-C5-1 & 1.51 $\pm$ 0.02 & 6.40 $\pm$ 0.01 & 0.50 $\pm$ 0.01 & 1.01 $\pm$ 0.04 &  & 0.27 $\pm$ 0.01 & 6.17 $\pm$ 0.01 & 0.58 $\pm$ 0.04 & 0.10 $\pm$ 0.06 \\
B213-C6-1 & 3.49 $\pm$ 0.01 & 6.97 $\pm$ 0.01 & 0.37 $\pm$ 0.01 & 1.51 $\pm$ 0.01 &  & 0.96 $\pm$ 0.02 & 6.83 $\pm$ 0.01 & 0.48 $\pm$ 0.02 & 0.10 $\pm$ 0.03 \\
B213-C7-1 & 2.34 $\pm$ 0.05 & 6.84 $\pm$ 0.01 & 0.37 $\pm$ 0.01 & 1.29 $\pm$ 0.08 &  & 0.40 $\pm$ 0.02 & 6.65 $\pm$ 0.01 & 0.41 $\pm$ 0.02 & 0.10 $\pm$ 0.09 \\
\hline 
\end{tabular}}
\tablefoot{\tablefoottext{a}{Optically thick emission line and self-absorbed.} $\tau_{main}$ is the opacity of the main component. The optical depth of the most optically thin component is $\tau_{main}$ multiplied by 0.2 (in the case of H$^{13}$CN, HCN and N$_2$H$^+$), 0.01 (DCN) or 0.11 (N$_2$D$^+$).}
\end{table*}

\begin{table}[h!]
\caption{Density of H$_2$, column density and opacity of DNC and HN$^{13}$C.} \label{tab:nNtau-DNC}
\centering
\begin{tabular}{lccccc}
\hline\hline
Starless core & $n({\rm H_2})$            & $N({\rm DNC})$               & $\tau_{\rm DNC}$ & N(HN$^{13}$C)                & $\tau_{\rm HN^{13}C}$ \\
              & ($\times 10^5$ cm$^{-3}$) & ($\times 10^{12}$ cm$^{-2}$) &                  & ($\times 10^{12}$ cm$^{-2}$) &                       \\ \hline
B213-C16-1    & 1.0 $\pm$ 0.3             & 6.93 $\pm$ 2.24              & 2.59             & 1.72 $\pm$ 0.17              & 0.57                  \\
B213-C12-1    & 1.0 $\pm$ 0.3             & 3.58 $\pm$ 0.94              & 1.62             & 1.42 $\pm$ 0.14              & 0.52                  \\
B213-C1-1     & 0.6 $\pm$ 0.1             & 8.0 $\pm$ 2.3                & 3.13             & 1.36 $\pm$ 0.14              & 0.59                  \\
B213-C6-1     & 1.2 $\pm$ 0.3             & 6.4 $\pm$ 1.8               & 1.84             & 0.99 $\pm$ 0.10               & 0.34                  \\
B213-C2-1     & 0.8 $\pm$ 0.2             & 5.0 $\pm$ 1.3               & 2.36             & 1.34 $\pm$ 0.14              & 0.55                  \\
B213-C7-1     & 0.6 $\pm$ 0.4             & 3.80 $\pm$ 0.95               & 1.97             & 0.66 $\pm$ 0.07              & 0.34                  \\
\hline
\end{tabular}
\end{table}

\begin{table}[h!]
\caption{Column densities of molecules with unresolved hyperfine structure, N (in cm$^{-2}$).}
\label{tab:N-nohfs}
\resizebox{\textwidth}{!}{
\begin{tabular}{lccccccc|} 
\hhline{ =  =  =  =  =  =  = }
Starless core  & N(DNC) & N(DN$^{13}$C) & N(HN$^{13}$C)  & N(HC$^{15}$N) & N(H$^{15}$NC) & N(HNC)\\ \hline 
B213-C16-1  & (6.9 $\pm$ 2.2)  $10^{12}$ & (1.51 $\pm$ 0.21)  $10^{11}$ & (1.72 $\pm$ 0.17)  $10^{12}$ & (2.82 $\pm$ 0.42)  $10^{11}$ & (4.12 $\pm$ 0.42)  $10^{11}$ & (4.87 $\pm$ 0.49)  $10^{12}$ \\
B213-C12-1 & (3.58 $\pm$ 0.94)  $10^{12}$ & (7.2 $\pm$ 1.4)  $10^{10}$ & (1.42 $\pm$ 0.14)  $10^{12}$ & (1.17 $\pm$ 0.26)  $10^{11}$ & (2.87 $\pm$ 0.31)  $10^{11}$ & (6.99 $\pm$ 0.70)  $10^{12}$ \\
B213-C1-1 & (8.0 $\pm$ 2.3)  $10^{12}$ & (1.24 $\pm$ 1.6)  $10^{11}$ & (1.36 $\pm$ 0.14)  $10^{12}$ & (2.70 $\pm$ 0.50)  $10^{11}$ & (2.81 $\pm$ 0.30)  $10^{11}$ & (5.54 $\pm$ '.56)  $10^{12}$ \\
B213-C6-1  & (6.4 $\pm$ 1.8)  $10^{12}$ & (1.21 $\pm$ 0.19)  $10^{11}$ & (9.9 $\pm$ 1.0)  $10^{11}$ & (9.7 $\pm$ 2.2)  $10^{10}$ & (2.24 $\pm$ 0.24)  $10^{11}$ & (2.53 $\pm$ 0.25)  $10^{12}$ \\
B213-C2-1  & (5.0 $\pm$ 1.3)  $10^{12}$ & (1.03 $\pm$ 0.23)  $10^{11}$ & (1.34 $\pm$ 0.14)  $10^{12}$ & (3.26 $\pm$ 0.39)  $10^{11}$ & (3.34 $\pm$ 0.34)  $10^{11}$ & (4.31 $\pm$ 0.43)  $10^{12}$ \\
B213-C7-1  & (3.80 $\pm$ 0.95)  $10^{12}$ &$<$ 3.97 $10^{10}$ & (6.61 $\pm$ 0.69)  $10^{11}$ & (1.19 $\pm$ 0.40)  $10^{11}$ & (1.61 $\pm$ 0.19)  $10^{11}$ & (2.77 $\pm$ 0.28)  $10^{12}$ \\
B213-C10-1  & (5.3 $\pm$ 1.6)  $10^{12}$ & (6.5 $\pm$ 1.1)  $10^{10}$ & (1.26 $\pm$ 0.13)  $10^{12}$ & (1.53 $\pm$ 0.25)  $10^{11}$ & (2.19 $\pm$ 0.28)  $10^{11}$ & (5.62 $\pm$ 0.56)  $10^{12}$ \\
\hline
\end{tabular}}
\end{table}

\begin{table}[h!]
\centering
\caption{Column densities of molecules with resolved hyperfine structure, N (in cm$^{-2}$).} \label{tab:N-hfs} 
\begin{tabular}{lccccc}
\hline\hline
Starless core &
  N(H$^{13}$CN) &
  N(DCN) &
  N(HCN) &
  N$_2$H$^+$ & 
  N$_2$D$^+$\\
\hline
B213-C16-1 & (1.73 $\pm$ 0.17) 10$^{12}$ & (5.96 $\pm$ 0.60) 10$^{12}$ &  \tablefootmark{a}*** & (6.67 $\pm$ 0.67) 10$^{12}$ & (1.83 $\pm$ 0.19) 10$^{12}$ \\
B213-C12-1 & (1.20 $\pm$ 0.12) 10$^{12}$ & (1.71 $\pm$ 0.17) 10$^{12}$ & *** & (7.15 $\pm$ 0.72) 10$^{12}$ & (3.88 $\pm$ 0.42) 10$^{11}$ \\
B213-C1-1 & (1.68 $\pm$ 0.17) 10$^{12}$ & (3.83 $\pm$ 0.38) 10$^{12}$ & (3.47 $\pm$ 0.35) 10$^{13}$ & (9.46 $\pm$ 0.95) 10$^{12}$ & (3.87 $\pm$ 0.40) 10$^{12}$ \\
B213-C6-1 & (7.92 $\pm$ 0.79) 10$^{11}$ & (3.31 $\pm$ 0.77) 10$^{12}$ & *** & (6.34 $\pm$ 0.63) 10$^{12}$ & (2.29 $\pm$ 0.24) 10$^{12}$ \\
B213-C2-1 & (1.24 $\pm$ 0.12) 10$^{12}$ & - &  & (6.56 $\pm$ 0.66) 10$^{12}$ & (6.56 $\pm$ 0.82) 10$^{11}$ \\
B213-C7-1 & (1.00 $\pm$ 0.13) 10$^{12}$ & (1.12 $\pm$ 0.11) 10$^{13}$ & *** & (6.74 $\pm$ 0.67) 10$^{12}$ & (8.59 $\pm$ 0.99) 10$^{11}$ \\
B213-C10-1 & (1.11 $\pm$ 0.11) 10$^{12}$ & (3.23 $\pm$ 0.32) 10$^{12}$ & *** & (6.63 $\pm$ 0.66) 10$^{12}$ & (8.91 $\pm$ 0.91) 10$^{11}$ \\
\hline
\end{tabular}
\tablefoot{\tablefoottext{a}{Optically thick emission line and self-absorbed.}}
\end{table}

\begin{table}[h!]
\caption{$^{12}$C/$^{13}$C and isomeric ratios, with the mean and standard deviation values.} \label{tab:12-13Cratio+isomers}
\begin{tabular}{lccccccccc}
\hhline{ =  =  =  =  =  =  =  =  =  = }
 & \multicolumn{3}{c}{$^{12}$C/$^{13}$C} &  &  & \multicolumn{4}{c}{Isomers} \\ \cline{2-4} \cline{7-10} 
\multirow{2}{*}{Starless core} & \multirow{2}{*}{$\frac{HCN}{H^{13}CN}$} & \multirow{2}{*}{$\frac{DNC}{DN^{13}C}$} & \multirow{2}{*}{$\frac{HNC}{HN^{13}C}$} &  &  & \multirow{2}{*}{$\frac{H^{13}CN}{HN^{13}C}$} & \multirow{2}{*}{$\frac{HC^{15}N}{H^{15}NC}$} & \multirow{2}{*}{$\frac{DCN}{DNC}$} & \multirow{2}{*}{$\frac{DCN}{(DN^{13}C\times 68)}$} \\
 &  &  &  &  &  &  &  &  &  \\ \hline
B213-C16-1                                 & -                                       & 46 $\pm$ 16                             & 2.83 $\pm$ 0.40                         &  &  & 1.01 $\pm$ 0.14                                  & 0.68 $\pm$ 0.12                                  & 0.86 $\pm$ 0.29                        & 0.58 $\pm$ 0.10                                        \\
B213-C12-1                                 & -                                       & 50 $\pm$ 16                             & 4.92 $\pm$ 0.70                         &  &  & 0.85 $\pm$ 0.12                                  & 0.41 $\pm$ 0.10                                  & 0.48 $\pm$ 0.13                        & 0.351 $\pm$ 0.076                                        \\
B213-C1-1                                  & 20.7 $\pm$ 2.9                          & 65 $\pm$ 20                             & 4.07 $\pm$ 0.58                         &  &  & 1.24 $\pm$ 0.18                                  & 0.96 $\pm$ 0.21                                  & 0.48 $\pm$ 0.15                        & 0.454 $\pm$ 0.075                                        \\
\hline
\end{tabular}
\end{table}

\begin{table}[h!]
\caption[l]{Deuterium fraction ratios, with the mean and standard deviation values.} \label{tab:deut}
\centering
\resizebox{\textwidth}{!}{
\begin{tabular}{lcccccc}
\hline\hline
\multirow{2}{*}{Starless core}  & \multirow{2}{*}{$\frac{DNC}{(HN^{13}C\times 68)}$} & \multirow{2}{*}{$\frac{DNC}{HNC}$} &  \multirow{2}{*}{$\frac{DN^{13}C}{HN^{13}C}$}  & \multirow{2}{*}{$\frac{DCN}{(H^{13}CN\times 68)}$} & \multirow{2}{*}{$\frac{DCN}{HCN}$} & \multirow{2}{*}{$\frac{N_2D^+}{N_2H^+}$} \\ 
    &  &  &  &  &  &  \\ \hline
B213-C16-1 & 0.06 $\pm$ 0.02 & 1.4 $\pm$ 0.5 & 0.09 $\pm$ 0.01 & 0.051 $\pm$ 0.007 &  & 0.27 $\pm$ 0.04 \\
B213-C12-1 & 0.04 $\pm$ 0.01 & 0.5 $\pm$ 0.1 & 0.05 $\pm$ 0.01 & 0.021 $\pm$ 0.003 &  & 0.054 $\pm$ 0.008 \\
B213-C1-1 & 0.09 $\pm$ 0.03 & 1.4 $\pm$ 0.4 & 0.09 $\pm$ 0.01 & 0.034 $\pm$ 0.005 & 0.11 $\pm$ 0.02 & 0.41 $\pm$ 0.06 \\
B213-C6-1 & 0.10 $\pm$ 0.03 & 2.5 $\pm$ 0.8 & 0.12 $\pm$ 0.02 & 0.06 $\pm$ 0.02 &  & 0.36 $\pm$ 0.05 \\
B213-C2-1 & 0.06 $\pm$ 0.01 & 1.2 $\pm$ 0.3 & 0.08 $\pm$ 0.02 &  &  & 0.10 $\pm$ 0.02 \\
B213-C7-1 & 0.09 $\pm$ 0.02 & 1.4 $\pm$ 0.4 & $<$ 0.060  & 0.16 $\pm$ 0.03 &  & 0.13 $\pm$ 0.02 \\
\hline
\end{tabular}}
\end{table}

\begin{table}[h!]
\caption{$^{14}$N/$^{15}$N ratios, with the mean and standard deviation values.} \label{tab:14-15N}
\centering
\begin{tabular}{lcccccc}  \hhline{ =  =  =  =  =  =  = }
\multirow{2}{*}{Starless core} & \multirow{2}{*}{$\frac{HN^{13}C}{H^{15}NC}$} & \multirow{2}{*}{$\frac{(HN^{13}C\times 68)}{H^{15}NC}$} & \multirow{2}{*}{$\frac{HNC}{H^{15}NC}$} & \multirow{2}{*}{$\frac{H^{13}CN}{HC^{15}N}$} & \multirow{2}{*}{$\frac{(H^{13}CN\times 68)}{HC^{15}N}$} & \multirow{2}{*}{$\frac{HCN}{HC^{15}N}$} \\
 &  &  &  &  &  &  \\ \hline
\hline
B213-C16-1 & 4.20 $\pm$ 0.60 & 280 $\pm$ 40 & 11.8 $\pm$ 1.7 & 6.1 $\pm$ 1.1 & 420 $\pm$ 70 &  \\
B213-C12-1 & 4.90 $\pm$ 0.70 & 340 $\pm$ 50 & 24 $\pm$ 4 & 10 $\pm$ 2 & 700 $\pm$ 170 &  \\
B213-C1-1 & 4.80 $\pm$ 0.70 & 330 $\pm$ 50 & 20 $\pm$ 3 & 6.2 $\pm$ 1.3 & 420 $\pm$ 90 & 130 $\pm$ 30 \\
B213-C6-1 & 4.40 $\pm$ 0.70 & 300 $\pm$ 40 & 11.3 $\pm$ 1.7 & 8 $\pm$ 2 & 560 $\pm$ 140 &  \\
B213-C2-1 & 4.00 $\pm$ 0.60 & 270 $\pm$ 40 & 12.9 $\pm$ 1.9 & 3.80 $\pm$ 0.60 & 260 $\pm$ 40 &  \\
B213-C7-1 & 4.10 $\pm$ 0.60 & 280 $\pm$ 40 & 17 $\pm$ 3 & 8 $\pm$ 3 & 600 $\pm$ 200 &  \\
\hline
\end{tabular} 
\end{table}

\begin{table}[]
\caption{Main production and destruction reactions of N$_2$D$^+$, DNC, and DCN at 2$\times$10$^5$ yr and 10$^6$ yr for models C and D.}
\label{tab:model-reactions}
\resizebox{\textwidth}{!}{
\begin{tabular}{llllll} \hhline{ =  =  =  =  =  = }
      & \multicolumn{1}{c}{}        & \multicolumn{2}{c}{t = 2$\times$10$^5$yr}                                                                                            & \multicolumn{2}{c}{t = 10$^6$yr}                                                                                                 \\
Model & \multicolumn{1}{c}{Species} & \multicolumn{1}{c}{production}                          & \multicolumn{1}{c}{destruction}                                   & \multicolumn{1}{c}{production}                             & \multicolumn{1}{c}{destruction}                                   \\ \hline
C     & N$_2$D$^+$                  & N$_2$ + o-H$_2$D$^+$ $\rightarrow$ o-H$_2$ + N$_2$D$^+$   & N$_2$D$^+$ + e$^-$ $\rightarrow$ D + N$_2$                       & N$_2$ + mD$_3^+$ $\rightarrow$ oD$_2$ + N$_2$D$^+$         & N$_2$D$^+$ + e$^-$ $\rightarrow$ D + N$_2$                       \\
      &                             & N$_2$ + p-H$_2$D$^+$ $\rightarrow$ p-H$_2$ + N$_2$D$^+$    & N$_2$D$^+$ + CO $\rightarrow$ DCO$^+$ + N$_2$                    & N$_2$ + o-D$_2$H$^+$ $ \rightarrow$ HD + N$_2$D$^+$     &                                                                   \\
      &                             &                                                         &                                                                   &                                                            &                                                                   \\
      & DNC                         & HCND$^+$ + e$^-$ $\rightarrow$ H + DNC                  & DNC + p-H$_3^+$ $\rightarrow$ HD + HCNH$^+$                                      & HCND$^+$ + e$^-$ $\rightarrow$ H + DNC                     & DNC + H$^+$ $\rightarrow$ H + DNC$^+$                        \\
\hline                                                                
\end{tabular}}
\end{table}

\twocolumn
\section{Additional figures}

\begin{figure}[h!]
    \centering
    \includegraphics[width = 0.5\textwidth]{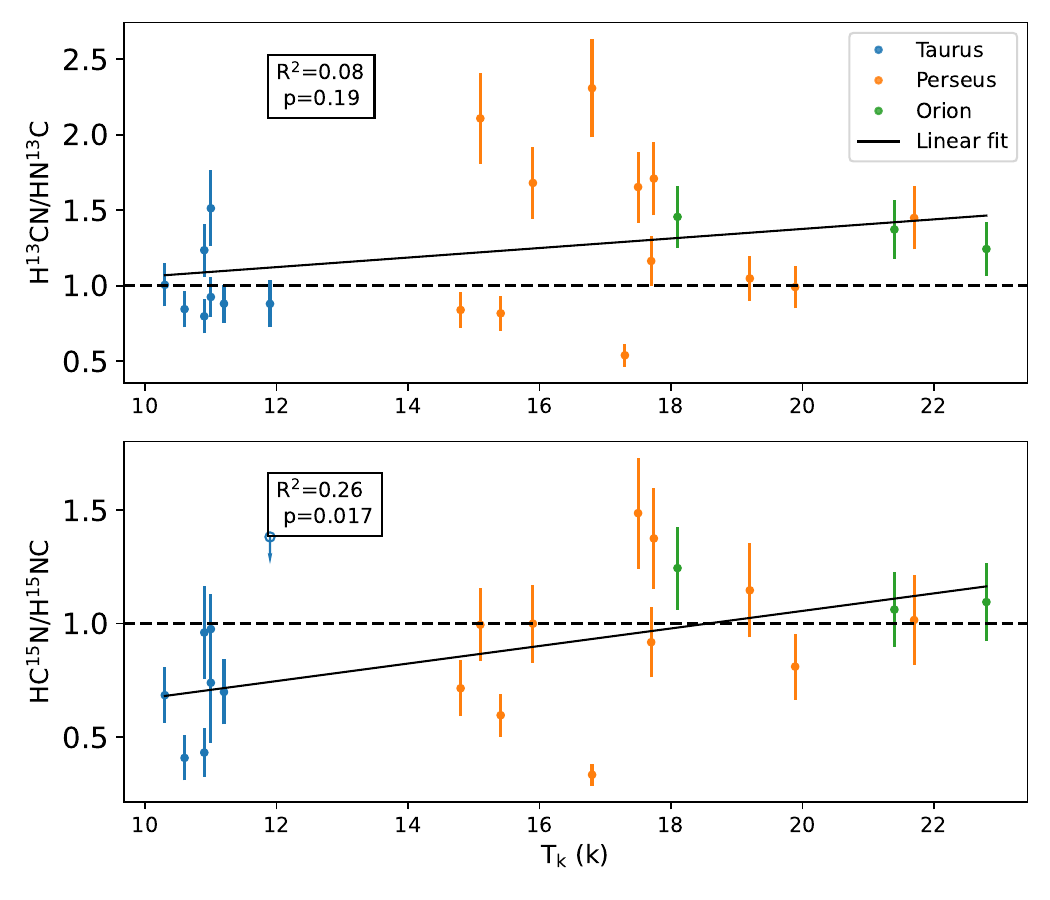}
    \caption{Comparison between isomeric ratios and gas temperature. A linear fit, the correlation coefficient R$^2$ and the p-value  are also represented.}
    \label{fig:isom-T}
\end{figure}

\begin{figure}[h!]
    \centering
    \includegraphics[width = 0.5\textwidth]{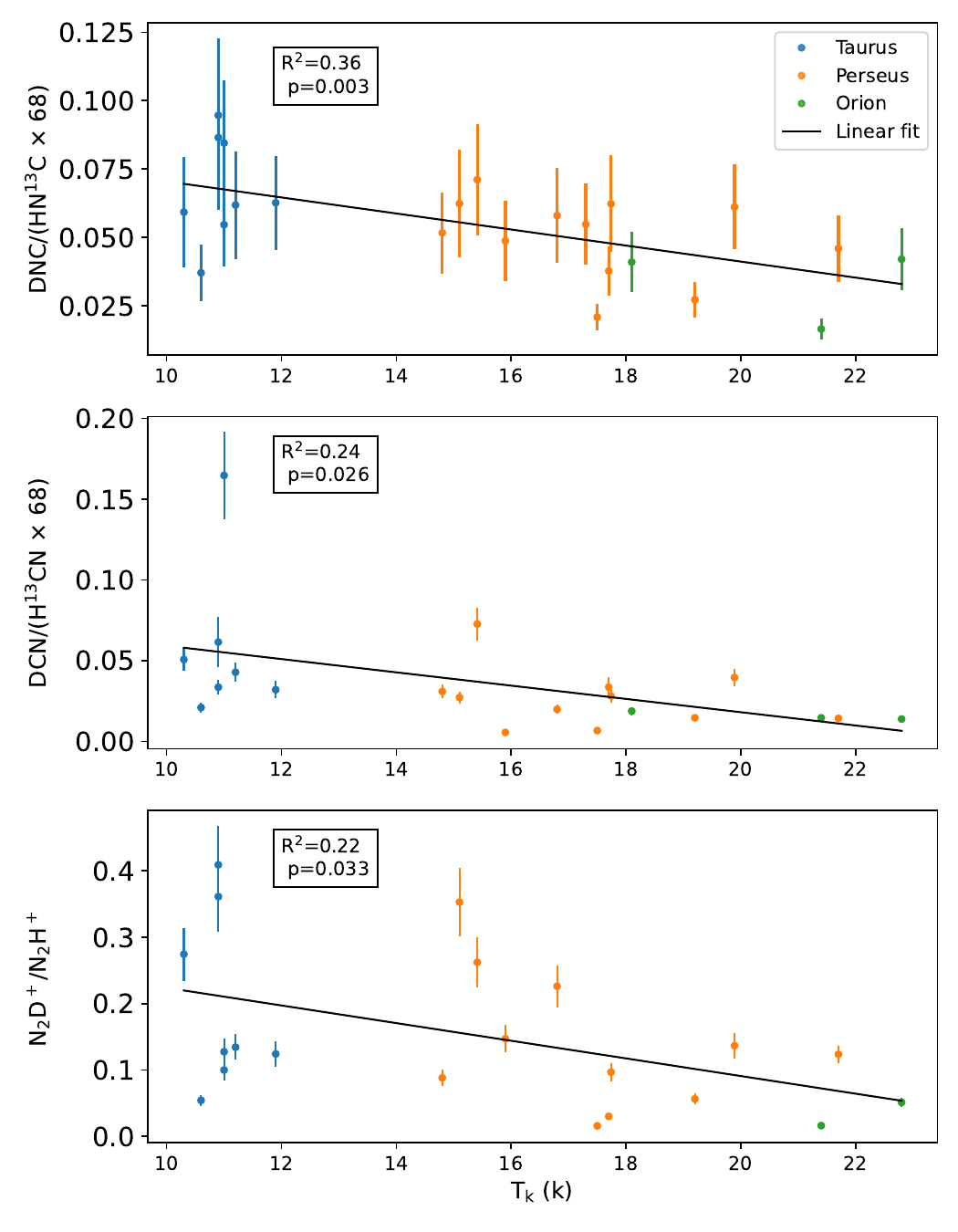}
    \caption{Comparison between the deuterium fractionation and gas temperature. A linear fit, the correlation coefficient R$^2,$ and the p-value  are also represented.}
    \label{fig:Dfrac-T}
\end{figure}

\begin{figure}[h!]
    \centering
    \includegraphics[width = 0.5\textwidth]{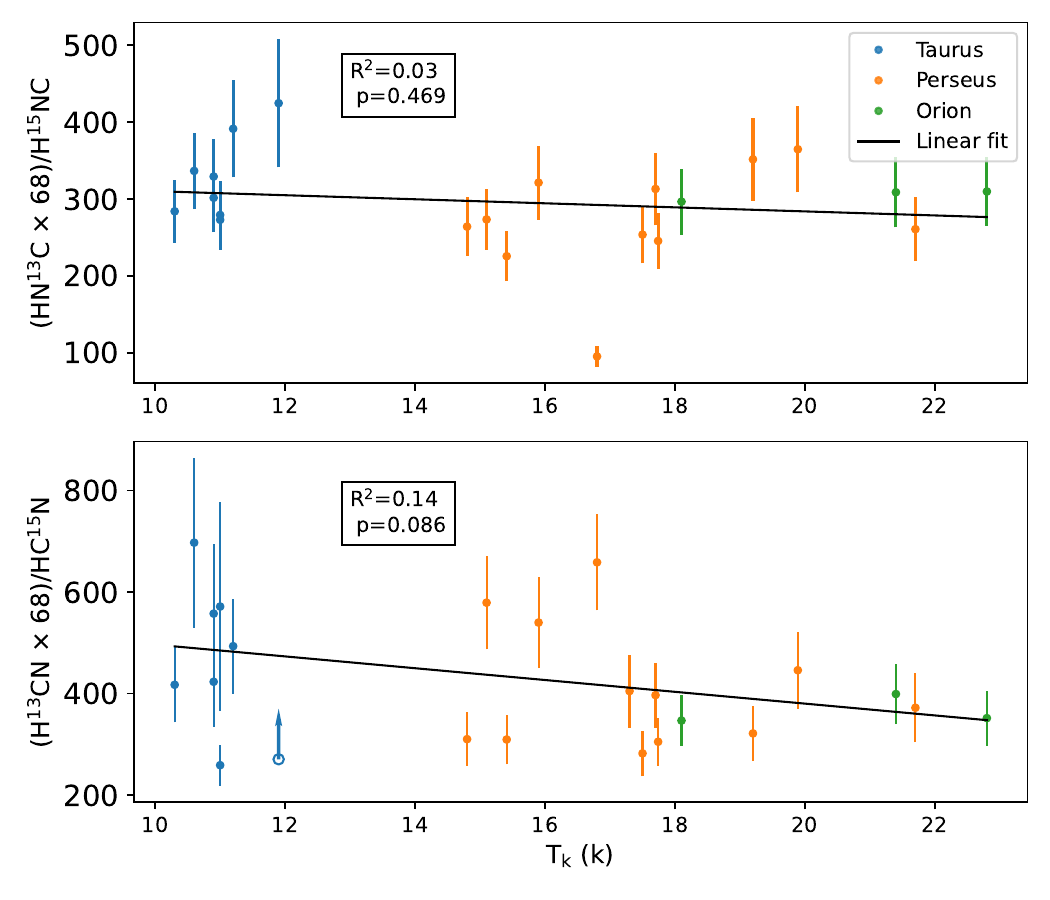}
    \caption{Comparison between the $^{14}$N/$^{15}$N ratio and gas temperature. A linear fit, the correlation coefficient R$^2,$ and the p-value  are also represented.}
    \label{fig:14N/15N-T}
\end{figure}

\begin{figure}[h!]
    \centering
    \includegraphics[width = 0.5\textwidth]{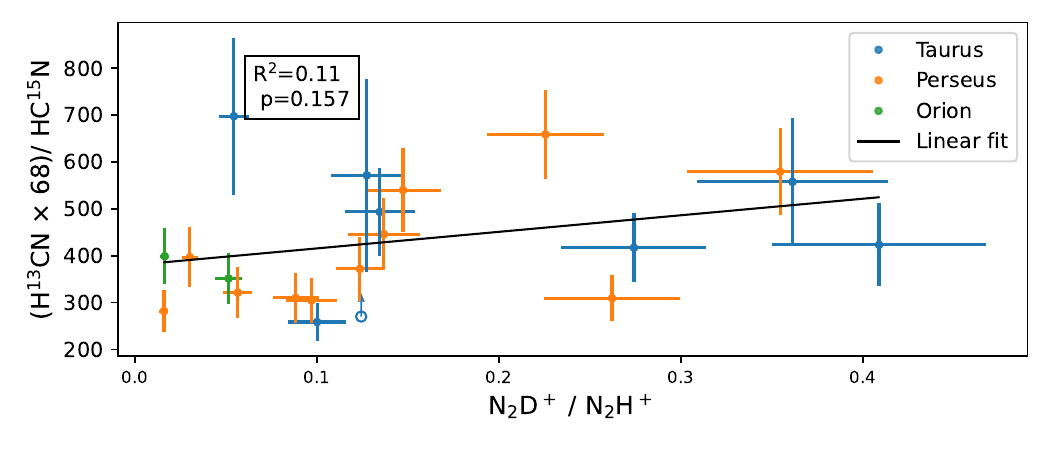}
    \caption{Comparison between the $^{14}$N/$^{15}$N ratio and deuterium fractionation. A linear fit, the correlation coefficient R$^2,$ and the p-value  are also represented.}
    \label{fig:Dfrac-14N/15N}
\end{figure}

\begin{figure}[h!]
    \centering
    \includegraphics[width = \hsize]{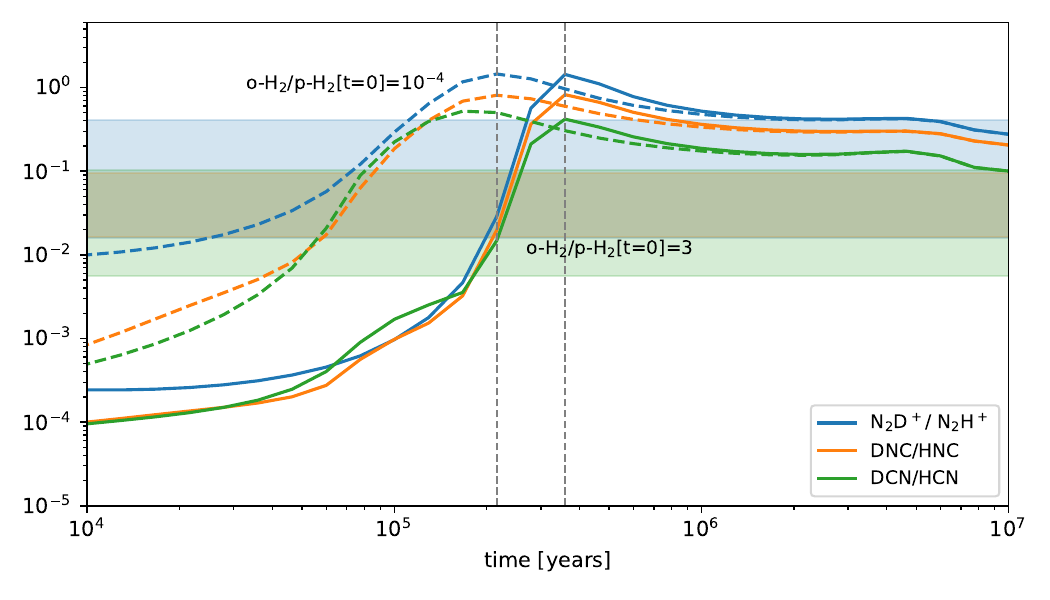}
    \caption{Comparison of the evolution of D$_{frac}$ up to 10 Myr according to the initial ortho-to-para ratio of H$_2$. Shaded horizontal bands indicate the observed range of the deuteration values for N$_2$H$^+$ (blue), HNC (orange), and HCN (green). Vertical dashed lines indicate the time when steady state is reached.}
    \label{fig:o-pH2}
\end{figure}

\begin{figure*}[h!]
    \centering
    \includegraphics[width = \hsize]{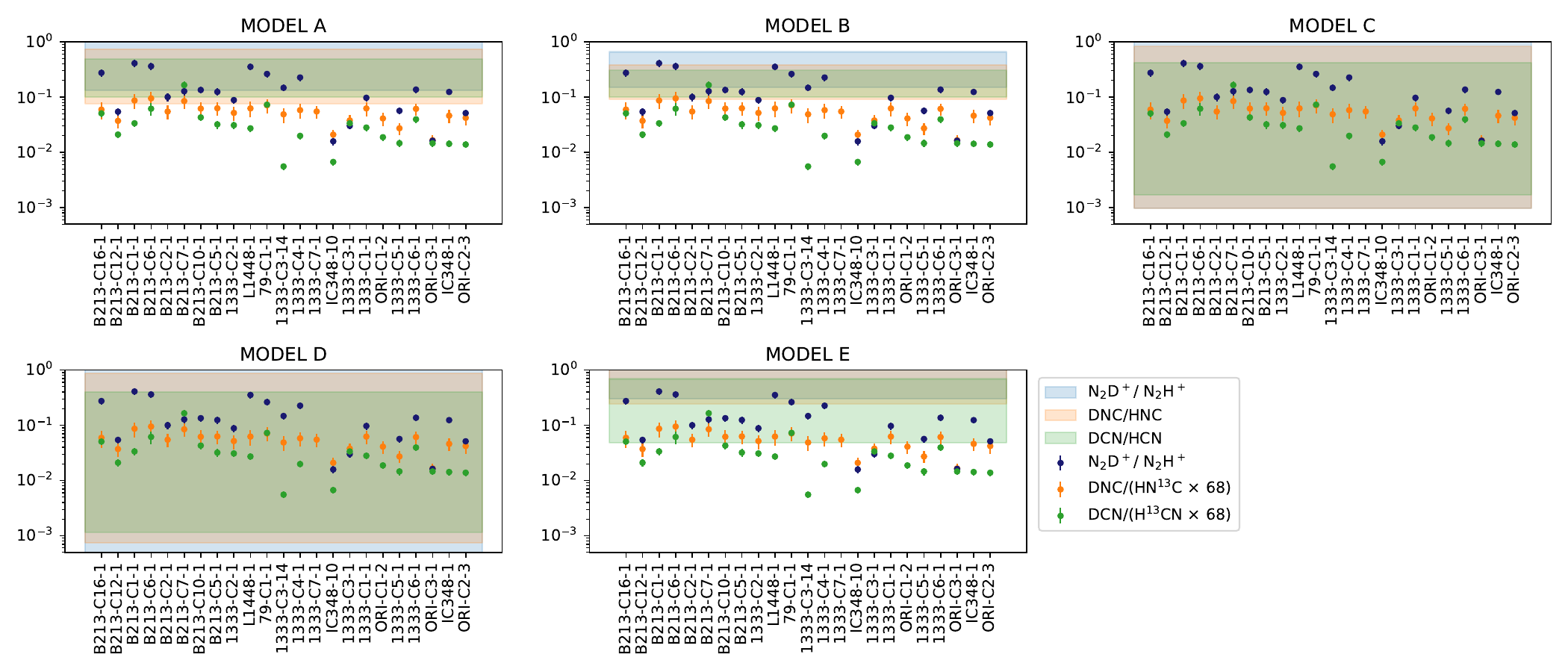}
    \caption{
    Comparison between observations and chemical model predictions for chemical ages between 0.1 and 10 Myr. Blue, orange, and green dots represent the different observed deuterium fraction ratios across the core sample. The shadowed areas show the predicted range of values of the ratios from 0.1 Myr to 10 Myr. The sources are ordered by increasing kinetic temperature.}
    \label{fig:models_obs}
\end{figure*}

\begin{figure*}[h!]
    \centering
\includegraphics[width = \textwidth]{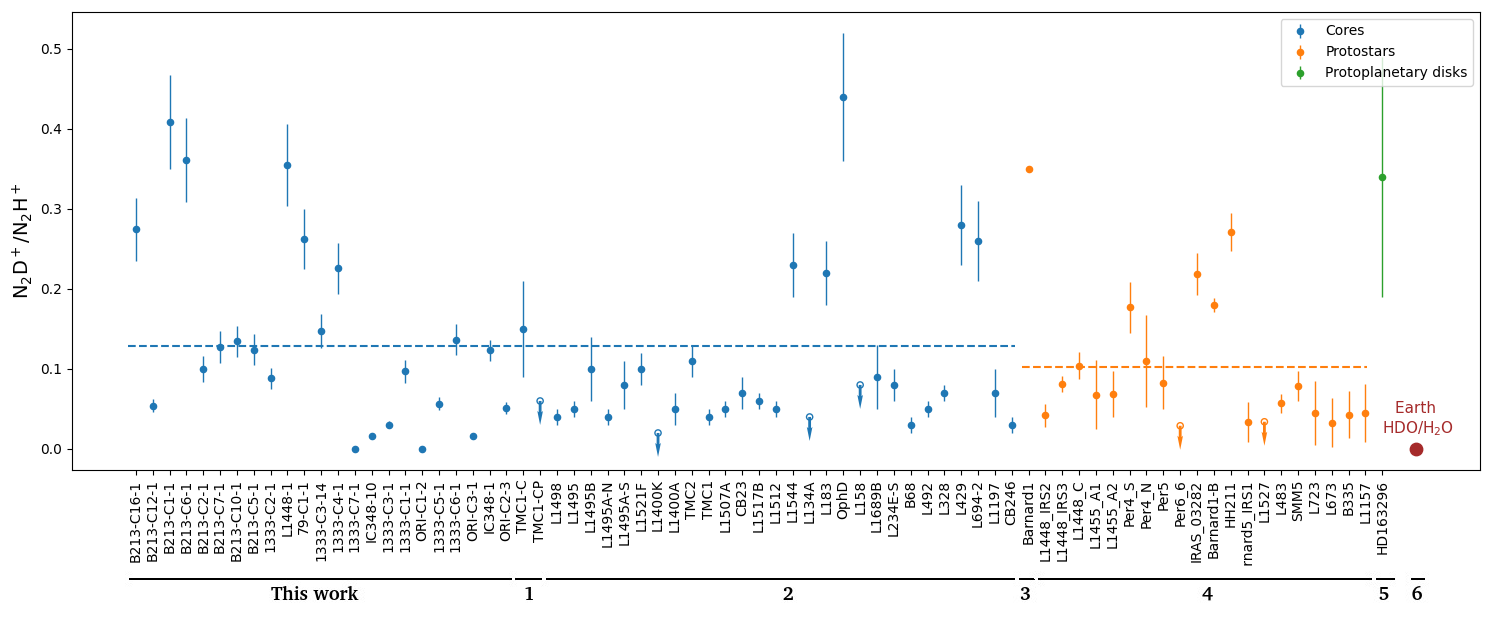}
    \caption{N$_2$D$^+$/N$_2$H$^+$ at different evolutionary stages from cores to protoplanetary disks. Measure of HDO/H$_2$O from Earth is also considered. Dashed lines represent the average values of each state. \textbf{References:} (1) \citealp{Navarro2021}; (2) \citealp{Crapsi_2005}; (3) \citealp{Daniel2013}; (4) \citealp{M.Emprechtinger2009}; (5) \citealp{Salinas2017}; (6) \citealp{Lecuyer1998}.}
    \label{fig:deut_comp_sour}
\end{figure*}

\begin{figure*}[h!]
    \centering
\includegraphics[width = 0.98\textwidth]{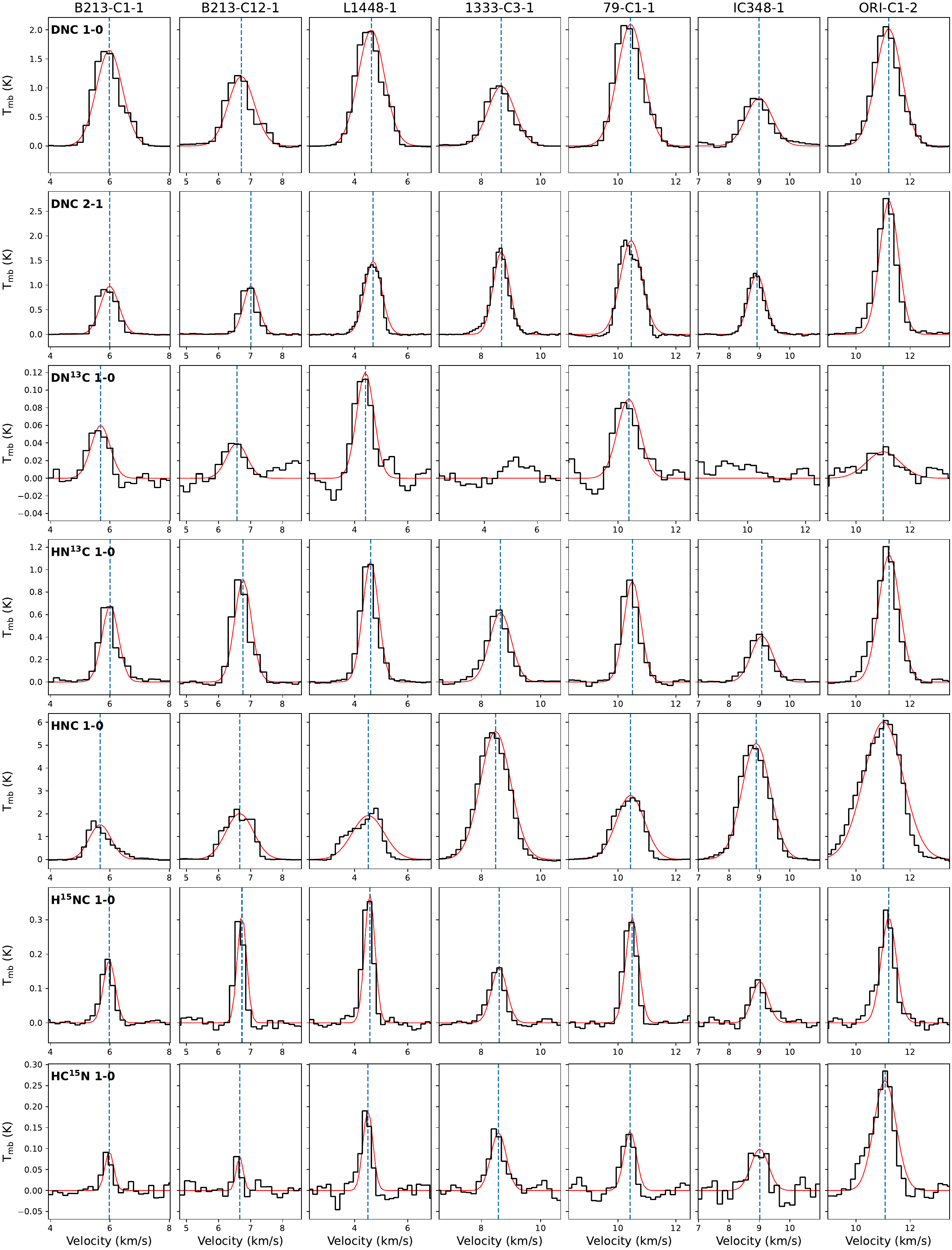} 
\caption{Line spectra of molecules with unresolved hyperfine structure toward some of the starless cores. The solid red lines show the RADEX best fit to compute column densities.}\label{fig:radex-fits}
\end{figure*}

\end{appendix}

\end{document}